\documentclass[
reprint, 
amsmath, 
amssymb, 
aps, 
prb, 
floatfix,
superscriptaddress
]{revtex4-2}
\pdfoutput=1

\usepackage{soul}

\usepackage[T1]{fontenc} 
\usepackage{lmodern}
\usepackage{amsmath}
\usepackage{array}
\usepackage{dsfont}
\usepackage{mathtools}
\usepackage{graphicx}
\usepackage{dcolumn}
\usepackage{bm}
\usepackage{float}
\usepackage{hyperref}
\hypersetup{
    colorlinks=true,
    linkcolor=blue,
    citecolor=blue,
    urlcolor=blue       
}
\usepackage{physics}
\usepackage{empheq}
\usepackage{color}
\usepackage{multirow}
\usepackage{orcidlink}
\usepackage{afterpage}
\usepackage{mathrsfs}
\usepackage{tikz}
\usetikzlibrary{shapes}
\usetikzlibrary{tikzmark,calc}

\newcommand{\bs}[1]{\boldsymbol{#1}}

\allowdisplaybreaks

\begin{document}

\title{
%Spontaneous and Explicit Continuous Translation Symmetry Breaking in the Presence of 
Electronic Crystal Phases in the Presence of
Non-Uniform Berry Curvature and Tunable Berry Flux
: The \texorpdfstring{$\boldsymbol{\lambda}_{\boldsymbol{N}}$}{$\lambda_N$}-Jellium model
}

\author{F\'elix Desrochers\orcidlink{0000-0003-1211-901X}}
\altaffiliation{These authors contributed equally to this work.}
\affiliation{%
Department of Physics, University of Toronto, Toronto, Ontario M5S 1A7, Canada
}%
\affiliation{%
Department of Physics, Harvard University, Cambridge, MA 02138, USA
}%

%%%
\author{Joe Huxford\orcidlink{0000-0002-4857-0091}}
\altaffiliation{These authors contributed equally to this work.}
\affiliation{%
Department of Physics, University of Toronto, Toronto, Ontario M5S 1A7, Canada
}%
\affiliation{Department of Physics and Astronomy, The University of Manchester, Oxford Road, Manchester M13 9PL, United Kingdom}

\author{Mark R. Hirsbrunner\orcidlink{0000-0001-8115-6098}}
\affiliation{%
Department of Physics, University of Toronto, Toronto, Ontario M5S 1A7, Canada
}%

\author{Yong Baek Kim}
\affiliation{%
Department of Physics, University of Toronto, Toronto, Ontario M5S 1A7, Canada
}%

\date{\today}

\begin{abstract}
    Recent experiments on multilayer graphene systems have rekindled interest in electronic crystal phases in two dimensions---but now for phases enriched by non-trivial quantum geometry. In this work, we introduce a simple continuum model with tunable Berry curvature distribution and total flux, enabling systematic study of crystallization in geometrically nontrivial bands. In the noninteracting limit, the addition of a $C_6$-symmetric periodic potential yields a rich phase diagram, for which we provide several analytical insights. Notably, we derive a general formula for the Chern number in the weak-potential regime that is broadly applicable to single-band projected models. Removing the periodic potential and treating Coulomb interactions self-consistently at the Hartree–Fock level, the resulting phase diagrams host a variety of crystalline states, including anomalous Hall crystals, halo Wigner crystals in which localized electrons spontaneously acquire orbital angular momentum leading to depleted electron occupation at the zone center, and a novel halo anomalous Hall crystal that combines these properties with a finite Chern number. We identify why these phases are energetically favorable through analytical and energetic considerations. Our results provide insight into the interplay between crystallization and band geometry, while also offering a simple toy model amenable to numerical methods beyond mean-field.
\end{abstract}

\maketitle

%%%%%%%%%%%%%%%%%%%%%%%%%%%%%%%%%%%%%%%%%%%%%%%%%%%%%%%%%%%%%
%                     Introduction                          %
%%%%%%%%%%%%%%%%%%%%%%%%%%%%%%%%%%%%%%%%%%%%%%%%%%%%%%%%%%%%%
\section{Introduction}

The two-dimensional electron gas is a cornerstone of modern condensed matter physics. It serves as the starting point for understanding the emergence of band insulators from the explicit breaking of continuous translation symmetry by a periodic potential, and Wigner crystallization via the spontaneous breaking of translation symmetry by strong interactions~\cite{wigner1934interaction, pines1952collective, bohm1953collective}. Despite being an old and well-studied subject~\cite{ceperley1978ground, tanatar1989ground, rapisarda1996diffusion, varsano2001spin, gori2004pair, drummond2009phase, drummond2009quantum, smith2024unified, azadi2024quantum, azadi2025quantum, valenti2025critical, bonsall1977some, fukuyama1979two, cote1991collective, fogler2000dynamical, chitra2001pinned, cote2008dynamical, monarkha2012two, mahan2013many, andrei1988observation, stormer1989comment, goldman1990evidence, williams1991conduction, paalanen1992electrical, yoon1999wigner, qiu2012connecting}, electron crystallization has reappeared as a topic of crucial importance in understanding a large number of recent experimental results on two-dimensional materials~\cite{tsui2024direct, hossain2020observation, smolenski2021signatures, falson2022competing, munyan2024evidence, xiang2025imaging, seiler2022quantum, seiler2024signatures, Su2025, munyan2024evidence, xiang2025imaging, lu2024fractional, lu2025extended}. Many of these experimental results differ significantly from crystallization described by the conventional jellium model, as the systems of interest host non-trivial band geometry (Berry curvature, Fubini-Study metric, and form factors)~\cite{xiao2010berry, parameswaran2013fractional, Roy2014, yu2024quantum, verma2025quantum, liu2025quantum}. Non-trivial band geometry allows for new possibilities such as the emergence of anomalous Hall crystals (AHCs)~\cite{halperin1986compatibility, kivelson1986cooperative, kivelson1987cooperative, tesanovic1989hall, dong2024theory, zhou2024fractional, dong2024anomalous, soejima2024anomalous, dong2024stability, kwan2023moire, yu2024moire, tan2024parent, tan2025variational, patri2024extended, zheng2024sublattice, zhou2025new, zeng2024berry, bernevig2025berry, desrochers2025elastic} (also referred to as Wigner-Hall crystals in~\cite{seiler2022quantum, seiler2024signatures}), that are topological analogs of WCs carrying a non-zero Chern number, as well as halo or chiral Wigner crystals (HWCs)~\cite{soejima2025jellium} which semi-classically correspond to a lattice of localized electrons that acquire spontaneous orbital angular momentum~\cite{joy2023wigner, joy2025chiral}. 

Notably, the interplay between crystallization and band geometry appears to play an essential role across several graphene-based platforms. Signatures of AHCs have been observed in transport measurements on Bernal bilayer graphene~\cite{seiler2022quantum, seiler2024signatures}, and recent work on twisted bilayer–trilayer graphene reported ``generalized anomalous Hall crystals'' with enlarged unit cells, the formation of which is driven by the cooperation of the moir\'e potential and strong interactions~\cite{Su2025}. Reports of an extended integer quantum anomalous Hall state in rhombohedral pentalayer graphene --- robust across a wide range of filling and displacement field ---  have further been interpreted as potentially arising from a weakly pinned AHC~\cite{dong2024anomalous, zhou2024fractional, dong2024theory, soejima2024anomalous, dong2024stability, patri2024extended}. 

Such a situation undoubtedly calls for the development of a more holistic theoretical picture of the interplay between (explicit and spontaneous) crystallization and band geometry. To tackle such challenges, the development of tunable, analytically tractable toy models appears necessary to shed light on the mechanisms behind the formation of these topological electronic crystals. Several toy models have already been introduced in the literature~\cite{soejima2024anomalous, dong2024stability, zheng2024sublattice, tan2024parent, soejima2025jellium}. Importantly, Ref.~\cite{tan2024parent} introduced the so-called ideal parent band or infinite Chern band model describing a quadratic band with constant Berry curvature $\mathcal{B}$, whereas Ref.~\cite{soejima2025jellium} introduced the $\lambda$-jellium model with a tunable concentrated Berry curvature profile that encloses a total Berry flux of $2\pi$. Despite their utility, both models have their limitations. Neither has a total Berry flux that can be tuned, and the infinite Chern band model contains an infinite number of internal degrees of freedom, thereby hindering the possibility of performing standard variational and diffusion Monte-Carlo.

In this work, we introduce an extension of these constructions: the $\lambda_N$-jellium model. This model has $N$ internal degrees of freedom and results in a lowest isolated ideal quadratic band with a tunable Berry curvature profile that encloses a total Berry flux of $2\pi(N-1)$ (see Fig.~\ref{fig:fig_band_geometry_lambda_n}). This model interpolates between $\lambda$-jellium ($N=2$) and the infinite Chern band ($N\to\infty$) models, and can therefore be used to better understand what is specific to each limit and how the generic behavior changes as the total flux is varied. 

We first study this model with an additional $C_6$-symmetric periodic potential. Numerical results reveal a highly complex phase diagram that features band insulators with large positive and negative Chern numbers, as well as phase transitions that cannot be captured by symmetry eigenvalues of filled single-particle states at high-symmetry points. We provide some analytical understanding of the observed features by studying the model in the weak potential limit. Following ideas outlined in previous works~\cite{dong2024stability, bernevig2025berry}, our analytical exploration of the weak potential regime culminates in an expression for the full Chern number (Eq.~\eqref{Equation_Chern_low_U_capped_mt}) applicable to any band projected model with a monotonically increasing kinetic energy. This result provides a simple explanation for why band insulators with only Chern numbers opposite to the sign of the occupied band's Berry curvature arise from a weak potential in the $N\to\infty$ limit, without resorting to the mappings introduced in Ref.~\cite{tan2024parent}. We further explain why a $\mathcal{C}=0$ insulator is generically expected for a highly concentrated Berry curvature close to the $\Gamma$ point and a sufficiently large periodic potential.

We then proceed to study spontaneous crystallization induced by strong electronic repulsion. As the Berry curvature profile and total flux are varied, we find AHCs with a range of Chern numbers that can be understood in the weak interaction limit based on the ``Berry flux rounding'' argument outlined in Refs.~\cite{dong2024stability, bernevig2025berry}. At larger interactions, we observe a HWC as in Ref.~\cite{soejima2025jellium} and a previously unknown halo anomalous Hall crystal (HAHC) that combines a finite Chern number and a non-trivial irreducible representation for the filled state at the $\Gamma$ point. Our extensive Hartree-Fock calculations and analytical exploration pave the way for detailed numerical studies of the stability of these electronic crystals beyond the mean-field level, which remains an outstanding unresolved issue. The $\lambda_N$ model may be particularly promising for such studies, as its simplicity and limited spinor dimension make it amenable to variational and diffusion Monte Carlo methods~\cite{ceperley1978ground, tanatar1989ground, gori2004pair, varsano2001spin, de2005effects, attaccalite2002correlation, rapisarda1996diffusion, drummond2009phase, drummond2009quantum, smith2024unified, azadi2024quantum, valenti2025critical, azadi2025quantum, reddy2025quantum, becca2017quantum}. Furthermore, the $N>2$ models may be required for future numerical studies to unambiguously provide evidence for the stability of AHCs in microscopic models, given that they support the AHC phase over a wider parameter regime at large interactions in Hartree-Fock than the $N=2$ model. One may conjecture that such models may be more fertile for hosting AHCs.

The rest of this paper is organized as follows. Section~\ref{sec:lambda_N} introduces the $\lambda_N$ model and describes its band geometry. Section~\ref{sec:periodic_potential} investigates the effect of a $C_6$ periodic potential and various analytically tractable regimes of the resulting phase diagram. Section~\ref{sec:spontaneous_crystallization} explores the Hartree-Fock phase diagram in the presence of interactions. We close by discussing our results in light of other existing works and comment on possible future directions in Sec.~\ref{sec:conclusion}.

\section{The \texorpdfstring{$\boldsymbol{\lambda}_{\boldsymbol{N}}$}{$\lambda_N$} Model}\label{sec:lambda_N}

\subsection{Constructing The Model}

To study the role of quantum geometry on the phases stabilized by a periodic potential or strong electronic interaction, one first needs a non-interacting model with tunable Berry curvature. In this work, we study these effects by introducing the $\lambda_N$-jellium model, an $N$-band continuum model with internal degrees of freedom labelled by $\ket{0}, \ldots \ket{N-1}$ and the Hamiltonian 
\begin{align}
    H^{\lambda_N}_{m,n}(\bs{q}) = \frac{\abs{\bs{q}}^2}{2m} \delta_{m,n} + \Delta \bra{m} \hat{L}(\bs{q}) \ket{n}.
\end{align}
Here $\hat{L}(\bs{q})$ is a momentum-dependent positive semi-definite operator
\begin{align}
    \hat{L}(\bs{q}) &= \hat{\mathcal{D}}_{\bs{q}}^\dagger \hat{\mathcal{D}}_{\bs{q}},
\end{align}
with $\hat{\mathcal{D}}_{\bs{q}}$ denoting a shifted ladder operator
\begin{align}
    \hat{\mathcal{D}}_{\bs{q}} = \hat{b} - \sqrt{\mathcal{B}} q_{\bar{z}} \mathcal{P}^{\perp}_{N-1},
\end{align}
$\hat{b}$ is a conventional ladder operator (i.e., $\hat{b}\ket{n}=\sqrt{n}\ket{n-1}$ for $n\ge 1$ and $\hat{b}\ket{0}=0$), $\sqrt{\mathcal{B}}$ is a real number that has units of length, $q_{\bar{z}}=(q_x-iq_y)/\sqrt{2}$ is the antiholomorphic momentum, and $\mathcal{P}^{\perp}_{N-1} = \sum_{n=0}^{N-2}\dyad{n}$ is a projector in the subspace orthogonal to the one spanned by $\ket{N-1}$. The above model reproduces the (complex conjugate of the) $\lambda$-jellium model of Ref.~\cite{soejima2025jellium} for $N=2$ and the infinite Chern band model of Ref.~\cite{tan2024parent} in the $N\to\infty$ limit.

The $\hat{L}(\bs{q})$ operator has a zero eigenvalue at every momentum, with eigenstate $\ket{\psi_{\bs{q}}^{(0)}} = e^{i\bs{q}\cdot\hat{\bs{r}}} \ket{s_{\bs{q}}^{(0)}}$, defined by 
\begin{align} \label{eq:lambda_N_nullspace_condition}
    \hat{\mathcal{D}}_{\bs{q}} \ket{s_{\bs{q}}^{(0)}} &= 0. 
\end{align}
Solving Eq.~\eqref{eq:lambda_N_nullspace_condition} yields the zero-mode structure 
\begin{align} \label{eq:lambda_N_model_coefficients_wf}
    \ket{s_{\bs{q}}^{(0)}} &= \mathcal{N}_{\bs{q}} \sum_{n=0}^{N-1} \frac{\left( \sqrt{\mathcal{B}} q_{\bar{z}} \right)^n}{\sqrt{n!}}\ket{n},
\end{align}
where 
\begin{align}
    \mathcal{N}_{\bs{q}}^{-2} 
    &= \left( \sum_{n=0}^{N-1} \frac{\left( \frac{\mathcal{B} \abs{\bs{q}}^{2}}{2} \right)^{n}}{n!}\right) 
    = \frac{ \exp(\frac{\mathcal{B} \abs{\bs{q}}^2}{2}) \Gamma\left(N,\frac{\mathcal{B}}{2} \abs{\bs{q}}^2\right)}{(N-1)!}
\end{align}
is a normalization constant and $\Gamma(a,b)=\int_{b}^{\infty}dt e^{-t}t^{a-1}$ is the upper incomplete gamma function. The $\lambda_N$ model has $N$ positive energy bands with dispersion 
\begin{align}
    E^{(n)}(\bs{q}) &= \frac{|\bs{q}|^2}{2m} + \Delta \varepsilon^{(n)}(\bs{q}),
\end{align}
where $\varepsilon^{(n)}(\bs{q})$ are the eigenvalues of $\hat{L}(\bs{q})$. The lowest band has a quadratic dispersion (since $\varepsilon^{(0)}(\bs{q})=0$) and is separated by a gap $\Delta$ from the second lowest one.

%%%%%
\subsection{Band Geometry}

%%%%%%%%%%%%%%%%%%%%%%%%%%%%%%%%%%%%%%%%%%%%%%%%%%%%%%%%%%%
\begin{figure}
    \centering
    \includegraphics[width=1.0\linewidth]{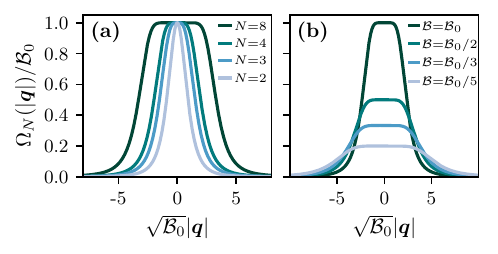}\vspace{-4mm}
    \caption{Berry curvature distribution of the $\lambda_N$-jellium model for (a) $\mathcal{B}=\mathcal{B}_0$ with $N=2,3,4,8$ and (b) $N=5$ with $\mathcal{B}=\mathcal{B}_0, \mathcal{B}_0/2, \mathcal{B}_0/3, \mathcal{B}_0/5$, where $\mathcal{B}_0$ is a reference value.}
\label{fig:fig_band_geometry_lambda_n}
\end{figure}
%%%%%%%%%%%%%%%%%%%%%%%%%%%%%%%%%%%%%%%%%%%%%%%%%%%%%%%%%%%

Sending $\Delta\to\infty$, we end up with a single isolated quadratic band. Hereafter, we restrict ourselves to this lowest band, whose quantum geometric properties are encoded in the form factors
\begin{align}
    \mathcal{F}_{N}(\bs{q}, \bs{q}') &= \bra{\psi_{\bs{q}}^{(0)}} e^{i(\bs{q} - \bs{q}')\cdot\bs{r}} \ket{\psi_{\bs{q}'}^{(0)}}= \braket{s_{\bs{q}}^{(0)}}{s_{\bs{q}'}^{(0)}}.
\end{align}
Using the expression for the eigenstates given in Eq.~\eqref{eq:lambda_N_model_coefficients_wf}, the form factor is
\begin{align}
    \mathcal{F}_{N}(\bs{q}, \bs{q}')
    &= \mathcal{N}_{\bs{q}} \mathcal{N}_{\bs{q}'} \sum_{n=0}^{N-1} \frac{(\mathcal{B} q_{z} q_{\bar{z}}^{\prime})^n}{n!}. \label{eq_form_factor_explicit}
\end{align}
It can be rewritten using the sum formula for the upper incomplete gamma function evaluated at positive integers as
\begin{align}
    \mathcal{F}_{N}(\bs{q}, \bs{q}')  
    &= \mathcal{M}_{N}(\bs{q}, \bs{q}') \mathcal{F}_{\text{LLL}}(\bs{q}, \bs{q}')  \label{eq:lambda_N_form_factor}
\end{align}
where the $N$-dependent prefactor is
\begin{align}
    \mathcal{M}_{N}(\bs{q}, \bs{q}') &= \frac{\Gamma(N, \mathcal{B} q_z q_{\bar{z}}^{\prime})}{ \sqrt{\Gamma\left(N,\frac{\mathcal{B}}{2} \abs{\bs{q}}^2\right) \Gamma\left(N,\frac{\mathcal{B}}{2} \abs{\bs{q}'}^2\right)}}, \label{eq:lambda_N_form_factor_prefactor}
\end{align}
and $\mathcal{F}_{\text{LLL}}(\bs{q}, \bs{q}')$ is the form factor associated with projection to the lowest Landau level (LLL) appearing in the Girvin-MacDonald-Platzman (GMP) algebra~\cite{girvin1986magneto}
\begin{align}
    \mathcal{F}_{\text{LLL}}(\bs{q}, \bs{q}') &= \exp(-\frac{\mathcal{B}}{4} \left( |\bs{q} - \bs{q}'|^2 + 2 i \bs{q}\wedge \bs{q}' \right)), \label{eq:LLL_form_factor}
\end{align}
with $\bs{q}\wedge \bs{q}'\equiv q_{x}q'_{y} - q_{y}q'_{x}$. The $N$-dependent prefactor becomes constant in the $N\to\infty$ limit
\begin{align}
    \lim_{N\to\infty} \mathcal{M}_{N}(\bs{q}, \bs{q}') &= 1,
\end{align}
in which case the form factor becomes identical to the LLL 
\begin{align}
    \lim_{N\to\infty} \mathcal{F}_{N}(\bs{q},\bs{q}') = \mathcal{F}_{\text{LLL}}(\bs{q},\bs{q}'),  \label{eq:form_factor_N_to_infty_limit}
\end{align}
as in the infinite Chern band model~\cite{tan2024parent}.

The form factor can be used to study the resulting band geometry as
defined by the quantum geometric tensor
\begin{equation}
    \begin{aligned}
        \mathcal{Q}_N^{a b}(\bs{q})
        &=
        \left\langle\partial_{q_a} u_{\bs{q}}^{(0)}\right| Q_{\bs{q}} \left|\partial_{q_b} u_{\bs{q}}^{(0)}\right\rangle \\
        &= g_{N}^{a b}(\bs{q}) - \frac{i}{2} \epsilon^{a b} \Omega_{N}(\bs{q}),
    \end{aligned}
\end{equation}
where $Q_{\bs{q}} = \mathds{1} - |u_{\bs{q}}^{(0)}\rangle\langle u_{\bs{q}}^{(0)}|$, $a$ and $b$ represent spatial dimensions, and $\epsilon^{a b}$ is the antisymmetric tensor. The real and imaginary parts of the quantum geometric tensor correspond to the Fubini-Study metric $g_{N}^{a b}(\bs{q})$ and Berry curvature $\Omega_{N}(\bs{q})$, respectively. Using the form factor~\eqref{eq_form_factor_explicit}, the Berry curvature of the $\lambda_N$-jellium model is 
\begin{align}
    \Omega_{N}(\bs{q}) &= \mathcal{B} + \mathcal{B} e^{-\frac{\mathcal{B} \abs{\bs{q}}^2 }{2}} \frac{E_{1-N}\left(\frac{\mathcal{B}}{2} \abs{\bs{q}}^2 \right)-E_{-N}\left(\frac{\mathcal{B}}{2} \abs{\bs{q}}^2 \right)}{\left(E_{1-N}\left(\frac{\mathcal{B}}{2} \abs{\bs{q}}^2 \right)\right)^2},\label{eq:berry_curvature_lambda_N}
\end{align}
with the exponential integral function $E_{n}(z) = z^{n-1} \Gamma(1-n, z)$. In the $N\to\infty$ limit, the second term in~\eqref{eq:berry_curvature_lambda_N} vanishes and we recover a completely flat Berry curvature 
\begin{align}
    \lim_{N\to\infty} \Omega_{N}(\bs{q}) &= \mathcal{B},
\end{align}
consistent with the LLL and Eq.~\eqref{eq:LLL_form_factor}\footnote{We also note that for $N=2$ the Berry curvature is $\Omega_{N=2}(\bs{q}) = 4 \mathcal{B}/\left(2 + \mathcal{B} \abs{\bs{q}}^2 \right)^2$, as for the $\lambda$-jellium model~\cite{soejima2025jellium} with the identification $\lambda=\sqrt{\mathcal{B}/2}$. The model of Ref.~\cite{soejima2025jellium} is the complex conjugate of our Hamiltonian for $N=2$. Still, the expression for the Berry curvature is the same since we use opposite conventions for the definition of the Berry connection. I.e., we define $A_{k_a}=i \langle s_{\boldsymbol{k}}^{(0)}| \partial_{k_a} |s_{\boldsymbol{k}}^{(0)}\rangle$.}. Within a disk of radius $\kappa$, the enclosed Berry curvature is 
\begin{align}
    I_{N}(\kappa) &= 2 \pi \int_{0}^{\kappa} d\abs{\bs{q}} \abs{\bs{q}} \Omega_{N}(|\bs{q}|) = \pi \mathcal{B} \kappa^2 -\frac{2 \pi e^{-\frac{\mathcal{B} \kappa^2}{2}}}{E_{1-N}\left(\frac{\mathcal{B} \kappa^2}{2}\right)}.
\end{align}
By sending $\kappa\to\infty$, we see that the total Berry flux contained in the band, 
\begin{align} \label{eq:total_berry_flux_in_lambda_n}
    \lim_{\kappa\to\infty} I_{N}(\kappa) =  2 \pi (N-1),
\end{align}
is fixed by $N$.

As a consequence of the antiholomorphic structure of the unnormalized lowest band wavefunction~\eqref{eq:lambda_N_model_coefficients_wf}, the Fubini-Study metric is given by~\cite{wang2021exact}
\begin{align}
   g^{ab}_{N}(\bs{q}) =\frac{1}{2} \Omega_{N}(\bs{q}) \delta^{a b}, \label{eq:FS_metric_lambda_N}
\end{align}
such that it saturates the trace condition $\Tr(g^{ab}_{N}(\bs{q})) = \left|\Omega_{N}(\boldsymbol{q})\right|$. That is, the $\lambda_N$ model has ideal quantum geometry~\cite{wang2021exact, ledwidth2023vortexability, liu2025theory}. 

The behavior of the Berry curvature (and of the Fubini-Study metric given Eq.~\eqref{eq:FS_metric_lambda_N}) is summarized in Fig.~\ref{fig:fig_band_geometry_lambda_n}. The Berry curvature is maximized at $|\bs{q}|=0$, where it reaches a value of $\mathcal{B}$ for all $N$ (Fig.~\ref{fig:fig_band_geometry_lambda_n}(b)), and decays algebraically at large momentum (i.e., $\Omega_N(|\bs{q}|)=4(N-1)/(\mathcal{B} |\bs{q}|^4)+\mathcal{O}(1/|\bs{q}|^6)$ as $|\bs{q}|\to\infty$). At small momenta, the Berry curvature can be expanded as
\begin{align}
    \Omega_{N}(|\bs{q}|) &= \mathcal{B} - \frac{N \mathcal{B}^N |\bs{q}|^{2(N-1)}}{2^{N-1} (N-1)!} + \mathcal{O}(|\bs{q}|^{2N}),
\end{align}
which results in an essentially flat Berry curvature plateau that gets wider as $N$ increases for a fixed $\mathcal{B}$ value (see Fig.~\ref{fig:fig_band_geometry_lambda_n}(a)). Indeed, for momenta $\bs{q}$ and $\bs{q}^\prime$ in this flat Berry curvature plateau, the form factor~\eqref{eq:lambda_N_form_factor} can be well approximated by $\mathcal{F}_N(\bs{q},\bs{q}')\approx \mathcal{F}_{\text{LLL}}(\bs{q},\bs{q}')$. This can be noted by observing that the prefactor~\eqref{eq:lambda_N_form_factor_prefactor} is roughly constant on this plateau, e.g., $\mathcal{M}_N(\bs{0},\bs{q})=\sqrt{(N-1)!/\Gamma(N, \mathcal{B}|\bs{q}|^2/2)} = 1 + \mathcal{B}^N |\bs{q}|^{2N}/(2^{N+1} N!) + \mathcal{O}(|\bs{q}|^{2(N+1)})$. In the $N \to\infty$ limit, the plateau becomes infinitely extended, and we recover the uniform Berry curvature and form factor of the LLL with magnetic field $\mathcal{B}$. The Berry curvature distribution of the $\lambda_N$ model is reminiscent of other models, such as $N$-layer rhombohedral graphene, where most of the Berry curvature in the lowest conduction band is also concentrated in a ring inside which the form factor can be approximated by the LLL version~\cite{bernevig2025berry}. This connection between rhombohedral multilayer graphene and the $\lambda_N$ model is outlined in Appendix~\ref{app:connection_lambdaN_RNG}, where we show that the $\hat{L}(\bs{q})$ operator of the $\lambda_N$ model can be obtained by modifying an effective model of rhombohedral $N$-layer graphene to have layer-dependent hoppings.

%%%%%%%%%%%%%%%%%%%%%%%%%%%%%%%%%%%%%%%%%%%%%%%%%%%%%%%%%%%%%
%                Periodic Potential                         %
%%%%%%%%%%%%%%%%%%%%%%%%%%%%%%%%%%%%%%%%%%%%%%%%%%%%%%%%%%%%%
\section{Periodic Potential}\label{sec:periodic_potential}

%%%%%%%%%%%%%%%%%%%%%%%%%%%%%%%%%%%%%%%%%%%%%%%%%%%%%%%%%%%
\begin{figure}
    \centering
    \includegraphics[width=1.0\linewidth]{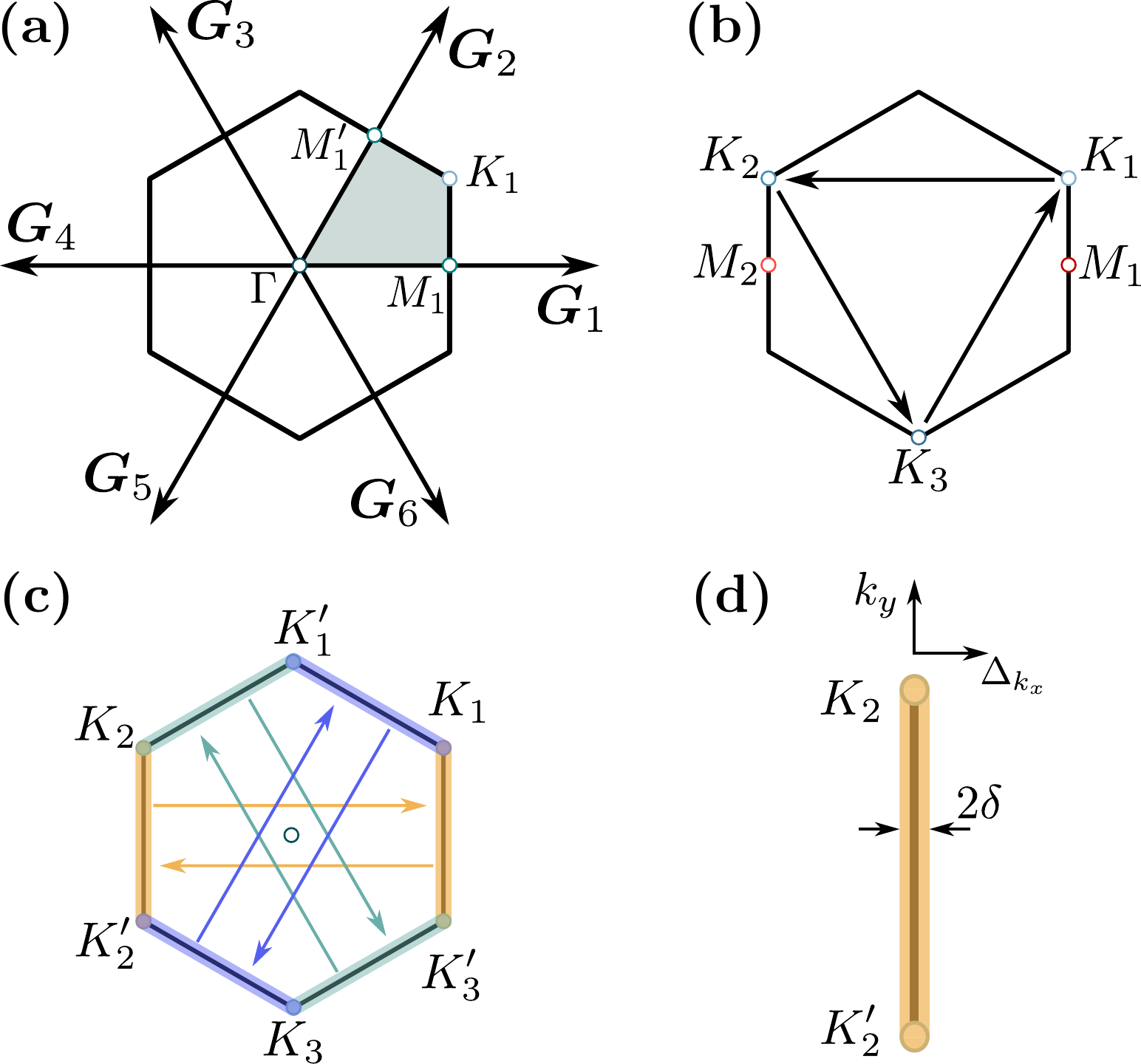}%\vspace{-4mm}
    \caption{
    (a) First Brillouin zone of the triangular lattice with its high-symmetry points and reciprocal lattice vectors. (b)  $K_1$, $K_2$, and $K_3$ are the three lowest kinetic energy states with crystal momentum $K$, and $M_1$ and $M_2$ are the lowest kinetic energy states with crystal momentum $M$. The free folded bands result in threefold degeneracy at $K$ and $K'$, and twofold degeneracy at $M$. (c) In the weak potential limit, only states along the boundary of the first Brillouin zone that are connected by a reciprocal lattice vector mix. (d) Focusing on the left edge of the first Brillouin zone, we define the width of states that mix as $2\delta$ and the momentum perpendicular to the wire measured from it as $\Delta_{k_x}$.
    }
\label{fig:1bz_schematics}
\end{figure}
%%%%%%%%%%%%%%%%%%%%%%%%%%%%%%%%%%%%%%%%%%%%%%%%%%%%%%%%%%%

We now study explicit breaking of continuous translation symmetry by subjecting the $\lambda_N$-jellium model to a $C_6$ symmetric periodic potential of the form 
\begin{align}
    U(\boldsymbol{r}) = - U_0\sum_{\bs{G}_j} e^{i \bs{G}_j \cdot \boldsymbol{r}},
\end{align}
where $U_0>0$ is the potential strength, and $\bs{G}_j = g (\cos((j-1)\pi/3), \sin((j-1)\pi/3))$ ($j=1,2,3,4,5,6$) are the six shortest reciprocal lattice vectors of the triangular lattice (see Fig.~\ref{fig:1bz_schematics}) with lattice constant $a_l$, such that $g=4\pi/(a_l \sqrt{3})$. Here we consider the problem at a fixed filling of one electron per unit cell ($\nu=1$). In its second-quantized form, the model is
\begin{subequations} \label{eq:model_lambda_n_pp}
\begin{align}
    \hat{\mathcal{H}} = \hat{\mathcal{H}}_{0} + \hat{\mathcal{H}}_{\mathrm{pot}},
\end{align}
with kinetic energy 
\begin{align} \label{eq:model_lambda_n_pp_kin_part}
    \hat{\mathcal{H}}_0=\sum_{\bs{k},\bs{g}} c_{\bs{k},\bs{g}}^{\dagger} \frac{|\bs{k}+\bs{g}|^2}{2m}  c_{\boldsymbol{k}, \bs{g}} = \sum_{\bs{k},\bs{g}} \mathcal{E}(\bs{k}+\bs{g}) c_{\bs{k},\bs{g}}^{\dagger}  c^{\phantom{\dagger}}_{\boldsymbol{k}, \bs{g}} 
\end{align}
and periodic potential
\begin{align} \label{eq:model_lambda_n_pp_potential_part}
    \hat{\mathcal{H}}_{\mathrm{pot}} = -U_0 \sum_{\boldsymbol{G}_i} \tilde{\rho}_{\bs{G}_i},
\end{align}
\end{subequations}
where $\mathcal{E}(\bs{q})=|\bs{q}|^2/2m$ is the free dispersion,  $c_{\bs{k},\bs{g}}^\dagger \equiv c_{\bs{k}+\bs{g}}^\dagger$ is the creation operator for the single-particle state $|\psi_{\bs{k}+\bs{g}}^{(0)}\rangle$, $\tilde{\rho}_{\bs{q}}$ is the density operator projected into the lowest band of the $\lambda_N$-jellium model 
\begin{align} \label{eq:model_lambda_n_projected_density}
    \tilde{\rho}_{\bs{q}} = \sum_{\bs{k},\bs{g}} c_{\boldsymbol{k}+\bs{g}+\boldsymbol{q}}^{\dagger} \mathcal{F}_N\left(\boldsymbol{k} + \bs{g} + \boldsymbol{q}, \boldsymbol{k} + \bs{g} \right) c^{\phantom{\dagger}}_{\bs{k} + \bs{g}},
\end{align}
$\boldsymbol{g} = n_1 \boldsymbol{G}_1+$ $n_2 \boldsymbol{G}_2$ ($n_1, n_2 \in \mathbb{Z}$) denote reciprocal lattice vectors, and $\bs{k}$ is taken throughout the article to be in the first Brillouin zone (1BZ).

A helpful perspective is to regard the above model as a momentum-space version of the Hofstadter model for every momentum $\bs{k}$~\cite{tan2024parent}. From this point of view, the kinetic term~\eqref{eq:model_lambda_n_pp_kin_part} plays the role of an on-site potential that energetically penalizes occupation of states at large momenta. The periodic potential~\eqref{eq:model_lambda_n_pp_potential_part} describes hopping between ``nearest-neighbor'' momenta $\bs{k}+\bs{g}$ and $\bs{k}+\bs{g}+\bs{G_i}$ with amplitude $U_0 |\mathcal{F}_N\left(\boldsymbol{k} + \bs{g} + \boldsymbol{G}_i, \boldsymbol{k} + \bs{g} \right)|$ 
and phase $\arg[\mathcal{F}_N\left(\boldsymbol{k} + \bs{g} + \boldsymbol{G}_i, \boldsymbol{k} + \bs{g} \right)]$. As such, a generic lowest-energy single-particle eigenstate of the Bloch Hamiltonian~\eqref{eq:model_lambda_n_pp} at momentum $\bs{k}$ 
\begin{align}
    \ket{\varphi_{\bs{k}}} = \sum_{\bs{g}} v_{\bs{g}}(\bs{k}) \ket{\psi^{(0)}_{\bs{k}+\bs{g}}}, \label{eq:generic_single_particle_state_at_high_symmetry_momentum}
\end{align}
has an energy 
\begin{subequations}
\begin{align}
    \bra{\varphi_{\bs{k}}} \hat{\mathcal{H}} \ket{\varphi_{\bs{k}}} = E_{\text{kin}}(\bs{k}) + E_{\text{pot}}(\bs{k}), \label{eq:energy_generic_single_particle_state}
\end{align}
where 
\begin{align}
    E_{\text{kin}}(\bs{k}) &= \sum_{\bs{g}} |v_{\bs{g}}(\bs{k})|^2 \frac{|\bs{k} + \bs{g}|^2}{2m} \label{eq:energy_generic_single_particle_state_kinetic} \\   
    E_{\text{pot}}(\bs{k}) &=  -U_0 \sum_{\bs{g},\bs{G}_i} v^*_{\bs{g}+\bs{G}_i}(\bs{k}) v_{\bs{g}}(\bs{k}) \nonumber \\
    &\qquad\quad\quad \times \mathcal{F}_{N}(\bs{k} + \bs{g} + \bs{G}_i, \bs{k} + \bs{g}) \label{eq:energy_generic_single_particle_state_potential} .
\end{align}
\end{subequations}
Physically, Eq.~\eqref{eq:energy_generic_single_particle_state} illustrates how the competition between kinetic and potential energy governs the structure of the single-particle states. Isolating the potential contribution from hopping between $\bs{k}+\bs{g}$ and $\bs{k}+\bs{g}+\bs{G}_i$
\begin{widetext}
\begin{align}
    E^{\bs{g}, \bs{g}+\bs{G}_i}_{\text{hop}}(\bs{k}) =&-U_0 v^*_{\bs{g}+\bs{G}_i}(\bs{k}) v_{\bs{g}}(\bs{k})  \mathcal{F}_{N}(\bs{k} + \bs{g} + \bs{G}_i, \bs{k} + \bs{g}) -U_0 v_{\bs{g}+\bs{G}_i}(\bs{k}) v^*_{\bs{g}}(\bs{k})  \mathcal{F}_{N}(\bs{k} + \bs{g} , \bs{k} + \bs{g}+ \bs{G}_i) \notag \\
    =& -2U_0 |v_{\bs{g}+\bs{G}_i}(\bs{k})| |v_{\bs{g}}(\bs{k})| |\mathcal{F}_{N}(\bs{k} + \bs{g} + \bs{G}_i, \bs{k} + \bs{g})| \notag \\
    &\qquad \times \cos\big(\arg[\mathcal{F}_{N}(\bs{k} + \bs{g} + \bs{G}_i, \bs{k} + \bs{g})] + \arg[v_{\bs{g}}(\bs{k})] - \arg[v_{\bs{g}+\bs{G}_i}(\bs{k})]\big), \label{eq:pp_hopping} 
\end{align}
\end{widetext}
we see that the potential promotes delocalization in momentum space and that minimization of the hopping energy between $\bs{k}+\bs{g}$ and $\bs{k}+\bs{g}+\bs{G}_i$ fixes the relative phase between coefficients $v_{\bs{g}}(\bs{k})$ and $v_{\bs{g}+\bs{G}_i}(\bs{k})$ to match the form factor connecting the two momenta. However, incorporating other reciprocal lattice vectors introduces competing constraints. These frustrate the phase conditions, preventing the form factor connecting the two momenta from uniquely fixing the relative phase.

The $N\to\infty$ limit with a periodic potential was studied in Ref.~\cite{tan2024parent}. Through exact mappings between models with $\mathcal{B}$ and $\mathcal{B}+2\pi/A_{\text{1BZ}}$, where $A_{\text{1BZ}}=8 \pi^2/(\sqrt{3} a_l^2)$ is the area of the first Brillouin zone, it was pointed out that for sufficiently large $\mathcal{B}$ one should expect Chern insulators with Chern number opposite to the sign of the Berry curvature of the non-interacting band (i.e., $\mathcal{C}<0$ for $\mathcal{B}>0$). Specifically for the $C_6$ symmetric potential, the prediction for $N\to\infty$ is that the resulting Chern number is independent of $U_0$ and given by $\mathcal{C} = -2 \lfloor \mathcal{B}A_{\text{1BZ}} / (4\pi) \rceil$, with $\lfloor x \rceil$ denoting rounding to the nearest integer. This naturally begs the question of the extent to which the above predictions are also applicable in the case of a non-uniform Berry curvature, and how the phase diagram evolves as $N$ increases and the Berry curvature becomes more uniform.

%%%%%%%%%%%%%%%%%%%%%%%%%%%%%%%%%%%%%%%%%%%%%%%%%%%%%%%%%%%
\begin{figure*}
    \centering
    \includegraphics[width=1.0\textwidth]{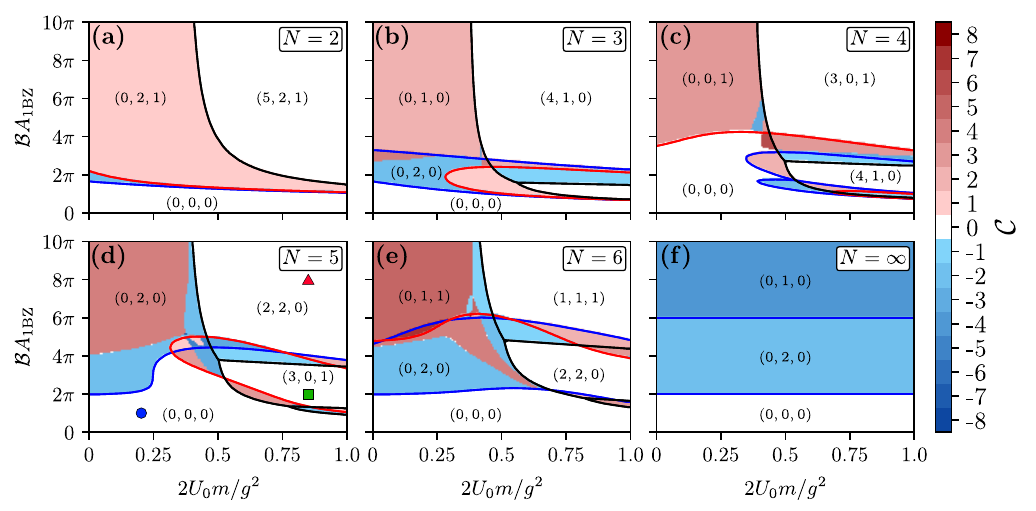}\vspace{-4mm}
    \caption{Evolution of the Chern number $\mathcal{C}$ of the band insulating phases in the model~\eqref{eq:model_lambda_n_pp} as a function of $U_0$ and $\mathcal{B}$ for (a) $N=2$, (b) $N=3$, (c) $N=4$, (d) $N=5$, (e) $N=6$, and (d) $N\to\infty$. The triplets $(\ell_\Gamma,\ell_K,\ell_M)$ denote the value of symmetry eigenvalues at high-symmetry points for the indicated phases. Transitions marked by black, blue, and red curves correspond to changes in the irreducible representation $\ell_\Gamma$, $\ell_K$, and $\ell_M$, signaling band inversions at the high-symmetry points $\Gamma$, $K$, and $M$, respectively.}
\label{fig:phase_diagram_periodic_potential}
\end{figure*}
%%%%%%%%%%%%%%%%%%%%%%%%%%%%%%%%%%%%%%%%%%%%%%%%%%%%%%%%%%%

%%%%%%%%%%%%%%%%%
\subsection{Numerical Results}

To address the above questions, we numerically solve the above model as a function of $\mathcal{B}$ and $U_0$ for different values of $N$. The results are summarized in Fig.~\ref{fig:phase_diagram_periodic_potential}, where we show the evolution of the Chern number of the resulting band insulating states. For any finite $N$, we observe a highly non-trivial behavior where the phase sensitively depends on the potential strength $U_0$ and $\mathcal{B}$. In particular, band insulators with negative and positive Chern numbers appear approximately as frequently, in sharp contrast with the $N\to\infty$ limit. We find that the features of the $N \to\infty$ phase diagram emerge only gradually as $N$ increases, while a richer set of structures persists in the phase diagram for any finite $N$.

To try to make sense of these results, we first recall from topological band theory that, in the presence of discrete rotational symmetry, partial information about the Chern number can be obtained from the irreducible representation of the occupied single-particle states at high-symmetry momentum points invariant under point group operations~\cite{Hughes2011,Fang2012}. For the model of interest with $\mathcal{B}\ne 0$, the form factor breaks time-reversal and inversion symmetry such that the model is invariant under elements of the $P6$ space group (No. 168). The point group is then $C_6$, and we have the high-symmetry momenta $\Gamma$, $K$, and $M$, which are invariant (modulo translation by reciprocal lattice vector) under the stabilizer groups $C_6$, $C_3$, and $C_2$, respectively. At these high-symmetry momenta, the occupied single-particle states therefore transform as irreducible representations of the stabilizers. Since $C_n$ with $n\in\mathbb{N}$ is an abelian group, all its irreducible representations are one-dimensional. This implies that a generic filled single-particle state~\eqref{eq:generic_single_particle_state_at_high_symmetry_momentum} at the high-symmetry momentum $\bs{\kappa}\in\{\Gamma,K,M\}$ has to transform under an element of $C_{n_{\kappa}}$ (i.e., a rotation by $2\pi m/n_{\bs{\kappa}}$, $\hat{R}_{2\pi m/n_{\bs{\kappa}}}$, with $m\in\{0,1,\ldots,n_{\bs{\kappa}}-1\}$) as
\begin{align}
    \hat{R}_{2\pi m/n_{\bs{\kappa}}} \ket{\varphi_{\bs{\kappa}}} 
    &= \sum_{\bs{g}} v_{\bs{g}}(\bs{\kappa}) \ket{\psi^{(0)}_{R_{2\pi m/n_{\bs{\kappa}}}(\bs{\kappa}+\bs{g})}} \nonumber \\
    &= \exp[\frac{i 2 \pi m \ell_{\bs{\kappa}}}{n_{\bs{\kappa}}}] \ket{\varphi_{\bs{\kappa}}}, \label{eq:definition_symmetry_indicators}
\end{align}
where $\ell_{\bs{\kappa}}\in\{0,1,\ldots n_{\kappa}-1\}$. 
The eigenvalue thus represents the phase difference between states in the linear superposition~\eqref{eq:generic_single_particle_state_at_high_symmetry_momentum} that are part of the same orbit and are exchanged under $\hat{R}_{2\pi m/n}$. For a $C_6$ symmetric model, the eigenvalues $\ell_{\Gamma}\in\{0,1,2,3,4,5\}$, $\ell_{K}\in\{0,1,2\}$, and $\ell_{M}\in\{0,1\}$ are related to the Chern number by~\cite{Fang2012}
\begin{align}\label{eq:chern_number_from_symmetry_indicators}
    \mathcal{C} = \ell_\Gamma + 2 \ell_K + 3 \ell_M \quad (\text{mod}~6).
\end{align}

The above observation offers a valuable starting point for rationalizing the complex phenomena observed in Fig.~\ref{fig:phase_diagram_periodic_potential}. Indeed, all band insulators can be identified beyond their Chern number by the triplet $(\ell_\Gamma,\ell_K,\ell_M)$, and any transition where this triplet changes must be accompanied by a closure of the single-particle gap~\cite{Hughes2011, po2017symmetry, song2018quantitative, kruthoff2017topological}. Most of the observed phase transitions correspond to boundaries where one of the symmetry eigenvalues $\ell_{\bs{\kappa}}$ changes, as labeled in Fig.~\ref{fig:phase_diagram_periodic_potential}. These transitions reflect band inversions where the gap at $\bs{\kappa}$ closes. Typically, only one indicator changes at a time, because different high-symmetry points are not directly coupled in \eqref{eq:model_lambda_n_pp}. The complexity of Fig.~\ref{fig:phase_diagram_periodic_potential} is considerably reduced when viewed through the lens of individual symmetry eigenvalues, as shown in Fig.~\ref{fig:fig_symmetry_indicators} of Appendix~\ref{app:periodic_potential}. Indeed, the majority of the features in Fig.~\ref{fig:phase_diagram_periodic_potential} can be rationalized by overlaying the three phase diagrams associated with the lowest-energy symmetry eigenvalues. However, it should be noted that an analysis based solely on symmetry eigenvalues still overlooks much of the physics. Such an analysis cannot predict the actual sign of the Chern number and misses some transitions where the symmetry eigenvalues remain the same but the Chern number changes by multiples of 6 (see, e.g., close to $\mathcal{B}A_{\text{1BZ}}=4\pi$ at small $2U_0 m/g^2$ for $N=5$). 

%%%%%%%%%%%%%%%%%%%%%%%%%%%%%%%%%%%%%%%%%%%%%%%%%%%%%%%%%%%
\begin{figure}
    \centering
    \includegraphics[width=1.0\linewidth]{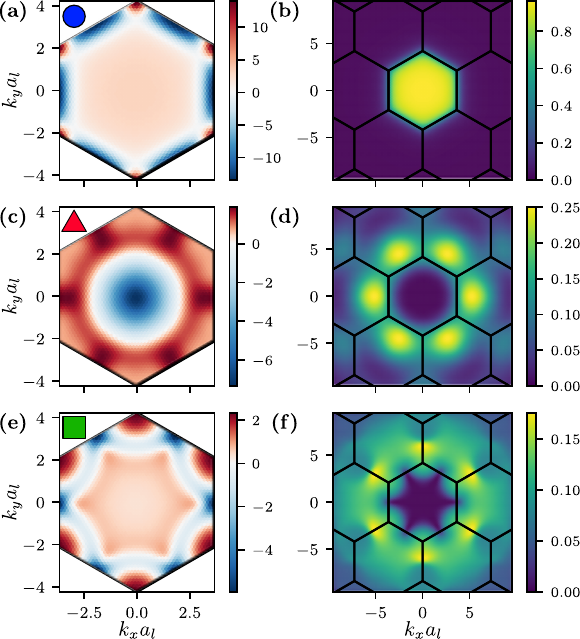}\vspace{-3mm}
    \caption{(a,c,e) Berry curvature $\Omega(\bs{k})$ and (b,d,f) momentum space occupation of the parent band states $\expval{c_{\bs{k},\bs{g}}^\dagger c_{\bs{k},\bs{g}}}$
    %of the single particle states $|\psi_{\bs{k}+\bs{g}}^{(0)}\rangle$ 
    (i.e., $|v_{\bs{g}}(\bs{k})|^2$) for the many-body ground state of the $N=5$ model with (a,b) $\mathcal{B}A_{\text{1BZ}}=\pi$ and $2U_0 m/g^2=0.2$, (c,d) $\mathcal{B}A_{\text{1BZ}}=8\pi$ and $2U_0 m/g^2=0.8$, and (e,f) $\mathcal{B}A_{\text{1BZ}}=2\pi$ and $2U_0 m/g^2 =0.8$ (see markers in panel (d) of Fig.~\ref{fig:phase_diagram_periodic_potential}). All states have $\mathcal{C}=0$, but the different rows have symmetry eigenvalues $(\ell_\Gamma,\ell_K,\ell_M)$ equal to (a,b) $(0,0,0)$, (c,d) $(2,2,0)$, and (e,f) $(3,0,1)$.}
\label{fig:bc_and_momentum_space_occupation_periodic_potential}
\end{figure}
%%%%%%%%%%%%%%%%%%%%%%%%%%%%%%%%%%%%%%%%%%%%%%%%%%%%%%%%%%%

The physical implication of the eigenvalue at the zone center is particularly interesting. A non-zero $\ell_\Gamma$ implies that the occupation of the parent band single-particle state  $|\psi_{\bs{0}}^{(0)}\rangle$ (i.e., $|v_{\bs{0}}(\Gamma)|^2$) must vanish. This is because the state $|\psi_{\bs{0}}^{(0)}\rangle$ is invariant under rotation. As such, any single-particle eigenstate at the $\Gamma$ with $v_{\bs{0}}(\Gamma)\ne 0$ will have to transform according to the trivial irreducible representation $\ell_\Gamma=0$. This is illustrated in Fig.~\ref{fig:bc_and_momentum_space_occupation_periodic_potential} where the Berry curvature $\Omega(\bs{k})$ of the reconstructed lowest band and momentum space occupation of the parent band are shown for different $\mathcal{C}=0$ band insulators of the $N=5$ model. We see that for a band insulator with $\ell_\Gamma=0$, the single-particle state $|\psi_{\bs{0}}^{(0)}\rangle$ is occupied (Fig.~\ref{fig:bc_and_momentum_space_occupation_periodic_potential}(b)), whereas its occupation vanishes for any state with $\ell_\Gamma\ne 0$ (Fig.~\ref{fig:bc_and_momentum_space_occupation_periodic_potential}(d) and (f)). Any phase with $\ell_{\Gamma}\ne 0$ must then have a significant occupation of single-particle states beyond the first Brillouin zone with large kinetic energy. These simple considerations lead us to the conclusion that band insulators with $\ell_{\Gamma}\ne 0$ only appear at large potential strength, so that the minimization of the potential energy can counterbalance the prohibitively large kinetic energy of such states, as seen in Fig.~\ref{fig:phase_diagram_periodic_potential}.

%%%%%%%%%%%%%%%%%
\subsection{Analytical Insights}

Now that we have established the phase diagram for various $N$, multiple questions naturally arise: how can we understand such phase diagrams, and what can we learn from them? While we have already pointed out that many features can be traced to the irreducible representations of filled states at high-symmetry points, this does not explain why specific transitions occur. As such, trying to analytically understand features of the phase diagrams and how they evolve with $N$ is arguably much more valuable than the phase diagrams themselves. Insights drawn from such investigations should help understand more generic behavior beyond the specific model under consideration. In what follows, we embark on an analytical exploration of the phase diagrams in two analytically tractable regimes. We first explore the weak potential limit and demonstrate how to derive an analytical expression for the full Chern number. Next, we provide explanations for the $\mathcal{C}=0$ band insulator that is present at large $U_0$ and large $\mathcal{B}A_{\text{1BZ}}$ for all finite $N$ models (see Fig.~\ref{fig:phase_diagram_periodic_potential}(a)-(e)) based on a simple approximation of the form factor.

%%%%%%%%%
\subsubsection{Weak Potential limit}
\label{Sec_weak_potential}

We begin our analytical examination by studying the weak potential limit. This should be a tractable limit since the lowest energy eigenstates of Eq.~\eqref{eq:model_lambda_n_pp} at a given $\bs{k}$ will not significantly mix states whose kinetic energy differs greatly when $U_0$ is small. Taking $U_0$ to be infinitesimal, we can approximate that the lowest energy eigenstates form a linear superposition comprising only states with the same lowest kinetic energy satisfying the nesting condition $\mathcal{E}(\bs{k}+\bs{g})=\mathcal{E}(\bs{k}+\bs{g}+\bs{G}_i)$. For the quadratic dispersion of the lowest band, this means that the parent band occupation is unity in the bulk of the first hexagonal Brillouin zone depicted in Fig.~\ref{fig:1bz_schematics}(a). States only mix on the boundary of this first Brillouin zone. Occupation is divided across two or three states with the same kinetic energy around the edges and corners of the first Brillouin zone, respectively (see Fig.~\ref{fig:1bz_schematics}(c)).

To analyze this limit, we first consider only high-symmetry points. Determination of the irreducible representation of the filled band at those high-symmetry points predicts the Chern number modulo 6 as in Eq.~\eqref{eq:chern_number_from_symmetry_indicators}. By computing the energy in each symmetry sector, we thus show how one can analytically capture phase transitions in Fig.~\ref{fig:phase_diagram_periodic_potential} where $\ell_K$ or $\ell_M$ change in the weak $U_0$ regime. However, this restriction to high-symmetry points is unsatisfactory in many respects, as it does not capture various transitions where the Chern number changes by multiples of 6 and remains agnostic to the sign of the Chern number. To go beyond this treatment, we then fully consider all states that mix along the edges of the first Brillouin zone. Making use of the gauge invariant projector formalism~\cite{Avron1983,Pozo2020, Mitscherling2025}, our treatment of the boundaries culminates in Eq.~\eqref{Equation_Chern_low_U_capped_mt}, which provides an analytical expression for the full Chern number that only depends on the form factor and applies to any single band-projected $C_6$-symmetric model (beyond the $\lambda_N$ model) with monotonically increasing kinetic energy.

\paragraph*{\underline{(i) High-Symmetry Points:}}
We thus begin by considering only eigenstates at high-symmetry points. At the $\Gamma$ point, there is a unique state at the center of the first Brillouin zone (i.e., $\bs{g}=\bs{0}$) that minimizes the kinetic energy. The single-particle states closest in energy have a much larger kinetic energy $|\bs{G}_i|^2/2m$. They should not mix significantly with $|\psi_{\bs{0}}^{(0)}\rangle$ for small $U_0$. In this regime, the lowest energy eigenstate at the zone center is then $\ket{\varphi_\Gamma} = |\psi^{(0)}_{\bs{0}}\rangle$, and we have $\ell_{\Gamma}=0$, as argued above. 

In contrast, at the $M$ and $K$ points, there are one and two additional exactly degenerate states, respectively, located on boundaries of the first Brillouin zone (see Fig.~\ref{fig:1bz_schematics}(b)). Requiring the single-particle states to transform as irreducible representations of the stabilizer at high-symmetry momenta, the lowest energy eigenstates at $K$ and $M$ can be approximated in the weak potential limit as 
\begin{subequations} \label{eq:form_eigenstates_high_symmetry_points_low_U0}
\begin{align}
    \ket{\varphi_{K}^{(\ell_K)}} &= \frac{1}{\sqrt{3}} \left( \ket{\psi^{(0)}_{K_1}} +  \bar{\omega}^{\ell_K} \ket{\psi^{(0)}_{K_2}} +  \bar{\omega}^{2\ell_K}  \ket{\psi^{(0)}_{K_3}} \right) \label{eq:form_eigenstates_high_symmetry_points_low_U0_K} \\
    \ket{\varphi_{M}^{(\ell_M)}} &= \frac{1}{\sqrt{2}} \left( \ket{\psi_{M_1}^{(0)}} + e^{i \pi \ell_M} \ket{\psi_{M_2}^{(0)}}  \right), \label{eq:form_eigenstates_high_symmetry_points_low_U0_M}
\end{align}
\end{subequations}
with $\bar{\omega}=\exp(-i 2 \pi / 3)$. Since we restrict the system to a three-dimensional subspace at the $K$ point, but block-diagonalize the Hamiltonian into three distinct irreducible representations, we automatically know that $|\varphi_{K}^{(\ell_K)}\rangle$ are eigenstates. Similar considerations also apply to $|\varphi_{M}^{(\ell_M)}\rangle$. To determine the lowest energy eigenstate, we have to determine which irreducible representation, as labelled by $\ell_K$ and $\ell_M$, minimize the potential energies $E_{\text{pot}}^{(\ell_K)}= \bra{\varphi_{K}^{(\ell_K)}} \hat{\mathcal{H}}_{\text{pot}} \ket{\varphi_{K}^{(\ell_K)}}$ and $E_{\text{pot}}^{(\ell_M)}= \bra{\varphi_{M}^{(\ell_M)}} \hat{\mathcal{H}}_{\text{pot}} \ket{\varphi_{M}^{(\ell_M)}}$.

Through direct substitution, we find that the energy of the eigenstate ~\eqref{eq:form_eigenstates_high_symmetry_points_low_U0_M} is 
\begin{align} \label{eq:var_energy_pp_M_point}
    E_{\text{pot}}^{(\ell_M)} &= (-1)^{\ell_M+1} U_0 \frac{ e^{-\frac{\mathcal{B} A_{\text{1BZ}}}{2 \sqrt{3}}} \Gamma \left(N,-\frac{\mathcal{B} A_{\text{1BZ}}}{4 \sqrt{3}}\right)}{\Gamma \left(N,\frac{\mathcal{B} A_{\text{1BZ}}}{4 \sqrt{3}}\right)}. 
\end{align}
This form predicts that the ground state wavefunction will have $\ell_{M}=0$ when $\Gamma \left(N,-\mathcal{B} A_{\text{1BZ}}/4 \sqrt{3}\right)>0$ and $\ell_{M}=1$ when $\Gamma \left(N,-\mathcal{B} A_{\text{1BZ}}/4 \sqrt{3}\right)<0$. However, the upper incomplete Gamma function $\Gamma(N,-x)$ is always positive on the domain $x\in[0,\infty)$ when $N$ is an odd positive integer. This is in agreement with Fig.~\ref{fig:phase_diagram_periodic_potential}, where we always find $\ell_M=0$ in the low $U_0$ limit as we increase $\mathcal{B}$ for $N$ odd. For $N$ even, the function $\Gamma(N,-x)$ is positive for $x=0$ but changes sign once as $x$ increases. The position of this zero predicts a transition of $\ell_M$ at $\mathcal{B}A_{\text{1BZ}}\approx 2.205\pi$ for $N=2$, $\mathcal{B}A_{\text{1BZ}}\approx 3.520 \pi$ for $N=4$, and $\mathcal{B}A_{\text{1BZ}}\approx 4.810 \pi$ for $N=6$, in agreement with results in Fig.~\ref{fig:phase_diagram_periodic_potential} (see also Fig.~\ref{fig:fig_symmetry_indicators} in Appendix~\ref{app:periodic_potential}). In the $N\to\infty$ limit, the ratio of the upper incomplete Gamma functions in Eq.~\eqref{eq:var_energy_pp_M_point} gives unity, and we are left with 
\begin{align}
    E_{\text{pot}}^{(\ell_M)} &= (-1)^{\ell_M+1} U_0 \exp(-\frac{\mathcal{B} A_{\text{1BZ}}}{2 \sqrt{3}}),  
\end{align}
which is minimized for $\ell_M=0$ independently of $\mathcal{B} A_{\text{1BZ}}$. 

At the $K$ point, although more complex, the energy of the eigenstates~\eqref{eq:form_eigenstates_high_symmetry_points_low_U0_K}
\begin{widetext}
\begin{align} \label{eq:energy_K_point_periodic_potential}
    E_{\text{pot}}^{(\ell_K)} &= -\frac{ 2 U_0 e^{-\frac{\mathcal{B}A_{\text{1BZ}}}{2 \sqrt{3}}} }{\Gamma \left(N, \frac{\mathcal{B}A_{\text{1BZ}}}{3 \sqrt{3}}\right)} \Re\left(
    \exp(i \left( \frac{2\pi\ell_K}{3} + \frac{\mathcal{B}A_{\text{1BZ}}}{6} \right)) \Gamma \left(N, \frac{\mathcal{B}A_{\text{1BZ}}}{18} \left(3 i - \sqrt{3}\right) \right)\right)
\end{align}
\end{widetext}
also correctly predicts the values of $\ell_K$ and transition points seen in Figs.~\ref{fig:phase_diagram_periodic_potential} and~\ref{fig:fig_symmetry_indicators} in the weak $U_0$ limit. The lowest-energy $\ell_{K}$ is determined by
\begin{align}
    \min_{\ell_{K}}\left[-\cos\left( \frac{2 \pi  \ell_K}{3} + \frac{\mathcal{B}A_{\text{1BZ}}}{6} + \phi(N,\mathcal{B}A_{\text{1BZ}}) \right)\right],
\end{align}
where $\phi(N,\mathcal{B}A_{\text{1BZ}})=\arg\left[\Gamma \left(N, \left(3 i - \sqrt{3}\right) \mathcal{B}A_{\text{1BZ}}/18 \right)\right]$. The energy~\eqref{eq:energy_K_point_periodic_potential} reduces to 
\begin{align}
    E_{\text{pot}}^{(\ell_K)} &= -2 U_0 e^{-\frac{\mathcal{B}A_{\text{1BZ}}}{2 \sqrt{3}}} \cos \left(\frac{1}{6}(\mathcal{B}A_{\text{1BZ}} + 4 \pi  \ell_K)\right)
\end{align}
in the $N\to\infty$ limit. Minimization of this expression for $\ell_K$ captures all transitions present in Fig.~\ref{fig:phase_diagram_periodic_potential}(f).

\paragraph*{\underline{(ii) Full Chern Number Determination: }} 

Although very useful, the above analysis based on symmetry eigenvalues at high-symmetry points falls short in many respects. An analysis based solely on high-symmetry momenta cannot provide any information about the actual sign of the Chern number and misses various transitions where the Chern number changes by multiples of 6. To remedy this, we directly construct the Chern number by considering the states along the entire boundary, not just the high-symmetry points. Using a projector formalism \cite{Avron1983, Pozo2020, Mitscherling2025}, we give expressions for the Berry curvature in both the bulk of the 1BZ and the boundary in terms of density matrices. Then, we show that the Chern number is determined by the form factor connecting opposite sides of the 1BZ.

As a preliminary, we discuss how to obtain the Chern number of a crystal band arising from translation symmetry breaking (spontaneous or explicit) in a continuum parent band. There are two primary, and ultimately equivalent, methods for directly extracting the Chern number of the crystal band. One can either construct the Berry connection and take a line integral around the BZ, or one can construct the Berry curvature and integrate it over the BZ. In this work, we find it convenient first to construct the Berry curvature. This allows us to use the projector formalism \cite{Avron1983,Pozo2020, Mitscherling2025} to obtain the Berry curvature directly from the one-body reduced density matrices, without the need to first determine a smooth gauge for the wavefunctions at different crystal momenta. We describe this process, and some general expressions for a crystal emerging from a continuum band, in Appendix~\ref{Section_proj_formalism}. Considering a parent band (such as $\lambda_N$) with Berry curvature $\Omega^{\text{P}}$ and Berry connection $\bs{A}^{\text{P}}$, the Berry curvature of the descendant crystal band is given by
\begin{widetext}
\begin{align}
	\Omega(\bs{k})	=&i\sum_{\bs{g}_1, \bs{g}_2, \bs{g}_3} \mathcal{P}_{\boldsymbol{g}_2\boldsymbol{g}_1}(\bs{k})\big(\partial_{k_x}+iA^{\text{P}}_x (\bs{k}+\bs{g}_3 )-iA^{\text{P}}_x(\bs{k}+\bs{g}_2)\big)\mathcal{P}_{\bs{g}_3 \bs{g}_2}(\bs{k}) \big( \partial_{k_y}+ i A^{\text{P}}_y(\bs{k}+\bs{g}_1)-i A^{\text{P}}_y (\bs{k}+\bs{g}_3)\big)\mathcal{P}_{\bs{g}_1 \bs{g}_3}(\bs{k}) - (x \leftrightarrow y)\notag \\
	&+ \sum_{\bs{g}} n(\bs{k}+\bs{g}) \Omega^{\text{P}}(\bs{k}+\bs{g}). \label{Eq_Berry_projector_main}
\end{align}
\end{widetext}
Here $\mathcal{P}$ is the one-body reduced density matrix
\begin{align}
    \mathcal{P}_{\boldsymbol{g}_1\boldsymbol{g}_2}\left(\boldsymbol{k}\right)= \left\langle c_{\boldsymbol{k} \boldsymbol{g}_{1}}^{\dagger} c^{\phantom{\dagger}}_{\boldsymbol{k} \boldsymbol{g}_{2}}\right\rangle,\label{eq:density_matrix_scc_earlier}
\end{align}
which is in one-to-one correspondence with a given Slater determinant state, and $n(\bs{k}+\bs{g})=\mathcal{P}_{\bs{g}\bs{g}}(\bs{k})$ is the momentum space occupation. We see that the Berry connection of the parent band appears together with a momentum space derivative, in a manner that closely resembles a covariant derivative. This ensures that the expression is invariant under changes to the gauge of the underlying parent band, as well as the gauge of the crystal state.

Another useful expression for the Berry curvature can be obtained by splitting the density matrix into its phase and magnitude: 
\begin{align}
	\mathcal{P}_{\bs{g} \bs{g}'}(\bs{k}) &=v^*_{\bs{g}}(\bs{k}) v_{\bs{g}'}(\bs{k}) \\
    &=\sqrt{n(\bs{k}+\bs{g}) n(\bs{k}+\bs{g}')} e^{i \theta_{\bs{g}'\bs{g}}(\bs{k})}.
\end{align}
The Berry curvature can then be written as 
\begin{widetext}
\begin{align}
	\Omega(\bs{k})&=- \sum_{\bs{g}_2, \bs{g}_3} n(\bs{k}+\bs{g}_2)\big[ 
	\big( \partial_{k_x} \theta_{\bs{g}_2 \bs{g}_3}(\bs{k}) +A^{\text{P}}_x(\bs{k}+\bs{g}_3)-A^{\text{P}}_x(\bs{k}+\bs{g}_2)\big) \partial_{k_y}  n(\bs{k}+\bs{g}_3) \notag \\
	& \hspace{2cm} -\big( \partial_{k_y} \theta_{\bs{g}_2 \bs{g}_3}(\bs{k}) +A^{\text{P}}_y(\bs{k}+\bs{g}_3)-A^{\text{P}}_y(\bs{k}+\bs{g}_2)\big) \partial_{k_x} n(\bs{k}+\bs{g}_3)	\big] \notag \\
	&\quad  +\sum_{\bs{g}} n(\bs{k}+\bs{g}) \Omega^{\text{P}}(\bs{k}+\bs{g}), \label{Eqn_Berry_curvature_generic_main}
\end{align}
\end{widetext}
as we show in Appendix~\ref{Section_proj_formalism}. We observe that the Berry curvature has a contribution from the occupation-weighted parent Berry curvature (the latter term), as well as a term that depends on the changes in the phase and magnitude of the density matrix.

We now apply this expression to the low $U_0$ limit. In this case, we find two contributions to the total Berry flux: a bulk contribution, which is determined by the parent Berry curvature inside the 1BZ, and a boundary contribution. For crystal momenta away from the boundary of the 1BZ (in the bulk), the parent band occupation is frozen at unity for the state within the 1BZ and zero for the corresponding momenta with higher kinetic energy. Because of this, the derivatives $\partial_{\bs{k}} n(\bs{k}+\bs{g})$ are zero for these crystal momenta. As a result, the first terms of Eq.~\eqref{Eqn_Berry_curvature_generic_main} vanish and the Berry curvature in the bulk of the 1BZ is
\begin{align}
	\Omega^{\text{bulk}}(\bs{k})\approx \sum_{\bs{g}} n(\bs{k}+\bs{g}) \Omega^{\text{P}}(\bs{k}+\bs{g}) \approx  \Omega^{\text{P}}(\bs{k}). \label{Eq_Berry_bulk}
\end{align}	
That is, the Berry curvature of the parent band is passed on to the crystal. This is illustrated in Fig.~\ref{fig:bc_and_momentum_space_occupation_periodic_potential}(a), where the Berry curvature near the zone center is essentially inherited from the parent band. In the same region, the occupation of the parent band states in the 1BZ is nearly unity (Fig.~\ref{fig:bc_and_momentum_space_occupation_periodic_potential}(b)).

On the other hand, consider a momentum $\bs{q}$ near the boundary of the 1BZ. As we approach the boundary, this occupation changes from unity inside the 1BZ to $1/2$ exactly on the boundary and then zero outside (as the boundary is crossed, a different reciprocal lattice vector $\bs{g}$ brings the momentum into the 1BZ). This means that the occupation changes significantly near the boundary, making the first terms in Eq.~\eqref{Eqn_Berry_curvature_generic_main} large. As the potential strength $U_0$ is decreased, the boundary region in which the occupation changes shrinks, with its width $2 \delta \ll g$ (see Fig.~\ref{fig:1bz_schematics}(c)-(d)). This results in very large $\nabla  n(\bs{k}+\bs{g})$ at the boundary. If $U_0$ is small enough,  $n(\bs{k}+\bs{g})$ is a step function, so that its derivative across the boundary is similar to a Dirac delta function. Then we can treat the boundary region as a 1D wire and evaluate the additional terms in the Berry curvature only at the boundary.

To determine this boundary contribution to the Berry flux, we need to integrate over the boundary region. We first use the $C_6$ symmetry to focus on one pair of edges of the 1BZ (the left and right edges in Fig.~\ref{fig:1bz_schematics}(c)), with the contribution from the other two pairs of edges being the same. In terms of crystal momentum, these two edges occupy the same region, which we call $W_1$ (for wire 1). On the other hand, they are separated in the parent momentum space, and we refer to the two edges as LW and RW, respectively (for left wire and right wire). We first integrate the Berry curvature perpendicular to the edge (along the $\Delta_{k_x}$ direction in Fig.~\ref{fig:1bz_schematics}(d)), across the boundary (which, due to the periodicity of the BZ, includes the contributions from both LW and RW). For definiteness, we will write the momentum in terms of the position on the left wire, $\bs{k}^{\text{LW}}$, with the corresponding position on the right wire given by $\bs{k}^{\text{LW}}+\bs{G}_1$ (even though the crystal momentum will ``jump'' from one wire to the other as we integrate across the wire). As we show in Appendix \ref{Section_weak_potential_projector}, integrating across the boundary near a point $\bs{k}^{\text{LW}}$ on the left edge results in
\begin{align}
	&\int_{-\delta}^{\delta} d\Delta_{k_x} \Omega^{W_1}(\bs{k}^{\text{LW}}+\Delta_{k_x}) \notag \\
    & \ \ =-\big( \partial_{k_y} \theta_{ \bs{0}\bs{G}_1 }(\bs{k}^{\text{LW}}) +A^{\text{P}}_y(\bs{k}^{\text{LW}}+\bs{G}_1)-A^{\text{P}}_y(\bs{k}^{\text{LW}})\big),
\end{align}
where $\theta_{ \bs{0}\bs{G}_1 }(\bs{k}^{\text{LW}})$ corresponds to the phase difference between the amplitudes for occupying the left and right edges of the 1BZ, and $2\delta$ is the infinitesimal width of the boundary region. This includes the contribution to the Berry flux from both the left and right edges at a single point along them. The total contribution from the two edges can then be obtained by integrating this contribution parallel to the edges (along $k_y$):
\begin{align}
	\Phi_{W_1}&=-\int_{\text{LW}} dk_y \big( \partial_{k_y} \theta_{ \bs{0}\bs{G}_1 }(\bs{k}) +A^{\text{P}}_y(\bs{k}+\bs{G}_1)-A^{\text{P}}_y(\bs{k})\big).
\end{align}

Note that $-[\int_{W_1} dk_y (A^{\text{P}}_y(\bs{k}) - A^{\text{P}}_y(\bs{k}+\bs{G}_1))]$ is one third the total integral of the parent Berry connection around the 1BZ, due to the sixfold symmetry. As a result, this integral evaluates to one-third of the total parent Berry flux through the 1BZ: $\int dk_y (A^{\text{P}}_y(\bs{k}^{\text{LW}}) - A^{\text{P}}_y(\bs{k}^{\text{LW}}+\bs{G}_1))=-\frac{1}{3}\int_{\text{1BZ}}d^{2}k \Omega^{P}(\bs{k}) = -\frac{1}{3} \Phi^{\text{parent}}_{1BZ}$. Adding the contributions from the other two wires, this will give a $- \Phi^{\text{parent}}_{BZ}$ contribution that cancels with the occupation-weighted parent Berry curvature contribution from the bulk. As such, the total Berry flux of the crystal is
\begin{align}
	\Phi&= 3 \Phi_{W_1} + \int d^2k \Omega_{\text{bulk}}\\
	&=-3 \int_{K_2'}^{K_2}  dk_y \partial_{k_y} \theta_{ \bs{0}\bs{G}_1 }(\bs{k}) . \label{Eq_Berry_flux_total}
\end{align}
Here, it is important to note that $\theta$ is a compact variable. So we cannot immediately use the fact that this is the integral of an exact derivative to evaluate it.

We see that, just as for the sliver-patch model for the interacting case \cite{bernevig2025berry}, the total Berry flux is given by the winding of a phase difference along the boundary of the 1BZ. We have not carefully treated the corners of the 1BZ here. However, Ref.~\cite{bernevig2025berry} shows that this conclusion still holds when the corners are considered (making use of $C_6$ rotational symmetry, and time-reversal followed by a mirror operation).

%%%%%%%%%%%%%%%%%%%%%%%%%%%%%%%%%%%%%%%%%%%%%%%%%%%%%%%%%%%
\begin{figure}
    \centering
    \includegraphics[width=1.0\linewidth]{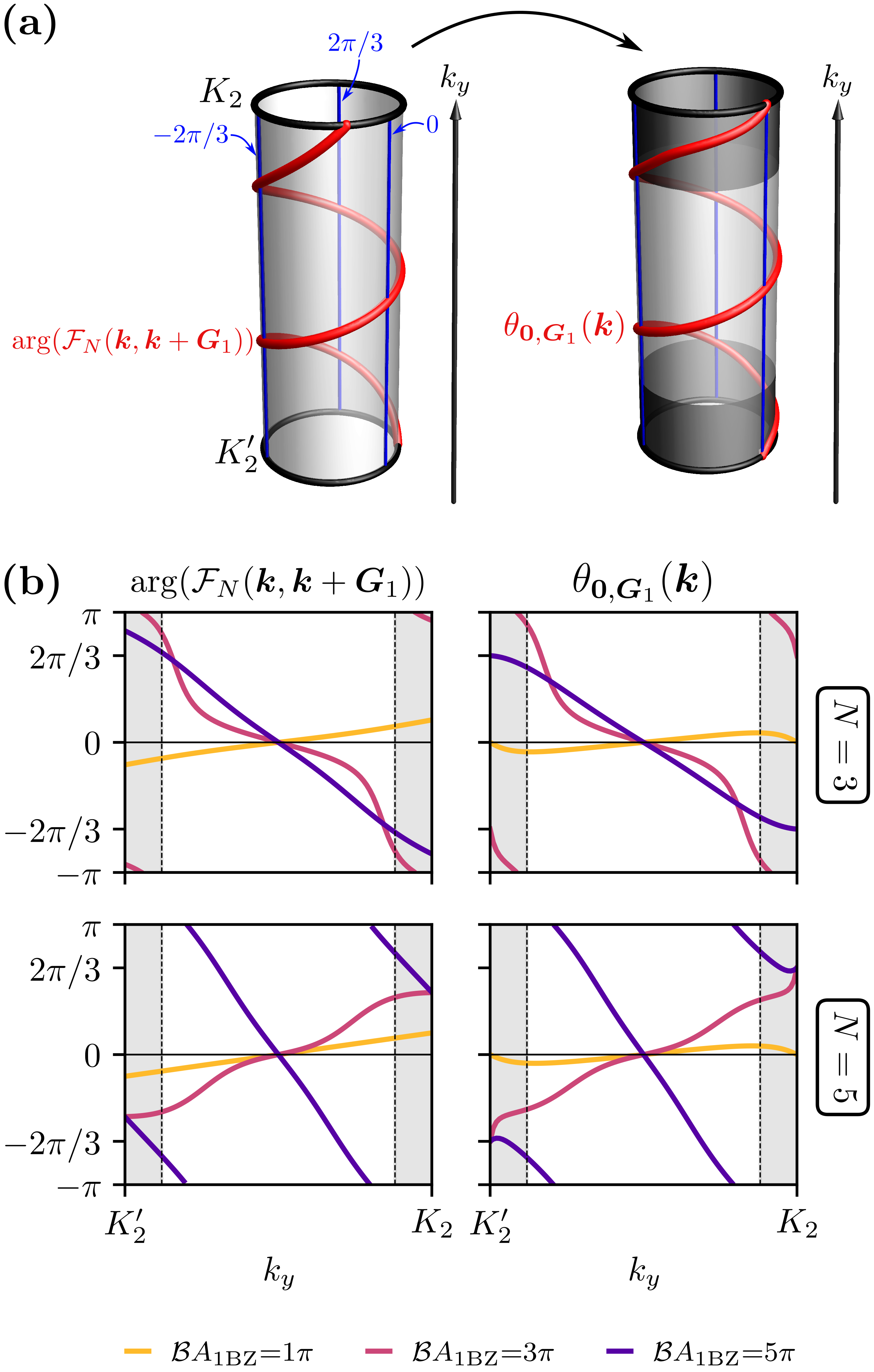}
    \caption{ 
    Form factor edge winding. (a) Schematic illustration of the mechanism where, along the left edge of the Brillouin zone, the phase difference $\theta_{\bs{0},\bs{G}_1}(\bs{k})$ (right) closely follows the phase of the form factor $\arg(\mathcal{F}_N(\bs{k},\bs{k}+\bs{G}_1))$ (left), but starts to deviate close to the corner of the Brillouin zone where three states almost have the same kinetic energy (dark region on the right) to lock in to the multiple of $2\pi/3$ (blue lines) closest to $\arg(\mathcal{F}_N(K_2, K_2+\bs{G}_1))$ at the top and $\arg(\mathcal{F}_N(K_2', K_2'+\bs{G}_1))$ at the bottom. (b) Evolution of the phase of the form factor between opposite edges $\arg(\mathcal{F}_N(\bs{k},\bs{k}+\bs{G}_1))$ (left column) and of the phase difference $\theta_{\bs{0},\bs{G}_1}(\bs{k})$ betweem $K_2'$ and $K_2$ (right column) for the $\lambda_N$ model with $N=3$ (first row) and $N=5$ (second row) for a weak potential of $2U_0m/g^2=0.15$ and different values of $\mathcal{B}$.
    }
\label{fig:ffew}
\end{figure}
%%%%%%%%%%%%%%%%%%%%%%%%%%%%%%%%%%%%%%%%%%%%%%%%%%%%%%%%%%%

\begin{figure}[b]
    \centering
    \vspace{2mm}
    \includegraphics[width=1.0\linewidth]{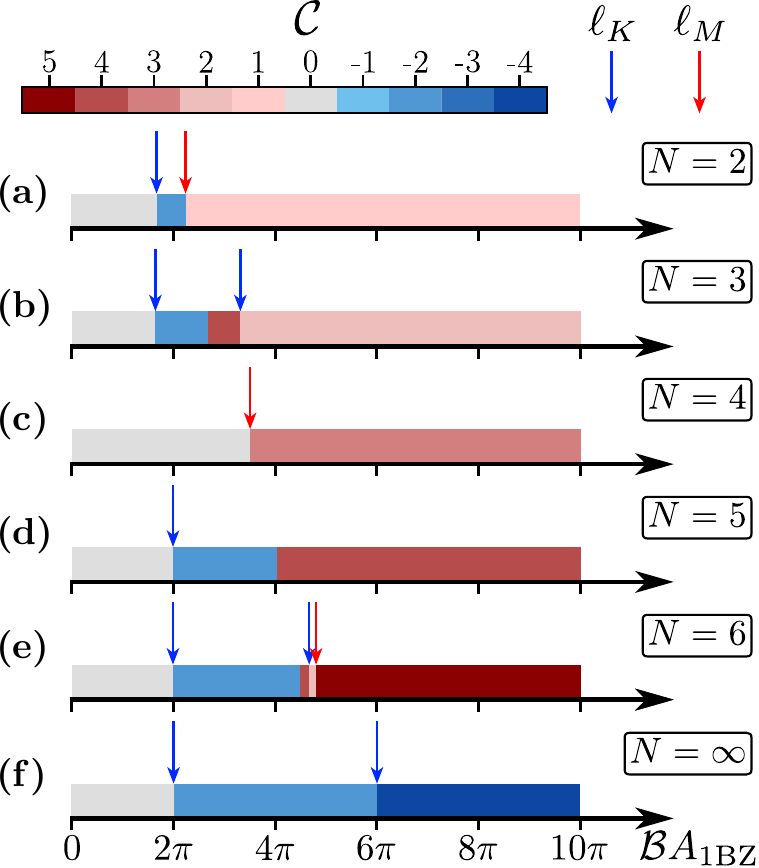}%\vspace{-4mm}
    \caption{Summary of the analytical results for the weak potential limit. Chern number of the band insulator as a function of $\mathcal{B}A_{\text{1BZ}}$ for (a) $N=2$, (b) $N=3$, (c) $N=4$, (d) $N=5$, (e) $N=6$, and (f) $N\to\infty$. The Chern number shown is the one predicted by Eq.~\eqref{Equation_Chern_low_U_capped_mt}. Red and blue arrows denote transitions predicted from minimizing the single-particle energy at the $M$ (Eq.~\eqref{eq:var_energy_pp_M_point}) and $K$ points (Eq.~\eqref{eq:energy_K_point_periodic_potential}), respectively.}
\label{fig:low_U_pp_analytic}
\end{figure}

To determine the total Berry flux, and thus the Chern number, we then need to construct the phase difference $\theta_{ \bs{0}\bs{G}_1 }$. Away from the 1BZ corner, only two states are mixed for each crystal momentum, which implies that we can directly minimize the hopping term Eq.~\eqref{eq:pp_hopping} to obtain $\theta_{\bs{0}\bs{G}_1}(\bs{k}) = \arg(\mathcal{F}_N(\bs{k}, \bs{k}+\bs{G}_1))$. However, at the corners, there is an interaction between three states, which generically frustrates the hopping terms. From the $C_6$ symmetry, the phase difference $\theta_{\bs{0} \bs{G}_1}$ is locked to a multiple of $2 \pi/3$ at the corner (for example, at $K'_2$ the phases $\theta_{\bs{0} \bs{G}_1}$, $\theta_{\bs{G}_1 \bs{G}_2}$, and $\theta_{\bs{G}_2 \bs{0}}$ are equal by symmetry, and must add to a multiple of $2 \pi$, see Eq.~\eqref{eq:form_eigenstates_high_symmetry_points_low_U0_K}). More specifically, the single-particle energy at the corners is minimized by setting $\theta_{\bs{0} \bs{G}_1}$ to the multiple of $2\pi/3$ closest to $\arg(\mathcal{F}_N(\bs{k}, \bs{k}+\bs{G}_1))$ evaluated at that point.
This form factor edge winding mechanism, including the corner rounding, is illustrated in Fig.~\ref{fig:ffew}. Fig.~\ref{fig:ffew}(b) shows how the phase difference $\theta_{\bs{0} \bs{G}_1}$ is locked to $\arg(\mathcal{F}_N(\bs{k}, \bs{k}+\bs{G}_1))$ in the middle of the wire (white region), but deviates smoothly closer to the corners (grey regions) to ultimately lock in to the multiple of $2\pi/3$ nearest to the angle of the form factor at the corners.
This tells us that the phase winding along the wire is
\begin{widetext}
\begin{align}
	\int_{K_2'}^{K_2}  dk_y \partial_{k_y} \theta_{ \bs{0}\bs{G}_1 } =& \int_{K_2'}^{K_2} \partial_{k_y}\arg(\mathcal{F}_N(\bs{k}, \bs{k}+\bs{G}_1))dk_y +  \arg(\mathcal{F}_N(\bs{K}'_{2}, \bs{K}'_{3})) - \lfloor \arg(\mathcal{F}_N(\bs{K}'_2, \bs{K}'_3))\rceil _{2 \pi /3} \notag \\
	&- \arg(\mathcal{F}_N(\bs{K_2}, \bs{K}_1)) + \lfloor\arg(\mathcal{F}_N(\bs{K}_2, \bs{K}_1 ))\rceil_{2 \pi /3}, 
\end{align}
where $\lfloor ...\rceil_{2\pi/3}$ denotes rounding to the nearest multiple of $2 \pi/3$. Here, the integral describes the phase winding along the main length of the edge, while the added terms are corrections that implement the phase rounding at the corners. From Eq.~\eqref{Eq_Berry_flux_total}, the Chern number is then
\begin{align} 
	\mathcal{C}	=\frac{\Phi}{2\pi}=& -\frac{3}{2\pi} \big[ \int_{K_2'}^{K_2} \partial_{k_y}\arg(\mathcal{F}_N(\bs{k}, \bs{k}+\bs{G}_1)) dk_y +  \arg(\mathcal{F}_N(\bs{K}'_2, \bs{K}'_3)) - \arg(\mathcal{F}_N(\bs{K}_2, \bs{K}_1)) \big] \notag \\
	&-\left\lfloor \frac{3}{2 \pi} \arg(\mathcal{F}_N(\bs{K}_2, \bs{K}_1))\right\rceil + \left\lfloor \frac{3}{2 \pi} \arg(\mathcal{F}_N(\bs{K}'_2, \bs{K}'_3))\right\rceil, \label{Equation_Chern_low_U_capped_mt}
\end{align}
\end{widetext}
where $\lfloor...\rceil$ indicates rounding to the nearest integer. As a final step, we substitute the expression for the form factor for the desired parent band. For the $N\rightarrow \infty$ limit, this results in $C=-2\lfloor\frac{\mathcal{B} A_{1BZ}}{4 \pi}\rceil$, recovering the expected value \cite{tan2024parent}. For finite values of $N$, the predicted Chern number, shown in Fig.~\ref{fig:low_U_pp_analytic}, agrees well with the exact diagonalization results presented in Fig.~\ref{fig:phase_diagram_periodic_potential}.

One feature of the phase diagram that this full Chern number determination can explain is the sign structure of the Chern number in the low $U_0$ limit. For the $N\rightarrow \infty$ limit, the phase winding of the form factor always has the same sign, resulting in a negative Chern number for all values of $\mathcal{B}$. On the other hand, for finite $N$, the derivative of the form factor can change sign as $\mathcal{B}$ increases, leading to the Chern number also swapping from negative to positive.

\subsubsection{Large $\mathcal{B}$ and $U_0$ limit}
\label{Section_potential_large_B}

We now move on to explore analytically another region of the phase diagrams of Fig.~\ref{fig:phase_diagram_periodic_potential}. In the limit of large $\mathcal{B}A_{\text{1BZ}}$, where the Berry curvature is only significant very close to the parent band zone center, a $\mathcal{C}=0$ band insulator is present in the large $U_0$ limit for any finite $N$ (see Fig.~\ref{fig:phase_diagram_periodic_potential}(a)-(e)). In particular, in the same large $\mathcal{B}A_{\text{1BZ}}$ limit, it appears that we systematically have a state with $(\ell_\Gamma, \ell_K,\ell_M)=(0,(-N+1) \mod 3, (-N+1) \mod 2)$ at small $U_0$ that then transitions to the $\mathcal{C}=0$ band insulator with  $(\ell_\Gamma, \ell_K,\ell_M)=((-N+1)\mod 6,(-N+1) \mod 3, (-N+1) \mod 2)$ as $U_0$ increases. There is a natural way to understand such generic features by treating the localized Berry curvature as a vortex at the origin, with the parent band states exhibiting a total phase winding of $2\pi(N-1)$ around the vortex.

To see this, we examine the form factor~\eqref{eq_form_factor_explicit}. When $|\mathcal{B} q_{z}q_{\bar{z}}'| \gg N$, the form factor and normalization constants are dominated by the highest-order contribution, such that 
\begin{align}	 \label{eq:approx_form_factor_large_B}
	\mathcal{F}_{N}(\bs{q}, \bs{q}') & \approx  \frac{(q_z q^{\prime}_{\bar{z}})^{N-1}}{(|q_z| |q^{\prime}_z|)^{N-1}}= e^{i(N-1)(\phi_{q}-\phi_{q'})}.
\end{align}
 The phase of this approximate form factor is given by $\phi_F(\boldsymbol{q}, \boldsymbol{q}') = (N-1) (\phi_{q} - \phi_{q'})$ where $\phi_q$ is the phase of $q_{z}$ (equivalent to the angle of vector $\bs{q}$ from the $x$-axis). The reason for this is that the spinor states far from the ``core'' (where the Berry curvature is present) in the underlying model are actually the same (polarized into the spinor state $\ket{N-1}$, i.e., see Eqs.~\eqref{eq:lambda_N_nullspace_condition} and~\eqref{eq:lambda_N_model_coefficients_wf}). However, the core inserts a $2(N-1) \pi$ Berry flux, which must be reflected by a phase winding of the states far from it to have a smooth manifold of states. In this sense, the parent band Bloch states far from the core are trivial apart from this phase, so their overlap (the form factor) is just a phase. 
 
If $\mathcal{B}A_{\text{1BZ}}$ is large, this simple form for the form factor will hold for all $\bs{q}=\bs{k}+\bs{g}$, $\bs{q'}=\bs{k}'+\bs{g}'$ except in a small region around the $\Gamma$ point of the parent band. As a result, the energy term describing the hopping between states $\bs{k}+\bs{g}$ and $\bs{k}+\bs{g}+\bs{G}_i$~\eqref{eq:pp_hopping} is given by
\begin{widetext}
\begin{align}
    E^{\bs{g}, \bs{g}+\bs{G}_i}_{\text{hopp}}(\bs{k})
    \approx& -2U_0 |v_{\bs{g}+\bs{G}_i}(\bs{k})| |v_{\bs{g}}(\bs{k})| 
     \cos\big(
    (N-1) (\phi_{k + g + G_i} - \phi_{k + g}) + \arg[v_{\bs{g}}(\bs{k})] - \arg[v_{\bs{g}+\bs{G}_i}(\bs{k})]\big),
\end{align}
\end{widetext}
which can be minimised everywhere that the limit holds, by taking the phase condition
\begin{equation}
    \arg[v_{\bs{g}}(\bs{k})]= (N-1) \phi_{k+g}. \label{Eq_phase_condition}
\end{equation}

We note that the above regime does not exist in the $ N \to \infty$ limit. Since the Berry curvature is constant in momentum space, it cannot be made to be concentrated only in a small area of the first Brillouin zone, no matter how large $\mathcal{B}$ is. As such, there does not exist any regime where the form factor~\eqref{eq:LLL_form_factor} can be approximated by~\eqref{eq:approx_form_factor_large_B}. However, the above-discussed limit exists for any finite $N$ model that encloses a finite amount of Berry curvature.

\paragraph*{\underline{(i) High-Symmetry Points: }} We can use this phase condition by first considering only high-symmetry points. Indeed, given Eq.~\eqref{Eq_phase_condition}, a single-particle state~\eqref{eq:generic_single_particle_state_at_high_symmetry_momentum} at high-symmetry point $\bs{\kappa}\in\{\Gamma,K,M\}$ transforms as
\begin{align}
    \hat{R}_{2\pi m/n_{\bs{\kappa}}} \ket{\varphi_{\bs{\kappa}}} &= \sum_{\bs{g}} |v_{\bs{g}}(\bs{\kappa})| e^{i\phi_{\kappa+g}} \ket{\psi^{(0)}_{R_{2\pi m/n_{\bs{\kappa}}}(\bs{\kappa}+\bs{g})}} \nonumber \\
    &= \exp[-i (N-1)\frac{ 2 \pi m }{n_{\bs{\kappa}}}] \ket{\varphi_{\bs{\kappa}}}, \label{eq:transformation_high_symmetry_point_high_B_limit}
\end{align}
from which we can directly read off the symmetry eigenvalues at the $K$ and $M$ points
$$ \ell_K = -(N-1) \text{ mod }3, \ \ell_M = -(N-1) \text{ mod }2.$$

At the $\Gamma$ point, if $U_0$ is small, the kinetic energy must be minimized, leading to a unity occupation of $\ket{\psi_{\bs{0}}^{(0)}}$. As discussed previously, this locks the irreducible representation to be trivial (i.e., $\ell_\Gamma=0$).

On the other hand, requiring the $\ell_\Gamma=0$ would generally frustrate the hopping terms between the $\Gamma$ points in further BZs. These points are away from the Berry curvature, and so the hopping terms would be minimized by the same phase condition Eq.~\eqref{Eq_phase_condition} described earlier. In the large $U_0$ limit, this leads to the indicator at the $\Gamma$ point becoming
$$\ell_{\Gamma}=-(N-1) \text{ mod }6$$
from Eq.~\eqref{eq:transformation_high_symmetry_point_high_B_limit}.
As a result, the Chern number~\eqref{eq:chern_number_from_symmetry_indicators} in the large $\mathcal{B}A_{\text{1BZ}}$ large $U_0$ limit is given by
\begin{align}
    \mathcal{C} \text{ mod }6 &= \big[-(N-1) - 2 [(N-1) \text{ mod 3}] \notag \\
    & \ \ - 3 [(N-1) \text{ mod 2}]\big] \text{ mod 6} \notag\\
    &=0.
\end{align}
We can also obtain the Chern number for the large $\mathcal{B}$ but low $U_0$ limit by locking the $\Gamma$ indicator to zero, giving $\mathcal{C}= 0- -(N-1) =N-1$. Both of these limits are clearly reproduced in Fig.~\ref{fig:phase_diagram_periodic_potential}(a)-(e). 

%%%%%%%%%%%%%%%%%%%%%%%%%%%%%%%%%%%%%%%%%%%%%%%%%%%%%%%%%%%
%%%%%%%%%%%%%%%%%%%%%%%%%%%%%%%%%%%%%%%%%%%%%%%%%%%%%%%%%%%

\paragraph*{\underline{(ii) Full Chern Number: }} We can once again go beyond high-symmetry points and obtain the full Chern number in this limit. To do so, we substitute the phase condition~\eqref{Eq_phase_condition} into the general expression~\eqref{Eqn_Berry_curvature_generic_main}. We have $\theta_{\bs{g} \bs{g}'} = (N-1) (\arg(\bs{k}+\bs{g})- \arg(\bs{k}+\bs{g}'))$. Therefore,
\begin{align*}
    \nabla \theta_{\bs{g} \bs{g}'} 
    &=\frac{N-1}{|\bs{k}+\bs{g}|} \hat{\phi}(\bs{k}+\bs{g}) -  \frac{N-1}{|\bs{k}+\bs{g}'|} \hat{\phi}(\bs{k}+\bs{g}'),
\end{align*}
where $\hat{\phi}(\bs{k}+\bs{g})$ is the azimuthal unit vector. Similarly, the parent Berry connection is $\bs{A}^{\text{P}}(\bs{k}+\bs{g})= \frac{ (N-1)}{|\bs{k}+\bs{g}|}\hat{\phi}(\bs{k}+\bs{g})$, using the fact that the Berry connection satisfies rotational symmetry and $\oint_{k\gg \sqrt{N/\mathcal{B}}} \bs{A}^{\text{P}} \cdot d\bs{k}= \Phi=2\pi(N-1)$. As a result, $\partial_{k_x} \theta_{\bs{g} \bs{g}'}  -A_x^{\text{P}}(\bs{k}+\bs{g}) + A_x^{\text{P}}(\bs{k}+\bs{g}') =0$ and the extra part of the Berry curvature in Eq.~\eqref{Eqn_Berry_curvature_generic_main} vanishes. Therefore, the Berry curvature away from the $\Gamma$ point is 
$$\Omega(\bs{k})= \sum_{\bs{g}} n(\bs{k}+\bs{g}) \Omega^{\text{P}}(\bs{k}+\bs{g}).$$ That is, the Berry curvature becomes the occupation-weighted parent Berry curvature. At large $U_0$, the parent band occupation is pushed out of the region with significant parent Berry curvature (which is the same region where the phase condition is not satisfied), so the Chern number is zero. At low $U_0$, on the other hand, the occupation is significant in the region of large Berry curvature, so there is a small occupied region where the phase condition is not satisfied. However, as we discussed in Section \ref{Sec_weak_potential}, the parent band occupation in the bulk of the 1BZ is near unity when $U_0$ is small. Because of this, the Berry curvature is still given by the occupation-weighted parent Berry curvature (see Eq.~\eqref{Eq_Berry_bulk}), even though the phase condition is not satisfied near the $\Gamma$ point. The total Berry flux is then the parent Berry flux through the 1BZ, giving $\mathcal{C}=N-1$.

%%%%%%%%%%%%%%%%%%%%%%%%%%%%%%%%%%%%%%%%%%%%%%%%%%%%%%%%%%%%%
%                Strong Interaction                         %
%%%%%%%%%%%%%%%%%%%%%%%%%%%%%%%%%%%%%%%%%%%%%%%%%%%%%%%%%%%%%
\section{Spontaneous Crystallization} \label{sec:spontaneous_crystallization}

%%%%%%%%%%%%%%%%%%%%%%%%%%%%%%%%%%%%%%%%%%%%%%%%%%%%%%%%%%%
\begin{figure*}
    \centering
    \includegraphics[width=1.0\textwidth]{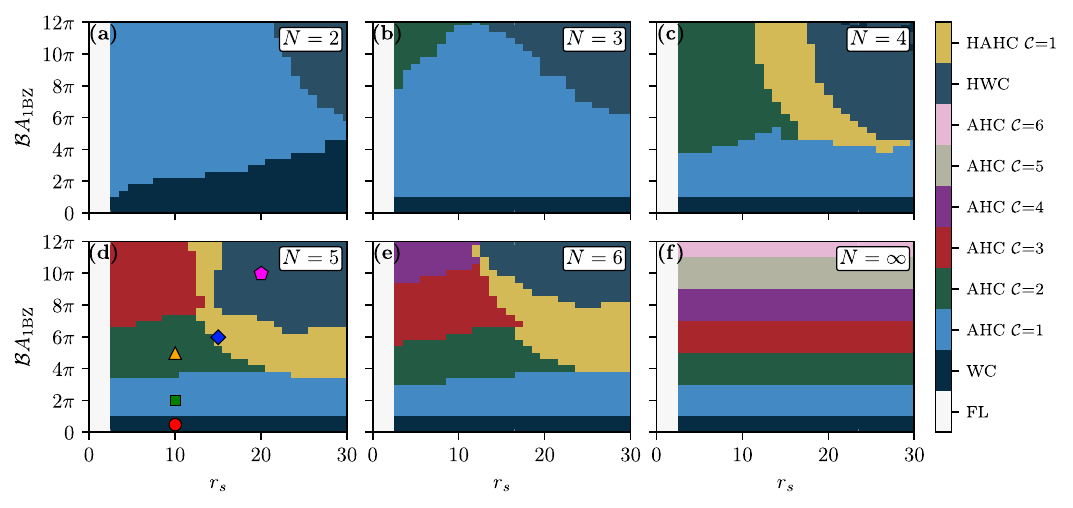}\vspace{-3mm}
    \caption{Hartree-Fock phase diagram of the $\lambda_N$ jellium model~\eqref{eq:lambda_n_int} for (a) $N=2$, (b) $N=3$, (c) $N=4$, (d) $N=5$, (e) $N=6$, (f) $N\to\infty$ as a function of $\mathcal{B}A_{\text{1BZ}}$ and $r_s$. The phase diagram identifies stable regions of Fermi liquid (FL), Wigner crystal (WC), anomalous Hall crystals with different Chern numbers $\mathcal{C}$, halo Wigner crystal (HWC) where $\mathcal{C}=0$ with $\ell_\Gamma\ne 0$, and a $\mathcal{C}=1$ halo anomalous Hall crystal (HAHC) with $\ell_\Gamma\ne 0$. Results are obtained by discretizing the first Brillouin zone with a $12 \times 12$ grid and keeping the 61 closest reciprocal lattice vectors.}
\label{fig:hf_phase_diagram}
\end{figure*}
%%%%%%%%%%%%%%%%%%%%%%%%%%%%%%%%%%%%%%%%%%%%%%%%%%%%%%%%%%%

We now turn to study the spontaneous breaking of continuous translation symmetry by electronic interactions~\cite{wigner1934interaction, pines1952collective, bohm1953collective, bonsall1977some, fogler2000dynamical, chitra2001pinned, tsui2024direct} in the presence of non-uniform Berry curvature. As for studies of crystallization in two-dimensional electron gas, it will be helpful to introduce the dimensionless Wigner-Seitz radius $r_s$ defined as the radius of a disk containing a single electron in units of the Bohr radius $a_B=4\pi\epsilon_0\hbar^2/(e^2 m)$. The average inter-electronic distance is then defined as $a_e = r_s a_B$, and $r_s$ characterizes the ratio of the average interaction to kinetic energy. As is conventional in numerical studies of the two-dimensional electron gas~\cite{ceperley1978ground, tanatar1989ground, gori2004pair, varsano2001spin, rapisarda1996diffusion, drummond2009phase, drummond2009quantum, azadi2024quantum, valenti2025critical, azadi2025quantum, soejima2025jellium}, we work in density-dependent units where the length is measured in units of $a_e$ (i.e., $a_e=1$) and we set $m=\hbar=e=1$, where $e$ is the charge of an electron. In such units, the continuum interacting model projected to the lowest band of the $\lambda_N$ model takes the form
\begin{subequations}\label{eq:hamiltonian_interacting_model}
\begin{align}
    \hat{\mathcal{H}} = \hat{\mathcal{H}}_0 + \hat{\mathcal{H}}_{\text{int}},
\end{align}
where
\begin{equation}
    \hat{\mathcal{H}}_0 = \sum_{\bs{q}} \frac{|\bs{q}|^2}{r_s^2} c_{\bs{q}}^\dagger c^{\phantom{\dagger}}_{\bs{q}}
\end{equation}
and
\begin{equation}
\begin{aligned}
    \hat{\mathcal{H}}_{\text{int}} &= \frac{1}{2 A} \sum_{\bs{q}} V(\bs{q}) :\tilde{\rho}_{\bs{q}} \tilde{\rho}_{-\bs{q}}: \\
    &= \frac{1}{2 A} \sum_{\boldsymbol{q}_1,\boldsymbol{q}_2, \boldsymbol{q}_3,\bs{q}_4} \tilde{V}_{\bs{q}_1\bs{q}_2\bs{q}_3\bs{q}_4} c_{\bs{q}_1}^{\dagger} c_{\boldsymbol{q}_2}^{\dagger} c_{\boldsymbol{q}_3} c_{\boldsymbol{q}_4}.
\end{aligned}
\end{equation}
\end{subequations}
In the above, $:\hat{A}:$ denotes normal ordering and 
\begin{align}\label{eq:projected_coulomb_interaction}
    \tilde{V}_{\bs{q}_1\bs{q}_2\bs{q}_3\bs{q}_4} =& V(\boldsymbol{q}_1 - \bs{q}_4) \mathcal{F}_{N}(\bs{q}_1, \bs{q}_4) \mathcal{F}_{N}(\bs{q}_2, \bs{q}_3) \nonumber \\
    &\quad \times \delta(\boldsymbol{q}_1 + \boldsymbol{q}_4 - \boldsymbol{q}_2 -\boldsymbol{q}_3)
\end{align}
is the band-projected interaction with the unscreened Coulomb potential $V(\boldsymbol{q}) = 4 \pi/(r_s |\bs{q}|)$. In our convention, the energy is measured in units of Rydberg (Ry) and the Fermi momentum is $k_F = 2/(r_s a_B) = 2$. We note that, in contrast to the usual jellium model where $a_e$ is the only length scale, the presence of the Berry curvature introduces a new ``\emph{Berry length}'' $\sqrt{\mathcal{B}}$, which has the same interpretation as the magnetic length in the $N\to\infty$ limit.

%%%%%%%%%%%%%%
\subsection{Numerical Results}

We study the above problem within the Hartree-Fock approximation, where the energy is optimized over the space of Slater determinants. In particular, we allow for spontaneous breaking of continuous translation symmetry. This results in the Hartree-Fock Hamiltonian
\begin{subequations} \label{eq:lambda_n_int}
\begin{align}
    \hat{\mathcal{H}}_{\text{HF}} &= \hat{\mathcal{H}}_0 + \hat{\mathcal{H}}_{\text{H}} + \hat{\mathcal{H}}_{\text{F}},
\end{align}
with
\begin{align}
    \hat{\mathcal{H}}_0 &= \sum_{\bs{k},\bs{g}_1} \frac{|\bs{k}+\bs{g}_1|^2}{r_s^2} c_{\bs{k}\bs{g}_1}^\dagger c^{\phantom{\dagger}}_{\bs{k}\bs{g}_1} \label{Eq_Kinetic} \\
    \mathcal{H}_{\mathrm{H}} &= \frac{1}{A} \sum_{\substack{\boldsymbol{k}_1 \boldsymbol{k}_2 \\\boldsymbol{g}_1 \boldsymbol{g}_2 \boldsymbol{g}_3 \boldsymbol{g}_4}} \tilde{V}_{\bs{k}_1+\bs{g}_1,\bs{k}_2+\bs{g}_2,\bs{k}_2+\bs{g}_3,\bs{k}_1+\bs{g}_4}   \nonumber \\
    &\hspace{1.5cm} \times \mathcal{P}_{\boldsymbol{g}_1 \boldsymbol{g}_4}(\boldsymbol{k}_1) c_{\boldsymbol{k}_2 \boldsymbol{g}_2}^{\dagger} c^{\phantom{\dagger}}_{\boldsymbol{k}_2 \boldsymbol{g}_3} \label{Eq_Hartree}  \\ %%%%%%%%
    \mathcal{H}_{\mathrm{F}} &= -\frac{1}{A} \sum_{\substack{\boldsymbol{k}_1 \boldsymbol{k}_2 \\\boldsymbol{g}_1 \boldsymbol{g}_2 \boldsymbol{g}_3 \boldsymbol{g}_4}} \tilde{V}_{\bs{k}_1+\bs{g}_1,\bs{k}_2+\bs{g}_2,\bs{k}_1+\bs{g}_3,\bs{k}_2+\bs{g}_4} \nonumber \\
    &\hspace{1.5cm} \times
    \mathcal{P}_{\boldsymbol{g}_1 \boldsymbol{g}_3}(\boldsymbol{k}_1) c_{\boldsymbol{k}_2 \boldsymbol{g}_2}^{\dagger} c^{\phantom{\dagger}}_{\boldsymbol{k}_2 \boldsymbol{g}_4}. \label{Eq_Fock}
\end{align}
\end{subequations}
Here, $\boldsymbol{g} = n_1 \boldsymbol{G}_1+$ $n_2 \boldsymbol{G}_2$ ($n_1, n_2 \in \mathbb{Z}$) once again denotes reciprocal lattice vectors of the lattice that spontaneously forms upon breaking of continuous translation by the electronic state. We assume hereafter that the system crystallizes into a triangular lattice, as in the conventional jellium model. It should be noted that this assumption may not necessarily be justified here~\cite{desrochers2025elastic, dong2025phonons}. For instance, the triangular lattice was already found to be unstable for the $N=2$ model in some parameter regime~\cite{dong2025phonons}. We defer a detailed investigation of lattice stability in the $\lambda_N$ model to future work. For a triangular lattice, the lattice constant is $a_l = \sqrt{2/(\sqrt{3} \pi)}$ as we have a unit cell of area $A_{\text{u.c.}}=\pi$ that encloses a single electron.

Solving the Hartree-Fock self-consistency conditions for different values of $N$, we obtain the phase diagrams in Fig.~\ref{fig:hf_phase_diagram}. At small interactions, a Berry fermi liquid~\cite{chen2016berry, chen2017berry, huang2024effective} (i.e., a metallic phase that preserves continuous translation symmetry in a Berry curvature background) is present for any $N$. As interaction increases beyond $r_s \approx 2$, a direct gap in the single-particle spectrum opens as continuous translation symmetry is spontaneously broken, resulting in a Wigner crystal (WC) with $\mathcal{C}=0$ or topological anomalous Hall crystals (AHC) with $\mathcal{C}\ne 0$ depending on values of $N$ and $\mathcal{B}A_{\text{1BZ}}$. In particular, we see how the $N\to\infty$ phase diagram, where crystals with $\mathcal{C}=\lfloor \mathcal{B}A_{\text{1BZ}} / (2\pi) \rceil$ are predicted independently of $r_s$~\cite{tan2024parent}, slowly emerges as $N$ is increased. 

\begin{figure*}
    \centering
    \includegraphics[width=1.0\textwidth]{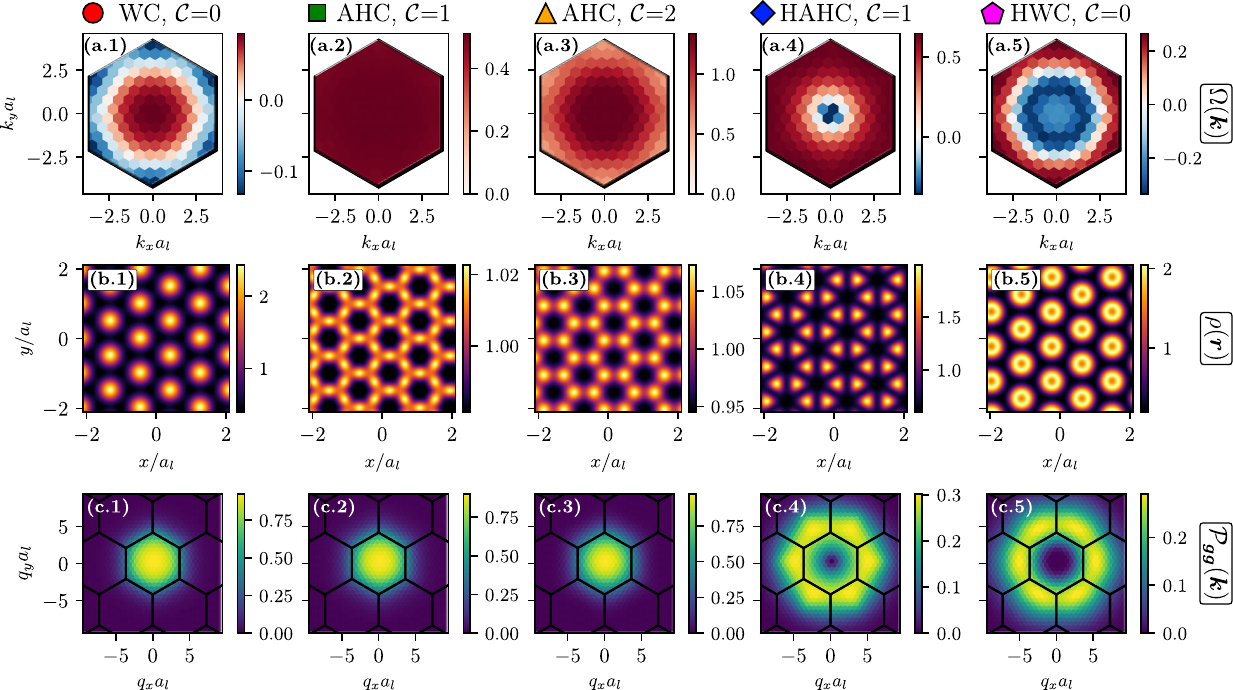}\vspace{-4mm}
    \caption{Observables of representative phases. (a) Berry curvature $\Omega(\bs{k})$, (b) real space density $\rho(\bs{r})$, and (c) momentum space occupations of the parent band states $\mathcal{P}_{\bs{g}\bs{g}}(\bs{k})$ for the (1) WC, (2) AHC with $\mathcal{C}=1$, (3) AHC with $\mathcal{C}=2$, (4) HAHC with $\mathcal{C}=1$, and (5) HWC. Representattive phases are taken for $N=5$ with (1) $r_s=10$ and $\mathcal{B}A_{\text{1BZ}}=\pi/2$, (2) $r_s=10$ and $\mathcal{B}A_{\text{1BZ}}=2\pi$, (3) $r_s=10$ and $\mathcal{B}A_{\text{1BZ}}=5\pi$, (4) $r_s=15$ and $\mathcal{B}A_{\text{1BZ}}=6\pi$, and (5) $r_s=20$ and $\mathcal{B}A_{\text{1BZ}}=10\pi$ (see markers in panel (d) of Fig.~\ref{fig:hf_phase_diagram}).}
\label{fig:hf_info_phases}
\end{figure*}

For finite $N$, however, the phase diagram remains considerably more intricate. In particular, we find at large $r_s$ and $\mathcal{B}A_{\text{1BZ}}$ a halo Wigner crystal (HWC) as in Ref.~\cite{soejima2025jellium} and predicted semiclassically for Bernal bilayer graphene in Refs.~\cite{joy2023wigner, joy2025chiral}. Although such a crystal also has $\mathcal{C}=0$ (Fig.~\ref{fig:hf_info_phases}(a.5)), it differs from WCs by its non-vanishing $C_6$ eigenvalue at the $\Gamma$ point (i.e., $\ell_\Gamma\ne 0$). As discussed above for the periodic potential case, $\ell_\Gamma\ne 0$ implies a vanishing occupation of the parent band single-particle state at the zone center (see Fig.~\ref{fig:hf_info_phases}(c.5)). Naturally, such phases arise only at large $r_s$, where interactions are strong enough to offset the substantial kinetic energy cost associated with significantly occupying single-particle states beyond the first Brillouin zone (Fig.~\ref{fig:hf_info_phases}(c)). This also explains why all electronic crystals have $\ell_\Gamma=0$ at smaller values of $r_s$. 

%%%%%%%%%%%%%%%%%%%%%%%%%%%%%%%%%%%%%%%%%%%%%%%%%%%%%%%%%%%
%%%%%%%%%%%%%%%%%%%%%%%%%%%%%%%%%%%%%%%%%%%%%%%%%%%%%%%%%%%

Starting at $N=4$, a notable addition to the phase diagram is the emergence of a new phase: a $\mathcal{C}=1$ halo anomalous Hall crystal (HAHC). In this phase, the crystal exhibits both $\mathcal{C}\ne 0$ and $\ell_\Gamma\ne 0$, as shown in Fig.~\ref{fig:hf_info_phases}(a.4-c.4). To the best of our knowledge, this is the first time such a phase has been reported. We discuss it in more detail in Sec.~\ref{subsec:HAHC}.

We emphasize that caution should be exercised when interpreting the above Hartree-Fock phase diagrams. Just as the correlation energy shifts the spinless Wigner crystal transition from $r_s \approx 2$ to $r_s \approx 30$~\cite{drummond2009phase, drummond2009quantum, azadi2024quantum, valenti2025critical, azadi2025quantum, mahan2013many}, beyond mean-field fluctuations should significantly modify the phase boundaries in Fig.~\ref{fig:hf_phase_diagram}. 
%In particular, since the presence of Berry curvature screens the interaction (see Eq.~\eqref{eq:projected_coulomb_interaction}), the critical value of $r_s$ for a transition between the Fermi liquid and electronic crystal should be larger in the $\lambda_N$ model with $\mathcal{B}\ne 0$ than for the bare jellium model, as in the presence of metallic gates~\cite{valenti2025critical} 
If we take the naive point of view that correlation effects significantly alter the balance between the liquid and crystal states, but do not modify substantially the energetic competition between different crystalline states, then $\lambda_N$ models with $N>2$ may be much more promising for future numerical work to confirm the stability of AHCs beyond mean-field. Indeed, the $\mathcal{C}=1$ AHC with $N\ge 3$ is stable over an extended range of $\mathcal{B}$ beyond $r_s=30$ (Fig.~\ref{fig:hf_phase_diagram}(b-f)), whereas the AHC with $N=2$ appears much more fragile and is only stable over a narrow range of $\mathcal{B}$ for beyond $r_s=30$ (Fig.~\ref{fig:hf_phase_diagram}(a)).

%%%%%%
\subsection{Analytical Insights}

Similar to the periodic potential case, we can gain some analytical insights into the behavior of the interacting crystal by taking appropriate limits. We first consider the case where the interaction is weak. In this case, the parent band occupation is largely confined to the 1BZ, and the Chern number is reasonably well predicted by the ``Berry flux rounding'' argument of Refs.~\cite{dong2024stability,bernevig2025berry} that explains the Chern number increasing in unit steps as $\mathcal{B}$ is increased. The other case we consider is the large $\mathcal{B}$, strong interaction case, for which we always see a $\mathcal{C}=0$ HWC for finite $N$.

\subsubsection{Small $r_s$ limit --- Berry flux rounding}
\label{Section_weak_interaction}

We first consider the weak interaction limit. The crystal state does not prevail over the Fermi liquid state, even at the HF level, until a finite interaction strength. However, we can still compute the Chern number of the HF state as long as a direct band gap is present. Then, because the Chern number is a property of the phase, it is preserved as we go from the zero interaction limit to the small but finite interaction case where the crystal state is present \cite{dong2024stability}.

In the self-consistent Hartree-Fock approximation, the Hamiltonian consists of the kinetic~\eqref{Eq_Kinetic}, Hartree~\eqref{Eq_Hartree}, and Fock~\eqref{Eq_Fock} terms. Of these, we expect the largest contribution to come from the kinetic term, followed by the Fock term. In particular, the small momentum transfer part of the Fock term should be the most significant of the interaction terms \cite{dong2024stability}. This is because the potential $V(\bs{q}) \sim 1/|\bs{q}|$, and the form factors also tend to decay at large momentum transfers, so small $q$ contributions dominate (and the Hartree term has no small $q$ part). However, we expect that neglecting other terms will be less justified for the $\lambda_N$ model compared to the infinite Chern band \cite{tan2024parent}, because the form factor does not decay exponentially with momentum transfer (see Eq.~\eqref{eq:LLL_form_factor}). This is particularly significant when the Berry curvature is localised near the gamma point, because the form factor is constant in magnitude for momenta well outside of the region with large Berry curvature (i.e., Eq.~\eqref{eq:approx_form_factor_large_B}).

If the small-$\bs{q}$ Fock term does indeed dominate over the other interaction terms, but is weak compared to the kinetic energy, we expect the ``Berry flux rounding'' discussed in Refs.~\cite{dong2024stability,bernevig2025berry} to hold. That is, the Chern number of the crystal can be obtained by rounding the parent Berry flux enclosed by the first Brillouin zone:
\begin{align}
    \mathcal{C}= \left\lfloor \int_{\text{1BZ}}d^2k \Omega_N(\bs{k}) / (2\pi) \right\rceil, \label{eq:berry_flux_rounding}
\end{align}
where $\Omega_N(\bs{k})$ is the Berry curvature of the parent band (Eq.~\eqref{eq:berry_curvature_lambda_N}), not of the Hartree-Fock band. In Table~\ref{tab:table_flux_rounding}, we show the values predicted by Berry flux rounding for the critical $\mathcal{B}A_{\text{1BZ}}$ where a transition between electronic crystal states with Chern numbers $a$ and $a+1$ should occur at small $r_s$ (i.e., $\mathcal{B}A_{\text{1BZ}}$ where $\int_{\text{1BZ}}d^2k \Omega_N(\bs{k}) = \pi(2a+1)$). The transitions predicted are in reasonable agreement with the numerical results of Fig.~\ref{fig:hf_phase_diagram}. In particular, we see that the $N=4$, $N=5$, and $N=6$ phase diagrams should contain AHC with $\mathcal{C}=3$, $\mathcal{C}=4$, and $\mathcal{C}=5$, respectively, at larger values of $\mathcal{B}A_{\text{1BZ}}$ than those presented in Fig.~\ref{fig:hf_phase_diagram}. In general, the Berry flux rounding argument predicts that the $\lambda_N$-jellium model should stabilize AHC with Chern number up to $\mathcal{C}=N-1$, considering the parent band contains a total Berry flux of $2\pi(N-1)$ (see Eq.~\eqref{eq:total_berry_flux_in_lambda_n}). Because the total parent Berry flux through the 1BZ increases smoothly with $\mathcal{B}$, the Chern number increases in steps of one from 0 to this maximum value of $N-1$. This is different from the weak periodic potential case discussed in Sec. \ref{Sec_weak_potential}, which depends on the winding of the form factor phase, because that winding can change sign as $\mathcal{B}$ changes.

\begin{table}
\caption{\label{tab:table_flux_rounding}%
Critical value of $\mathcal{B}A_{\text{1BZ}}$ for the transition between electronic crystal states with Chern number $a\leftrightarrow a+1$ predicted from Berry flux rounding at small $r_s$ for different $N$.
}
\begin{ruledtabular}
\begin{tabular}{c|ccccc}
&
$N=2$ & $N=3$ & $N=4$ & $N=5$ & $N=6$ \\
\colrule
$0\leftrightarrow 1$ & $2.001\pi$ & $1.10\pi$ & $1.01\pi$ & $1.00\pi$ & $1.00\pi$ \\
$1\leftrightarrow 2$ & --- & $7.34\pi$ & $3.70\pi$ & $3.18\pi$ & $3.05\pi$ \\
$2\leftrightarrow 3$ & --- & --- & $13.07\pi$ & $6.58\pi$ & $5.50\pi$ \\
$3\leftrightarrow 4$ & --- & --- & --- & $18.93\pi$ & $9.59\pi$ \\
$4\leftrightarrow 5$ & --- & --- & --- & --- & $24.86\pi$
\end{tabular}
\end{ruledtabular}
\end{table}

\subsubsection{Large $\mathcal{B}$ limit}

The other limit we consider is where the Berry curvature of the parent band is highly localized and the interaction is strong. In the HF phase diagram, we always observe a HWC in this limit, as shown in Fig.~\ref{fig:hf_phase_diagram}. As we now explain, the origin of this phase is similar to the $\mathcal{C}=0$ phase observed in the periodic potential case, in the analogous limit of strong potential and localized Berry curvature.

As we discussed for the periodic potential case in Section~\ref{Section_potential_large_B}, the form factor of the $\lambda_N$ model can be expressed as a pure phase (see Eq.~\eqref{eq:approx_form_factor_large_B}) for momenta away from the localized Berry curvature. Using this limit of the form factor, the Fock term takes the form
\begin{align}
    \mathcal{H}_{\mathrm{F}} =& -\frac{1}{A} \sum_{\substack{\boldsymbol{k}_1 \boldsymbol{k}_2 \\\boldsymbol{g}_1 \boldsymbol{g}_2 \boldsymbol{g}_3 \boldsymbol{g}_4}} V\left(\boldsymbol{k}_1+\boldsymbol{g}_1-\boldsymbol{k}_2-\boldsymbol{g}_4 \right) \notag \\
    & \times e^{i(N-1)(\phi_{k_1+g_1}-\phi_{k_2+g_4})} e^{i(N-1)(\phi_{k_2+g_2}-\phi_{k_1+g_3})} \notag\\
    & \times \mathcal{P}_{\boldsymbol{g}_1 \boldsymbol{g}_3}(\boldsymbol{k}_1) c_{\boldsymbol{k}_2 \boldsymbol{g}_2}^{\dagger} c_{\boldsymbol{k}_2 \boldsymbol{g}_4} \delta(\boldsymbol{g}_1+\boldsymbol{g}_2-\boldsymbol{g}_3 -\boldsymbol{g}_4). \label{Fock_term_term}
\end{align}
This energy can be minimized by taking the same phase condition we used for the analogous periodic potential case:
\begin{equation}
    \arg[v_{\bs{g}}(\bs{k})]= (N-1) \phi_{k+g}, \label{Eq_phase_condition_2}
\end{equation}
or in terms of the density matrices
\begin{align}
\mathcal{P}_{\boldsymbol{g}_1\boldsymbol{g}_2}\left(\boldsymbol{k}\right) &= v_{\bs{g}_1}(\bs{k})^*v_{\bs{g}_2}(\bs{k}) \\
&= |v_{\bs{g}_1}(\bs{k})v_{\bs{g}_2}(\bs{k})| e^{i(N-1) (\phi_{k+g_2}- \phi_{k+g_1})}. \label{Eq_phase_density_matrix}
\end{align}
If the density matrix satisfies this phase condition, then the phase from the density matrices cancels with the phase from the form factors, leaving every part of the Fock term negative (provided that $V(\bs{q})$ is positive for all $\bs{q}$).

This is a solution to minimize the Fock term, but not the only one. Indeed, we can generate a family of solutions by translating the crystal by an in-plane vector $\bs{x}$ through the addition of a momentum-dependent phase factor $v_{\boldsymbol{p}} \rightarrow v_{\boldsymbol{p}} e^{-i \boldsymbol{p} \cdot \boldsymbol{x}}$. Without loss of generality, we consider the untranslated solution, Eq.~\eqref{Eq_phase_density_matrix}, which preserves the six-fold rotation symmetry about the origin and is therefore convenient for calculating symmetry indicators.

Given this phase structure, we can make the same arguments that we made for the symmetry eigenvalues at the high-symmetry points in the periodic potential case in Section~\ref{Section_potential_large_B}. Namely, the eigenvalues at $K$ and $M$ have $\ell_K= -(N-1) \text{ mod }3$ and $\ell_M=-(N-1) \text{ mod }2$ respectively. If the interaction is weak, the $\Gamma$ indicator is locked to $\ell_\Gamma=0$ because the kinetic energy forces occupation of the parent $\Gamma$ point, which transforms trivially under symmetry. This results in a total Chern number of $\mathcal{C}=N-1 \text{ mod }6$, matching our result from the low $r_s$ limit. If the interaction is strong, the kinetic energy cost for occupying the states at further reciprocal lattice vectors is outweighed by the benefit from satisfying the Fock term. So the irreducible representation at the $\Gamma$ point is determined by the phase condition. This gives $\ell_\Gamma=-(N-1) \text{ mod }6$, giving $\mathcal{C}=0$.

This argument also works for the full Chern number determination, similar to the periodic potential case discussed in Section \ref{Section_potential_large_B}. Because the phase structure of the crystal state is the same, the argument is unchanged. This means that the results $\mathcal{C}=0$ for large interaction and $\mathcal{C}=N-1$ for small interaction hold (not just modulo 6). 

One way in which the interacting case differs from the periodic case in the large $\mathcal{B}$ limit is the origin of why the occupation is expelled from the parent band $\Gamma$ point. In the interacting case, two effects conspire to do this \cite{soejima2025jellium,joy2023wigner, joy2025chiral}.

Firstly, the occupation of the continuum $\Gamma$ point is suppressed due to the large quantum metric near the $\Gamma$ point \cite{abouelkomsanquantum2023, abouelkomsan2020particle} (which makes the Fock term less negative). To understand why a large quantum metric makes the Fock term less negative, consider the small momentum transfer component of the Fock term \eqref{Eq_Fock}. The energy associated with this term is given by
\begin{widetext}
\begin{align}
    E_F^{\text{small }\bs{q}}= -\frac{1}{A} \sum_{\bs{k}} \sum_{\bs{q} \ll g}\sum_{\bs{g}_1,\bs{g}_2} V(\bs{q}) \mathcal{F}_N(\bs{k}+\bs{q}+ \bs{g}_1, \bs{k}+\bs{g}_1)  \mathcal{F}_N (\bs{k}+\bs{g}_2, \bs{k}+\bs{q}+ \bs{g}_2)
    \mathcal{P}_{\bs{g}_1 \bs{g}_2}(\bs{k}+\bs{q}) \mathcal{P}_{\bs{g}_2 \bs{g}_1}(\bs{k}). 
\end{align}
In the small $\bs{q}$ limit, we can expand the form factor in terms of the quantum metric and Berry connection, following Ref. \cite{dong2024stability} \footnote{Note that there is a sign difference on the Berry connection compared to Ref.~\cite{dong2024stability} due to a difference in convention}:
\begin{align}
    \mathcal{F}_N&\left(\boldsymbol{k}+\bs{q}+\boldsymbol{g}_1, \boldsymbol{k}+\boldsymbol{g}_1\right) \approx  e^{-\frac{1}{2} g_{ij}(\boldsymbol{k}+\boldsymbol{g}_1) q_i q_j}  e^{i \int^{\boldsymbol{k}+\bs{q}+\boldsymbol{g}_1}_{\boldsymbol{k}+\boldsymbol{g}_1} \bs{A}(\bs{k}') \cdot d\bs{k}'},
 \end{align}
 so that the small $\bs{q}$ part of the Fock energy is
\begin{equation}
    \begin{aligned}
        E_F^{\text{small }\bs{q}} = -\frac{1}{A} \sum_{\bs{k}} \sum_{\bs{q} \ll g} \sum_{\bs{g}_1,\bs{g}_2}V(\bs{q}) &e^{-\frac{1}{2} (g_{ij}(\boldsymbol{k}+\boldsymbol{g}_1)+g_{ij}(\boldsymbol{k}+\boldsymbol{g}_2))q_iq_j} e^{i \int^{\boldsymbol{k}+\bs{q}}_{\boldsymbol{k}} [\bs{A}(\bs{k}'+\boldsymbol{g}_1)- \bs{A}(\bs{k}'+\boldsymbol{g}_2)]\cdot d\bs{k}'} \\
        &\times \mathcal{F}_N (\bs{k}+\bs{g}_2, \bs{k}+\bs{q}+ \bs{g}_2)
        \mathcal{P}_{\bs{g}_1 \bs{g}_2}(\bs{k}+\bs{q}) \mathcal{P}_{\bs{g}_2 \bs{g}_1}(\bs{k}). 
    \end{aligned}
\end{equation}
\end{widetext}
We see that the quantum metric, which controls the size of the form factor at small $\bs{q}$, limits the magnitude of the Fock energy. This reduces the energy that can be saved by satisfying the small $\bs{q}$ Fock term. This has the effect of pushing the electrons out of regions with large quantum metrics when the small $\bs{q}$ component of the Fock term is important. This includes the $\Gamma$ point of the $\lambda_N$ parent band, $\Gamma^{\text{P}}$, when $\mathcal{B}$ is large.

In addition, the phase of the Fock term can be perfectly satisfied when occupying the outlying states (where the Berry curvature is small) by satisfying the phase condition described above. These two effects increase the energy cost for occupying parent band states near the zone center relative to occupying states near other reciprocal lattice vectors. This leads to this region being depleted, and so the symmetry eigenvalue at the $\Gamma$ point is not locked to the trivial one. As a result, by the argument above based on the phase condition, the Chern number becomes zero at large interaction strengths for strongly concentrated Berry curvature.

%%%%%%
\subsection{The Halo Anomalous Hall Crystal}\label{subsec:HAHC}

%%%%%%%%%%%%%%%%%%%%%%%%%%%%%%%%%%%%%%%%%%%%%%%%%%%%%%%%%%%
\begin{figure}
    \centering
    \includegraphics[width=1.0\linewidth]{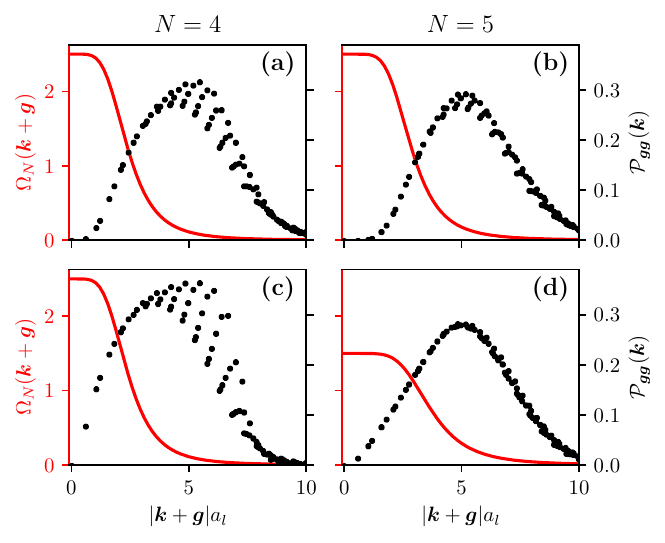}\vspace{-4mm}
    \caption{Berry curvature of the parent $\lambda_N$ model (red curve and left axis) and momentum space occupation (black dots and right axis) for the (a-b) HWC and (c-d) HAHC with (a)-(c) $N=4$ and (b)-(d) $N=5$. Results are for (a) $N=4$, $r_s=20$ and $\mathcal{B}A_{\text{1BZ}}=10\pi$, (b) $N=5$, $r_s=20$ and $\mathcal{B}A_{\text{1BZ}}=10\pi$, (c)  $N=4$, $r_s=15$ and $\mathcal{B}A_{\text{1BZ}}=10\pi$, and (d) $N=5$, $r_s=20$ and $\mathcal{B}A_{\text{1BZ}}=6\pi$. We note that $\Tr(g^{ab}_{N}(\bs{q})) = \left|\Omega_{N}(\boldsymbol{q})\right|$ since the band is ideal (see Eq.~\eqref{eq:FS_metric_lambda_N}).}
\label{fig:geometric_expulsion}
\end{figure}
%%%%%%%%%%%%%%%%%%%%%%%%%%%%%%%%%%%%%%%%%%%%%%%%%%%%%%%%%%%

The previously discussed ``geometric expulsion'' in which electronic occupation of the parent band states is depleted at the zone center to minimize its overlap with the metric is illustrated in Fig.~\ref{fig:geometric_expulsion}. We see that in the Halo phases (HWC and HAHC), electrons mostly occupy states on the shoulder and tail of the Berry curvature distribution (or of the metric given the band is ideal~\eqref{eq:FS_metric_lambda_N}) where it decays. 

This helps explain why halo crystal phases ($\ell_\Gamma \neq 0$) are observed with only zero or small Chern number. As the parent band occupation shifts to regions of small Berry curvature, the net Berry curvature of the crystal is likewise suppressed. In fact, we numerically find that the Chern number is well approximated over a large portion of the phase diagram for $N\ge 3$ by the following ``generalized occupation-weighted flux rounding'' (see Appendix~\ref{app:generalized_flux_rounding}):
\begin{align} \label{eq:occupation_weighted_flux_rounding}
    \left\lfloor \sum_{\bs{g}} \int_{\text{1BZ}}d^2k \Omega_N(\bs{k} + \bs{g}) \mathcal{P}_{\bs{g}\bs{g}}(\bs{k})  / (2\pi) \right\rceil, 
\end{align}
which reduces to Berry flux rounding~\eqref{eq:berry_flux_rounding} in the weak-interaction limit when only the first Brillouin zone is occupied. The above generalized rounding formula helps rationalize a few observations. In the phase diagrams of Fig.~\ref{fig:phase_diagram_periodic_potential}, we observe that some phase boundaries between crystals with different Chern numbers are often pushed to larger values of $\mathcal{B}A_{\text{1BZ}}$ than predicted from the Berry flux rounding argument at larger $r_s$. The interpretation of this phase boundary bending from the perspective of the above generalized flux rounding is that electrons are pushed to occupy states with larger momenta and weaker Berry curvature as the interaction is increased. This may result in a decrease of the Chern number close to the phase boundary as $r_s$ is increased for a fixed $\mathcal{B}$. In contrast, for the $N\to\infty$ limit, the above formula would predict an $r_s$-independent Chern number as observed in Fig.~\ref{fig:phase_diagram_periodic_potential}: a completely flat Berry curvature results in a fixed value of Eq.~\eqref{eq:occupation_weighted_flux_rounding} independent of occupation.

%~\footnote{We note the connection between the generalized momentum-weighted flux rounding and the formula given in~\cite{neupert2012elementary} whose validity was subsequently disproved~\cite{simon2014comment, neupert2014reply, simon2015fractional}.}.

Despite its usefulness, we stress that this remains a heuristic relation and does not hold numerically across the entire parameter space. Nonetheless, this heuristic further clarifies why only $\mathcal{C}=1$ HAHC phases appear. Since electrons are expelled from the high-curvature plateau in the halo phases, the momentum-weighted Berry curvature cannot grow arbitrarily large, given its rapid decay beyond the plateau. HAHC phases with larger Chern numbers may nevertheless be realized in alternative models with more slowly decaying Berry curvature, provided the decay of the metric remains fast enough that a crystal with $\ell_\Gamma \neq 0$ is energetically competitive.

%%%%%%%%%%%%%%%%%%%%%%%%%%%%%%%%%%%%%%%%%%%%%%%%%%%%%%%%%%%%%
%                     Conclusion                            %
%%%%%%%%%%%%%%%%%%%%%%%%%%%%%%%%%%%%%%%%%%%%%%%%%%%%%%%%%%%%%
\section{Conclusion} \label{sec:conclusion}

In this work, we have introduced the $\lambda_N$-jellium model, a minimal continuum model that realizes an isolated ideal band with a Berry curvature profile and total Berry flux that are both tunable. This model interpolates between the $\lambda$-jellium ($N = 2$) and infinite Chern band ($N\to \infty$) models, thereby providing a more flexible platform for studying how quantum geometry influences crystalline order. We used this model to investigate both explicit crystallization induced by a periodic potential and spontaneous crystallization driven by Coulomb interactions. In both regimes, we identified a wide variety of electronic crystal phases, including anomalous Hall crystals, halo Wigner crystals, and a novel halo anomalous Hall crystal. Importantly, we complemented our numerical results with analytical insights in several limits: the weak periodic potential regime, where we derived a general formula for the Chern number; the weak-interaction regime, where we connected the observed Chern numbers to Berry flux rounding~\cite{dong2024stability, bernevig2025berry}; and the large-$\mathcal{B}$ limit, where the Berry curvature can be treated as a vortex insertion. Our work thus provides a clear analytic understanding of why certain crystalline phases are favored in different regimes and how their topology is determined.

A standout feature of our analysis is the stark contrast between the phase diagrams obtained in the presence of a periodic potential (Fig.~\ref{fig:phase_diagram_periodic_potential}) and Coulomb interactions (Fig.~\ref{fig:hf_phase_diagram}). Ref.~\cite{tan2024parent} had already identified that, in the $N\to\infty$ limit, one expects different Chern numbers for crystal states in the periodic potential and interacting cases. Our work significantly extends this picture by showing that the behavior becomes even more complex and distinct in the presence of finite Berry flux. In the presence of a periodic potential (Fig.~\ref{fig:phase_diagram_periodic_potential}), we find a dense sequence of transitions across which the Chern number can jump from positive to negative and by seemingly arbitrary amounts. In contrast, in the interacting case (Fig.~\ref{fig:hf_phase_diagram}), the Chern numbers remain positive and vary only incrementally across transitions. In particular, in the interacting case, we find heuristically that the Chern number of the crystal state coincides over a large portion of the phase diagram with that predicted by rounding the occupation-weighted Berry curvature of the parent band (Eq.~\eqref{eq:occupation_weighted_flux_rounding}).

Such differences originate from the distinct mechanisms that determine the electronic crystal topology in both cases. In the periodic potential case, we have independent Hostadter models at each crystal momentum. Those momentum hoppings promote delocalization in momentum space, and the form factor sets the preferred relative phase between the states constituting the single-particle states of the filled band. By comparison, the energy of crystals formed through the spontaneous breaking of continuous translation appears to be significantly determined by the low-$\bs{q}$ part of the Fock term. This dominant Fock contribution leads to geometric expulsion, where electronic occupation avoids the region of concentrated metric~\cite{abouelkomsanquantum2023, abouelkomsan2020particle}, which is at the origin of the observed Halo crystal phases. The difference is perhaps most transparent when comparing the weak periodic potential and weak interaction limits: while the former leads to large and irregular jumps in the Chern number (see Fig.~\ref{fig:low_U_pp_analytic}) that are set by the form factor edge winding (Eq.~\eqref{Equation_Chern_low_U_capped_mt}), the latter produces a monotonic evolution of the Chern number dictated by Berry flux rounding.

The results presented here also offer broadly applicable insights into recent experiments. Indeed, the $\lambda_N$ model can serve as an analytically tractable approximate description of spin- and valley-polarized bands that enclose a finite amount of Berry curvature, such as those found in rhombohedral multilayer graphene. As a concrete example, consider rhombohedral multilayer graphene subjected to a weak hexagonal moir\'e potential (e.g., from aligned boron nitride) on one side, with electrons doped to integer filling and polarized away from the moir\'e layer. In this setting, our analysis of weak periodic potentials becomes directly relevant. For multilayers with three or more layers, the large band dispersion suppresses occupation of high-momentum states even more strongly than in the quadratic band studied here, making the weak-potential predictions applicable over a wide parameter regime. 

If the moir\'e Brillouin zone is fully contained within the broad Berry curvature plateau where the form factor is well approximated by $\mathcal{F}_{\text{LLL}}(\bs{q},\bs{q}')$\cite{lu2024fractional, lu2025extended}, the band insulator in the absence of interactions is predicted to carry a negative Chern number (i.e., opposite in sign to the parent Berry curvature), as in the $N\to\infty$ limit~\cite{tan2024parent}. In contrast, if the moir\'e Brillouin zone overlaps with the region where the Berry curvature decays rapidly, the form factor on the boundary is no longer well approximated by $\mathcal{F}_{\text{LLL}}(\bs{q},\bs{q}')$ and the Chern number predicted by the moir\'e potential can become positive (i.e., with the same sign as the parent Berry curvature). If the Berry becomes even more concentrated, such that essentially all of the Berry flux is contained in a small region of the first Brillouin zone, we would then expect a small periodic potential and weak interactions to both favor the same positive Chern number. The relative size of the Berry curvature plateau compared to the first Brillouin zone, which sets the above condition, can be tuned experimentally by varying the twist angle (i.e., the size of the first Brillouin zone). It can also be tuned by varying the displacement field, which controls the extent of the region containing the Berry curvature. In particular, if anomalous Hall crystals in rhombohedral multilayer graphene are confirmed in future experiments~\cite{seiler2022quantum, seiler2024signatures}, the displacement field may similarly be used to concentrate the Berry curvature and drive transitions between anomalous Hall crystals and halo Wigner crystals or halo anomalous Hall crystals.

Looking forward, several directions naturally emerge. First, it will be essential to study the competition between interactions and periodic potentials. While in trivial bands these two mechanisms cooperate to stabilize trivial crystals, the presence of nontrivial form factors may cause them to compete. Counterintuitively, even a weak periodic potential may then destabilize a spontaneously formed crystal. Second, the assumption that the system crystallizes into a triangular lattice deserves closer scrutiny~\cite{desrochers2025elastic}. Earlier work has shown that for $N = 2$ the triangular lattice is locally unstable to small deformation of the lattice in certain regimes~\cite{soejima2025jellium, dong2025phonons}. More broadly, it would also be important to study the global stability of these lattices and compare the energy of systems that crystallize with one electron per unit cell to other ones with enlarged unit cells~\cite{zhou2025new} and even self-doped systems that have a small density of holes~\cite{pankov2008self, kim2024dynamical}. Indeed, the electrons may be able to lower their overlap with the metric and thus lower their energy more efficiently by enlarging the unit cell or introducing holes than by forming halo crystals. Third, the question of stability beyond mean-field theory remains wide open. Hartree–Fock can identify possible crystalline phases but is not reliable for locating phase boundaries, particularly for the competition between Fermi liquids and electronic crystals. Studies using variational Monte Carlo, diffusion Monte Carlo, and other beyond-mean-field approaches would be highly desirable to test the robustness of the AHC and halo phases identified in this work. Such studies may also uncover fractionalized anomalous Hall crystals --- phases that combine the topological character of fractional quantum Hall states with spontaneously broken translational symmetry --- which have been recently proposed~\cite{soejima2024anomalous, tan2025variational}. More broadly, the $\lambda_N$ model may serve as a versatile tool for investigating other phenomena beyond crystallization in spin and valley-polarized systems with Berry curvature, such as the recently reported chiral superconductivity in rhombohedral tetralayer graphene~\cite{han2024signatures}.

\begin{acknowledgments}
    We acknowledge helpful conversations with Ashvin Vishwanath, Tomohiro Soejima, and Zhaoyu Han. This was supported by the Natural Sciences and Engineering Research Council of Canada (NSERC) and the Centre of Quantum Materials at the University of Toronto. Computations were performed on the Cedar and Fir clusters, which are hosted by the Digital Research Alliance of Canada. F.D. is further supported by the Vanier Canada Graduate Scholarship (CGV-186886). 
\end{acknowledgments}

%%%%%%%%%%%%%%%%%%%%%%%%%%%%%%%%%%%%%%%%%%%%%%%%%%
%                 Bibliography                   %
%%%%%%%%%%%%%%%%%%%%%%%%%%%%%%%%%%%%%%%%%%%%%%%%%%
%\bibliography{apssamp}
%\bibliography{output}
%\include{output.bbl}

%

%%%%%%%%%%%%%%%%%%%%%%%%%%%%%%%%%%%%%%%%%%%%%%%%%%
%                   Appendix                     %
%%%%%%%%%%%%%%%%%%%%%%%%%%%%%%%%%%%%%%%%%%%%%%%%%%
\vfill
\onecolumngrid
\newpage
\appendix

\tableofcontents

\section{Relation Between the \texorpdfstring{$\boldsymbol{\lambda}_{\boldsymbol{N}}$}{$\lambda_N$} Model and Rhombohedral \texorpdfstring{$\boldsymbol{N}$}{$N$}-layer Graphene} \label{app:connection_lambdaN_RNG}

We briefly comment here on a connection between the $\lambda_N$ model and a deformation of a minimal model for rhombohedral $ N$-layer graphene. Such a connection was already pointed out in the $N\to\infty$ limit in Appendix~F of Ref.~\cite{tan2024parent}. Here we use an alternative route~\footnote{F\'elix Desrochers thanks Zhaoyu Han for pointing out this way of presenting the connection.} to outline this connection in the finite $N$ case, following a procedure explained in Ref.~\cite{han2025exact}.

\begin{figure*}
    \centering
    \includegraphics[width=0.85\textwidth]{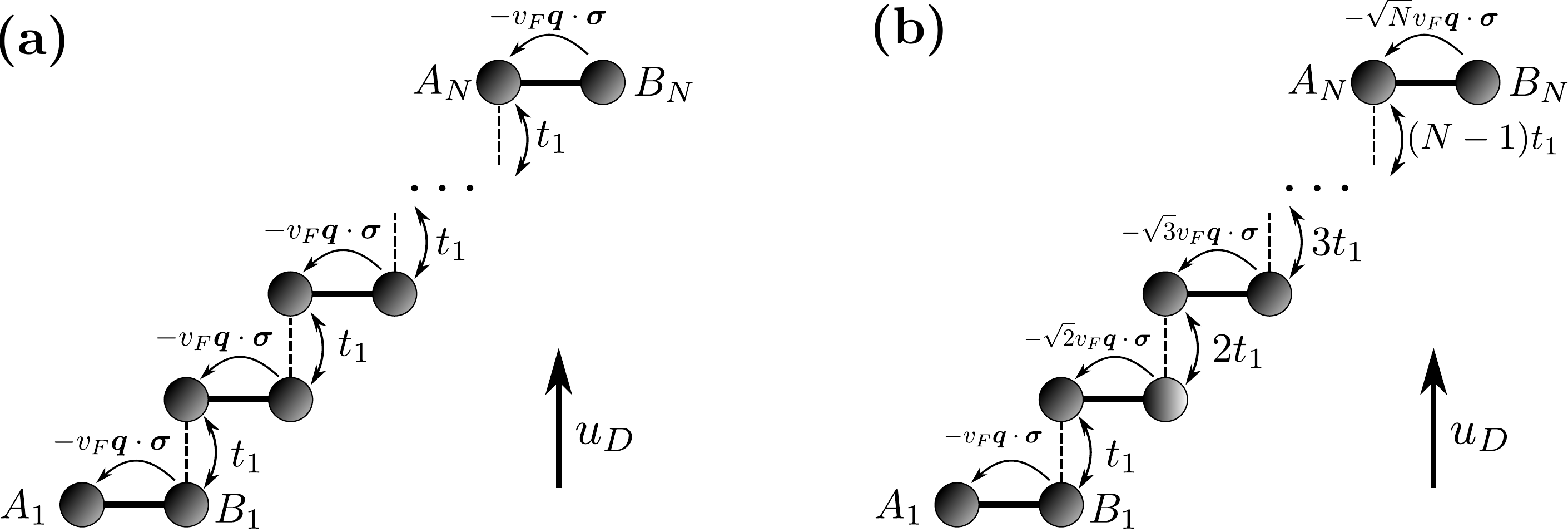}%\vspace{-4mm}
    \caption{(a) Simplified effective model for rhombohedral $N$-layer graphene yielding the hopping matrix~\eqref{eq:rng_effective_model}, and (b) modified effective model with layer-dependent intra-layer and inter-layer hopping yielding the hopping matrix~\eqref{eq:rng_effective_model_modified}. We define $\bs{\sigma}\equiv(\sigma_x, \sigma_y)$.}
\label{fig:rng}
\end{figure*}

The effective continuum model of rhombohedral $N$-layer graphene has valley $\eta\in\{K,K'\}$, spin $\sigma\in\{\uparrow,\downarrow\}$, sublattice $\alpha\in\{A,B\}$ and layers $l\in\{1,2,\ldots,N\}$ degrees of freedom. A minimal hopping model for a single spin and single valley of rhombohedral $N$-layer graphene in a perpendicular displacement field is
\begin{align} \label{eq:rng_effective_model}
    h_{\text{RNG}}(\boldsymbol{q})=\left(\begin{array}{ccccccc}
    u_d & -v_F \sqrt{2}q_{z} & & & & & \\
    -v_F\sqrt{2}q_{\bar{z}}& u_d & t_1 & & & & \\
    & t_1 & 2 u_d & -v_F \sqrt{2}q_{z} & & & \\
    & & -v_F\sqrt{2}q_{\bar{z}}& 2 u_d & t_1 & & \\
    & & & \ddots & \ddots & \ddots & \\
    & & & & t_1 & N u_d & -v_F \sqrt{2}q_{z} \\
    & & & & & -v_F\sqrt{2}q_{\bar{z}}& N u_d
    \end{array}\right),
\end{align} 
where the above is written in the $(\alpha l) \in\{(A, 1),(B, 1),(A, 2),(B, 2), \ldots,(A, N),(B, N)\}$ basis, $v_F$ represents the Fermi velocity of a single graphene layer, and $t_1$ is the interlayer hopping between $B_l$ and $A_{l+1}$ (see Fig.~\ref{fig:rng}(a)). 

In the experimentally relevant regime $u_d\ll t_1$, the above model results in two lowest energy bands with a remarkably flat bottom for $|\boldsymbol{q}| \lesssim q_0\equiv t_1 / v_F$ that becomes dispersive at larger momenta. In the $|\boldsymbol{q}| \ll q_0$ regime, the two lowest roughly flat bands are mostly supported on a single sublattice and have the approximate structure 
\begin{subequations} \label{eq:RNG_approx_wf}
\begin{align} \label{subeq:RNG_approx_wf_e}
    & \psi^{(e)}_{(A l)}(\bs{q}) \approx \mathcal{N}_{\bs{q}}^{(e)} \left(\sqrt{2} q_{\bar{z}} / q_0\right)^{l-1} \\
    & \psi^{(e)}_{(B l)}(\bs{q}) \approx - \mathcal{N}_{\bs{q}}^{(e)} \mathcal{N}_{\bs{q}}^{(e)} l \frac{u_d}{t_1}\left( \sqrt{2} q_{\bar{z}} / q_0\right)^l,
\end{align}
and 
\begin{align} \label{subeq:RNG_approx_wf_h}
    & \psi^{(h)}_{(B l)}(\bs{q}) \approx \mathcal{N}_{\bs{q}}^{(h)} \left(\sqrt{2} q_{z} / q_0\right)^{N-l} \\
    & \psi^{(h)}_{(A N)}(\bs{q}) \approx - \mathcal{N}_{\bs{q}}^{(h)} (N-l+1) \frac{u_d}{t_1}\left(  \sqrt{2} q_{z}/ q_0\right)^{N-l+1},
\end{align}
\end{subequations}
for the hole and electron band, respectively. One can understand this eigenmode structure by viewing the effective model in the $|\boldsymbol{q}| \ll q_0$ regime as an analog of the Su–Schrieffer–Heeger model in its topologically non-trivial regime. The two lowest bands of the above effective models are then analogous to the algebraically localized edge modes. Since the wavefunctions are almost entirely supported on one of the sublattices in the $|\boldsymbol{q}| \ll q_0$, it appears reasonable to assume the electron wavefunctions are entirely on the $A$ sublattice (and the hole wavefunctions entirely on the $B$ sublattice). Within this assumption, one may then want to obtain effective models that act purely on a single sublattice and yield wavefunctions that closely approximate Eq.~\eqref{eq:RNG_approx_wf}. Focusing on the electronic band, this can be achieved by ``integrating out'' the $B$ sublattice and considering the Schur complement of the $B$-sublattice hopping matrix
\begin{align}
    h_{\text{Schur}}^{(e)}(\bs{q}) &= h_{AA}(\bs{q}) - h_{AB}(\bs{q}) h_{BB}^{-1}(\bs{q}) h_{BA}(\bs{q}),
\end{align}
and similarly for the hole band 
\begin{align}
    h_{\text{Schur}}^{(h)}(\bs{q}) &= h_{BB}(\bs{q}) - h_{BA}(\bs{q}) h_{AA}^{-1}(\bs{q}) h_{AB}(\bs{q}),
\end{align}
where 
\begin{subequations}
    \begin{align}
        h_{AA}(\bs{q}) &= h_{BB}(\bs{q}) = \mqty(
        u_d & & & \\
         & 2 u_d & & \\
         & & \ddots & \\
         & &  &  Nu_d
        )\\
        h_{AB}(\bs{q}) &= \mqty(
        -v_F \sqrt{2} q_{z} & & & \\
         t_1 & -v_F \sqrt{2} q_{z} & & \\
         & \ddots & \ddots & \\
         & & t_1 & -v_F \sqrt{2} q_{z}
        ) \\
        h_{BA}(\bs{q}) &= \mqty(
        -v_F \sqrt{2} q_{\bar{z}} & t_1 & & \\
          & -v_F \sqrt{2} q_{\bar{z}} & \ddots & \\
         & & \ddots &  t_1 \\
         & &  & -v_F \sqrt{2} q_{\bar{z}}
        )
    \end{align}
\end{subequations}
are the different sublattice blocks. Such a procedure is generally uncontrolled and is only exact for zero modes. However, since we are mostly interested in the lowest lying modes that are only separated from zero by something on the order of $u_d$, the above approximate procedure appears reasonable~\cite{han2025exact}.  

Focusing on the electron band, we find
\begin{align}
    h^{(e)}_{\text{Schur}}(\boldsymbol{q}) 
    &= u_d \left(\begin{array}{ccccc}
    1 &  & & &  \\
     & 2 &  & & \\
    &  & 3 & & \\
    & & & \ddots & \\
    & & & & N
    \end{array}\right)
    \nonumber \\
    &\quad -
    \frac{t_1^2}{u_d} 
    \left(\begin{array}{ccccc}
    (|\bs{q}|/q_0)^2   & -\sqrt{2}q_{z}/q_0 & & &  \\
    -\sqrt{2}q_{\bar{z}}/k_0 & (|\bs{q}|/q_0)^2 + 1 & -\sqrt{2}q_{z}/(2q_0) & &  \\
     & -\sqrt{2}q_{z}/(2q_0) & (|\bs{q}|/q_0)^2 + 1/2 & \ddots  &   \\
    & & \ddots & \ddots & -\sqrt{2} q_{z}/((N-1) q_0)  \\
    & & & -\sqrt{2} q_{\bar{z}}/((N-1) q_0) &  (|\bs{q}|/q_0)^2 + 1/N
    \end{array}\right). 
\end{align}

We then note that the spinor structure of the lowest electronic band of rhombohedral $N$-layer graphene~\eqref{subeq:RNG_approx_wf_e} closely resembles the zero-eigenmode of the $\lambda$-N jellium model~\eqref{eq:lambda_N_model_coefficients_wf}. The above $A$-sublattice effective model also closely resembles the $\hat{L}(\bs{q})$ operator that encapsulates the band geometry of the $\lambda_N$ model:
\begin{align}
    \bra{m} \hat{L}(\bs{q}) \ket{n} &= \mqty(
    \frac{|\bs{q}|^2}{2}  \mathcal{B}  & -k_{z} \sqrt{\mathcal{B}} & & & & \\
    -q_{\bar{z}} \sqrt{\mathcal{B}} & \frac{|\bs{q}|^2}{2} \mathcal{B} + 1 & \sqrt{2 \mathcal{B}} q_{z} & & & \\
     & \sqrt{2 \mathcal{B}} q_{\bar{z}} & \frac{|\bs{q}|^2}{2} \mathcal{B} + 2 & \ddots & & \\
    & & \ddots & \ddots &  & \sqrt{(N-1) \mathcal{B}} q_{z} \\
    & & & & \sqrt{(N-1) \mathcal{B}} q_{\bar{z}} &  (N-1)
    )_{mn}. \label{eq:L_matrix_explicit_form}
\end{align}
This close resemblance then begs the question of how one should modify the above hopping matrix~\eqref{eq:rng_effective_model} to recover exactly~\eqref{eq:L_matrix_explicit_form} as the one sublattice effective model (obtained by computing the Schur complement) and the exact lowest mode structure of Eq.~\eqref{eq:lambda_N_model_coefficients_wf}. 

%%%%%%%%%%%%%%%%%%%%%%%%%%%%%%%%%%%%%%%%%%%%%%%%%%%%%%%%%%%
%%%%%%%%%%%%%%%%%%%%%%%%%%%%%%%%%%%%%%%%%%%%%%%%%%%%%%%%%%%

We explain how one can follow such a procedure. To do so, we make the interlayer hopping and intralayer hopping layer-dependent. More specifically, we define the hopping between $B_l$ and $A_{l+1}$ to be $lt_1$ and the intralayer hopping amplitude between $A_l$ and $B_l$ to be $-v_F\sqrt{2 l}q_{z}$ (see Fig.~\ref{fig:rng}(b)). This results in the new effective model
\begin{align} \label{eq:rng_effective_model_modified}
\tilde{h}_{\text{RNG}}(\boldsymbol{q})=
\left(\begin{array}{ccccccc}
    u_d & -\sqrt{2} v_F q_{z} & & & & & \\
    -v_F \sqrt{2} q_{\bar{z}} & u_d & t_1 & & & & \\
    & t_1 & 2 u_d & - 2v_F q_{z} & & & \\
    & & -2 v_F q_{\bar{z}} & 2 u_d & 2 t_1 & & \\
    & & & \ddots & \ddots & \ddots & \\
    & & & & (N-1)t_1 & (N-1) u_d & -\sqrt{2 N} v_F q_{z} \\
    & & & & & -\sqrt{2 N} v_F q_{\bar{z}} & N u_d
\end{array}\right)
\end{align}
with the sublattice blocks
\begin{subequations}
    \begin{align}
        \tilde{h}_{AA}(\bs{q}) &= \tilde{h}_{BB}(\bs{q})=  \mqty(
        u_d & & & \\
         & 2 u_d & & \\
         & & \ddots & \\
         & &  &  Nu_d
        )\\
        \tilde{h}_{AB}(\bs{q}) &= \mqty(
        -v_F \sqrt{2} q_z & & & \\
         t_1 & -v_F 2 q_z & & \\
         & \ddots & \ddots & \\
         & & (N-1) t_1 & -v_F \sqrt{2N} q_z
        ) \\
        \tilde{h}_{BA}(\bs{q}) &= \mqty(
        -v_F \sqrt{2} q_{\bar{z}} & t_1 & & \\
          & -v_F 2 q_{\bar{z}} & \ddots & \\
         & & \ddots & (N-1) t_1 \\
         & &  & -v_F \sqrt{2 N} q_{\bar{z}}
        ).
    \end{align}
\end{subequations}
The Schur complement of the $B$ sublattice is then
\begin{align}
    \tilde{h}_{\text {Schur }}(\boldsymbol{q}) &= \tilde{h}_{AA}(\bs{q}) - \tilde{h}_{AB}(\bs{q}) \tilde{h}_{BB}^{-1}(\bs{q}) \tilde{h}_{BA}(\bs{q}) \nonumber \\
    &= u_d \left(\begin{array}{ccccc}
    1 &  & & &  \\
     & 2 &  & & \\
    &  & 3 & & \\
    & & & \ddots & \\
    & & & & N
    \end{array}\right)
    \nonumber \\
    &\quad -
    \frac{t_1^2}{u_d} 
    \left(\begin{array}{ccccc}
    (|\bs{q}|/q_0)^2   & -\sqrt{2}q_{z}/q_0 & & & \\
    -\sqrt{2}q_{\bar{z}}/k_0 & (|\bs{q}|/q_0)^2 + 1 & -2q_{z}/q_0 & & \\
     & -2 q_{\bar{z}}/q_0 & (|\bs{q}|/q_0)^2 + 2 & \ddots &   \\
    & & \ddots & \ddots & -\sqrt{2(N-1)} q_{z}/q_0 \\
    & & &  -\sqrt{2(N-1)} q_{\bar{z}}/q_0 &  (|\bs{q}|/q_0)^2 + (N-1)
    \end{array}\right) \nonumber \\
    &\approx -
    \frac{t_1^2}{u_d} 
    \left(\begin{array}{ccccc}
    (|\bs{q}|/q_0)^2  & -\sqrt{2}q_{z}/q_0 & & & \\
    -\sqrt{2}q_{\bar{z}}/k_0 & (|\bs{q}|/q_0)^2 + 1 & -2q_{z}/q_0 & & \\
     & -2 q_{\bar{z}}/q_0 & (|\bs{q}|/q_0)^2 + 2 & \ddots &   \\
    & & \ddots & \ddots & -\sqrt{2(N-1)} q_{z}/q_0 \\
    & & &  -\sqrt{2(N-1)} q_{\bar{z}}/q_0 &  (N-1)
    \end{array}\right),
\end{align}
where in the last line we have dropped the terms linear in $u_d$ since we are interested in the physically relevant regime $t_1\gg u_d$ (which amounts to turning off the displacement field potential on the $A$ sublattice). In the last line, we have also dropped the $(|\bs{q}|/q_0)^2$ term in the $[\tilde{h}_{\text {Schur }}(\boldsymbol{q})]_{NN}$ component, which is physicaly equivalent to turning off intralayer hopping in the last layer. This should not significantly affect the structure of the lowest eigenmode of interest, given a sufficiently large $N$, since it decays algebraically with $N$. We then notice that the matrix in this last line is exactly equal to $-t_1^2 \hat{L}(\bs{q})/u_d$ with the identification $\mathcal{B}=2/q_0^2=2v_F^2/t_1^2$. 

In summary, the $\lambda_N$ model can be viewed as an approximate effective one-sublattice model of rhombohedral $N$-layer graphene, where intra-layer and interlayer hoppings are made layer-dependent. It precisely corresponds to the Schur complement of the $B$-sublattice for this effective model with the displacement field on the $A$ sublattice and intralayer hopping on the $N^{\text{th}}$ layer set to zero. 

%%%%%%%%%%%%%%%%%%%%%%%%%%%%%%%%%%%%%%
\section{Irreducible Representation at High-Symmetry Points}\label{app:periodic_potential}

%%%%%%%%%%%%%%%%%%%%%%%%%%%%%%%%%%%%%%%%%%%%%%%%%%%%%%%%%%%
\begin{figure*}
    \centering
    \includegraphics[width=1.0\textwidth]{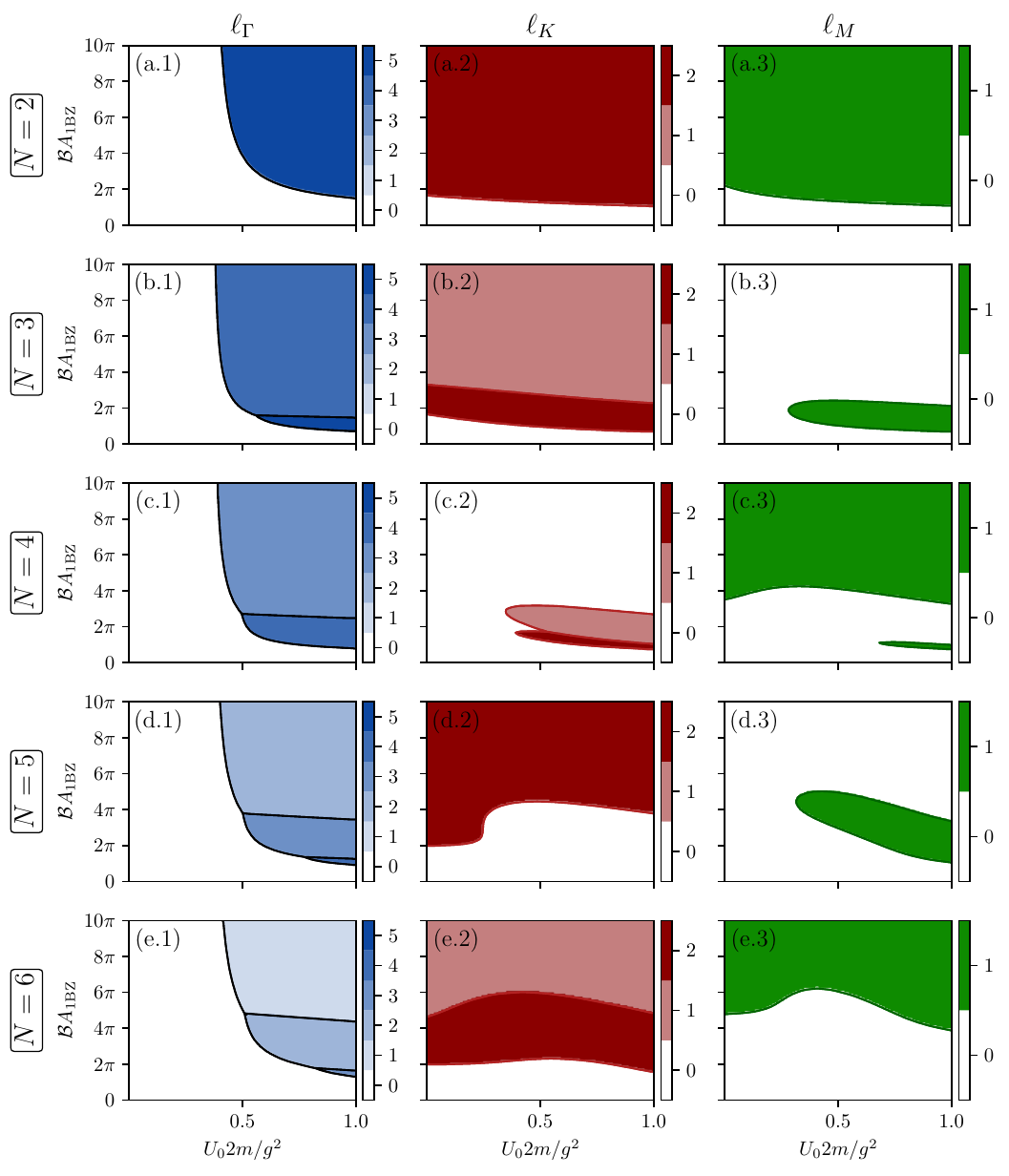}\vspace{-3mm}
    \caption{Symmetry eigenvalues $\ell_{\bs{\kappa}}$ at the (1) $\Gamma$, (2) $K$, and (3) $M$ points for the (a) $N=2$, (b) $N=3$, (c) $N=4$, (d) $N=5$, (e) $N=6$, and (f) $N\to\infty$ models.}
\label{fig:fig_symmetry_indicators}
\end{figure*}
%%%%%%%%%%%%%%%%%%%%%%%%%%%%%%%%%%%%%%%%%%%%%%%%%%%%%%%%%%%

Fig.~\ref{fig:fig_symmetry_indicators} shows the lowest energy irreducible representation at the $\Gamma$, $K$, and $M$ points (labelled by $\ell_\Gamma$, $\ell_K$, and $\ell_M$) obtained by diagonalizing the model~\eqref{eq:model_lambda_n_pp} at those high-symmetry points. Overlaying the phase diagrams for $\ell_\Gamma$, $\ell_K$, and $\ell_M$ on top of each other offers a simple way to understand most features present in the phase diagram of Fig.~\ref{fig:phase_diagram_periodic_potential}.

%%%%%%%%%%%%%%%%%%%%%%%%%%%%%%%%%%%%%%
\section{Projector Formalism For Berry Curvature}
\label{Section_proj_formalism}

In Sec.~\ref{Sec_weak_potential} of the main text, we used generic expressions for the Berry curvature of the crystal band (Eq.~\eqref{Eq_Berry_projector_main} and Eq.~\eqref{Eqn_Berry_curvature_generic_main}) to obtain the full Chern number in the weak potential limit. In this section, we provide derivations of these expressions using the projector formalism \cite{Avron1983, Pozo2020, Mitscherling2025}. Then, in Section \ref{Section_weak_potential_projector}, we provide a more detailed treatment of the weak potential limit, and in Section \ref{Appendix_projector_lambda_N}, we give some additional details about the form factor winding in the $\lambda_N$-jellium case.

As explained in Sec.~\ref{Sec_weak_potential}, the projector formalism is useful in our case because it avoids the need to adopt a smooth gauge choice for the single-particle states, and allows us instead to work directly with the gauge-invariant density matrices. In the projector formalism, the Berry curvature of a given band at crystal momentum $\bs{k}$ is given by \cite{Avron1983, Mitscherling2025} 
\begin{equation}
	\Omega(\bs{k}) = i \Tr (P(\bs{k}) \partial_{k_x} P(\bs{k}) \partial_{k_y} P(\bs{k})) - i \Tr (P(\bs{k}) \partial_{k_y} P(\bs{k}) \partial_{k_x} P(\bs{k})) .
\end{equation}
Here $P(\bs{k})$ is the projector to a given Bloch state at $\bs{k}$ (the argument $\bs{k}$ will be suppressed from here on unless it is necessary for clarity). This is given by
\begin{equation}
P= \ket{u_{\bs{k}}} \bra{u_{\bs{k}}} = \sum_{\bs{g}_1, \bs{g}_2} 	\mathcal{P}_{\boldsymbol{g}_2\boldsymbol{g}_1}(\bs{k})\ket{ u_{\bs{k}},\bs{g}_1} \bra{ u_{\bs{k}},\bs{g}_2},  
\end{equation}
where we have introduced the density matrix
\begin{align}
	\mathcal{P}_{\boldsymbol{g}\boldsymbol{g}'}\left(\boldsymbol{k}\right)= \left\langle c_{\boldsymbol{k} \boldsymbol{g}}^{\dagger} c_{\boldsymbol{k} \boldsymbol{g}'}\right\rangle \label{eq:density_matrix_scc_app}
\end{align}
and the states
$$\ket{ u_{\bs{k}},\bs{g}_1}= e^{-i \bs{k} \cdot \bs{r}} \ket{\bs{k}+\bs{g}_1} = e^{i \bs{g}_1 \cdot \bs{r}} \ket{s_{\bs{k}+\bs{g}_1}}.$$

Note that this state has an $e^{i\bs{g} \cdot \bs{r}}$ spatial dependence, so states at different $\bs{g}$ are orthogonal (but states at different $\bs{k}$ need not be). We can then obtain an expression for the Berry curvature in terms of the density matrices. This is convenient because the density matrices are manifestly invariant under changes to the overall phase of a crystal state $\ket{\psi_{\bs{k}}}$, which avoids the need to construct a smooth gauge for these states across the 1BZ \cite{Mitscherling2025}. 

We first evaluate the derivative of the projector as

	\begin{align*}
	\partial_{k_x}P = \sum_{\bs{g}_3, \bs{g}_4} \partial_{k_x}\mathcal{P}_{\bs{g}_4 \bs{g}_3} \ket{ u_{\bs{k}},\bs{g}_3} \bra{ u_{\bs{k}},\bs{g}_4}  +\sum_{\bs{g}_3, \bs{g}_4} \mathcal{P}_{\bs{g}_4 \bs{g}_3} \partial_{k_x}(\ket{ u_{\bs{k}},\bs{g}_3} \bra{ u_{\bs{k}},\bs{g}_4} ).
\end{align*}

As a result, $P \partial_{k_x} P$ is given by
\begin{align}
	P \partial_{k_x} P =  \sum_{\bs{g}_1, \bs{g}_2, \bs{g}_3, \bs{g}_4} 	\mathcal{P}_{\boldsymbol{g}_2\boldsymbol{g}_1}(\bs{k})\ket{ u_{\bs{k}},\bs{g}_1} \bra{ u_{\bs{k}},\bs{g}_2} \left( \partial_{k_x}\mathcal{P}_{\bs{g}_4 \bs{g}_3} \ket{ u_{\bs{k}},\bs{g}_3} \bra{ u_{\bs{k}},\bs{g}_4}  + \mathcal{P}_{\bs{g}_4 \bs{g}_3} \partial_{k_x}(\ket{ u_{\bs{k}},\bs{g}_3} \bra{ u_{\bs{k}},\bs{g}_4} )\right). \label{Equation_PXP_1}
\end{align}

This can be simplified by making use of the orthogonality of states at different $\bs{g}$:
\begin{align}
	\braket{u_{\bs{k}},\bs{g}_2}{u_{\bs{k}}, \bs{g}_3}&= \delta_{\bs{g}_2, \bs{g}_3}\\
	\bra{u_{\bs{k}},\bs{g}_2} \partial_{\bs{k}} \ket{{u_{\bs{k}}, \bs{g}_3}} &= -i \bs{A}^{\text{P}}(\bs{k}+\bs{g}_2) \delta_{\bs{g}_2, \bs{g}_3},
\end{align}
where $\bs{A}^{\text{P}}$ is the Berry connection of the parent band. A combination that occurs frequently in the following argument is 
\begin{align}
	\bra{ u_{\bs{k}},\bs{g}_2} \partial_{k_x}(\ket{ u_{\bs{k}},\bs{g}_3} \bra{ u_{\bs{k}},\bs{g}_4} ) &= -i \bs{A}^{\text{P}}(\bs{k}+\bs{g}_2) \delta_{\bs{g}_2, \bs{g}_3} \bra{ u_{\bs{k}},\bs{g}_4} + \delta_{\bs{g}_2, \bs{g}_3} \partial_{k_x} \bra{ u_{\bs{k}},\bs{g}_4}\\
	&= \delta_{\bs{g}_2, \bs{g}_3} 	\bra{ u_{\bs{k}},\bs{g}_2}\partial_{k_x}(\ket{ u_{\bs{k}},\bs{g}_2} \bra{ u_{\bs{k}},\bs{g}_4} )   .
\end{align}

With these relations, Eq.~\eqref{Equation_PXP_1} becomes

\begin{align}
	P \partial_{k_x} P =  \sum_{\bs{g}_1, \bs{g}_2, \bs{g}_4} 	\mathcal{P}_{\boldsymbol{g}_2\boldsymbol{g}_1}(\bs{k})\ket{ u_{\bs{k}},\bs{g}_1} (  \partial_{k_x}\mathcal{P}_{\bs{g}_4 \bs{g}_2}  \bra{ u_{\bs{k}},\bs{g}_4}  + \mathcal{P}_{\bs{g}_4 \bs{g}_2}  \bra{ u_{\bs{k}},\bs{g}_2}\partial_{k_x}(\ket{ u_{\bs{k}},\bs{g}_2} \bra{ u_{\bs{k}},\bs{g}_4} )).
\end{align}

Then 
\begin{align}
	P \partial_{k_x}P \partial_{k_y} P&=\sum_{\bs{g}_1, \bs{g}_2, \bs{g}_4, \bs{g}_5, \bs{g}_6} 	\mathcal{P}_{\boldsymbol{g}_2\boldsymbol{g}_1}(\bs{k})\ket{ u_{\bs{k}},\bs{g}_1} (  \partial_{k_x}\mathcal{P}_{\bs{g}_4 \bs{g}_2}  \bra{ u_{\bs{k}},\bs{g}_4}  + \mathcal{P}_{\bs{g}_4 \bs{g}_2}  \bra{ u_{\bs{k}},\bs{g}_2}\partial_{k_x}(\ket{ u_{\bs{k}},\bs{g}_2} \bra{ u_{\bs{k}},\bs{g}_4} ))\notag \\
	& \hspace{3cm} \times [\partial_{k_y}\mathcal{P}_{\bs{g}_6 \bs{g}_5} \ket{ u_{\bs{k}},\bs{g}_5} \bra{ u_{\bs{k}},\bs{g}_6}  + \mathcal{P}_{\bs{g}_6 \bs{g}_5} \partial_{k_y}(\ket{ u_{\bs{k}},\bs{g}_5} \bra{ u_{\bs{k}},\bs{g}_6} )]\\
	&= \sum_{\bs{g}_1, \bs{g}_2, \bs{g}_4, \bs{g}_6} \mathcal{P}_{\boldsymbol{g}_2\boldsymbol{g}_1}(\bs{k})\partial_{k_x}\mathcal{P}_{\bs{g}_4 \bs{g}_2} \partial_{k_y}\mathcal{P}_{\bs{g}_6 \bs{g}_4}\ket{ u_{\bs{k}},\bs{g}_1} \bra{ u_{\bs{k}},\bs{g}_6} \notag \\
	&+ \sum_{\bs{g}_1, \bs{g}_2, \bs{g}_4, \bs{g}_6} \mathcal{P}_{\boldsymbol{g}_2\boldsymbol{g}_1}(\bs{k})\partial_{k_x}\mathcal{P}_{\bs{g}_4 \bs{g}_2}  \mathcal{P}_{\bs{g}_6 \bs{g}_4} \ket{ u_{\bs{k}},\bs{g}_1} \bra{ u_{\bs{k}},\bs{g}_4} \partial_{k_y}(\ket{ u_{\bs{k}},\bs{g}_4} \bra{ u_{\bs{k}},\bs{g}_6} ) \notag\\
	&+ \sum_{\bs{g}_1, \bs{g}_2, \bs{g}_4, \bs{g}_6} \mathcal{P}_{\boldsymbol{g}_2\boldsymbol{g}_1}(\bs{k})\mathcal{P}_{\bs{g}_4 \bs{g}_2}  \partial_{k_y} \mathcal{P}_{\bs{g}_6 \bs{g}_4} \ket{ u_{\bs{k}},\bs{g}_1}  \bra{ u_{\bs{k}},\bs{g}_2} \partial_{k_x}(\ket{ u_{\bs{k}},\bs{g}_2} \bra{ u_{\bs{k}},\bs{g}_4} )  \ket{ u_{\bs{k}},\bs{g}_4}  \bra{ u_{\bs{k}},\bs{g}_6}  \notag\\
	&+ \sum_{\bs{g}_1, \bs{g}_2, \bs{g}_4, \bs{g}_5, \bs{g}_6} \mathcal{P}_{\boldsymbol{g}_2\boldsymbol{g}_1}(\bs{k})\mathcal{P}_{\bs{g}_4 \bs{g}_2}  \mathcal{P}_{\bs{g}_6 \bs{g}_5} \ket{ u_{\bs{k}},\bs{g}_1}  \bra{ u_{\bs{k}},\bs{g}_2}\partial_{k_x}(\ket{ u_{\bs{k}},\bs{g}_2} \bra{ u_{\bs{k}},\bs{g}_4} )   \partial_{k_y}(\ket{ u_{\bs{k}},\bs{g}_5} \bra{ u_{\bs{k}},\bs{g}_6} ) .
\end{align}

We can take the trace of this, which is

\begin{align}
	\Tr (	P \partial_{k_x}P \partial_{k_y} P) &= \sum_{\bs{g}} \bra{u_{\bs{k}},\bs{g}}P \partial_{k_x}P \partial_{k_y} P \ket{u_{\bs{k}},\bs{g}} \notag \\
	&= \sum_{\bs{g}, \bs{g}_2, \bs{g}_4} \mathcal{P}_{\boldsymbol{g}_2\boldsymbol{g}}(\bs{k})\partial_{k_x}\mathcal{P}_{\bs{g}_4 \bs{g}_2} \partial_{k_y}\mathcal{P}_{\bs{g} \bs{g}_4}+  \sum_{\bs{g}, \bs{g}_2, \bs{g}_4} \mathcal{P}_{\boldsymbol{g}_2\boldsymbol{g}}(\bs{k})\partial_{k_x}\mathcal{P}_{\bs{g}_4 \bs{g}_2}  \mathcal{P}_{\bs{g} \bs{g}_4}  \bra{ u_{\bs{k}},\bs{g}_4} \partial_{k_y}(\ket{ u_{\bs{k}},\bs{g}_4} \bra{ u_{\bs{k}},\bs{g}} ) \ket{u_{\bs{k}},\bs{g}} \notag \\
	&+ \sum_{\bs{g}, \bs{g}_2, \bs{g}_4} \mathcal{P}_{\boldsymbol{g}_2\boldsymbol{g}}(\bs{k})\mathcal{P}_{\bs{g}_4 \bs{g}_2}  \partial_{k_y} \mathcal{P}_{\bs{g} \bs{g}_4}  \bra{ u_{\bs{k}},\bs{g}_2} \partial_{k_x}(\ket{ u_{\bs{k}},\bs{g}_2} \bra{ u_{\bs{k}},\bs{g}_4} )  \ket{ u_{\bs{k}},\bs{g}_4}   \notag\\
	&+ \sum_{\bs{g}, \bs{g}_2, \bs{g}_4, \bs{g}_5} \mathcal{P}_{\boldsymbol{g}_2\boldsymbol{g}}(\bs{k})\mathcal{P}_{\bs{g}_4 \bs{g}_2}  \mathcal{P}_{\bs{g} \bs{g}_5} \bra{ u_{\bs{k}},\bs{g}_2}\partial_{k_x}(\ket{ u_{\bs{k}},\bs{g}_2} \bra{ u_{\bs{k}},\bs{g}_4} )   \partial_{k_y}(\ket{ u_{\bs{k}},\bs{g}_5} \bra{ u_{\bs{k}},\bs{g}} ) \ket{u_{\bs{k}},\bs{g}}. \label{Eq_Tr_PXPYP_1}
\end{align}

We then note that

\begin{align}
	\bra{ u_{\bs{k}},\bs{g}_4} \partial_{k_y}(\ket{ u_{\bs{k}},\bs{g}_4} \bra{ u_{\bs{k}},\bs{g}} ) \ket{u_{\bs{k}},\bs{g}} &= \bra{ u_{\bs{k}},\bs{g}_4} \partial_{k_y}\ket{ u_{\bs{k}},\bs{g}_4} + \partial_{k_y} (\bra{ u_{\bs{k}},\bs{g}} ) \ket{u_{\bs{k}},\bs{g}}\notag \\
	&= -i A_y^{\text{P}}(\bs{k}+\bs{g}_4) +i A_y^{\text{P}}(\bs{k}+\bs{g})
\end{align}
with a similar result holding for the matrix elements on the second line of Eq.~\eqref{Eq_Tr_PXPYP_1}. As a result, 
\begin{align}
	\Tr (	P \partial_{k_x}P \partial_{k_y} P)	&= \sum_{\bs{g}, \bs{g}_2, \bs{g}_4} \mathcal{P}_{\boldsymbol{g}_2\boldsymbol{g}}(\bs{k})\partial_{k_x}\mathcal{P}_{\bs{g}_4 \bs{g}_2} \big( \partial_{k_y}+ i A_y^{\text{P}}(\bs{k}+\bs{g})-i A_y^{\text{P}} (\bs{k}+\bs{g}_4))\mathcal{P}_{\bs{g} \bs{g}_4} \notag \\
	&+ \sum_{\bs{g}, \bs{g}_2, \bs{g}_4} \mathcal{P}_{\boldsymbol{g}_2\boldsymbol{g}}(\bs{k})\mathcal{P}_{\bs{g}_4 \bs{g}_2}  \partial_{k_y} \mathcal{P}_{\bs{g} \bs{g}_4} \big(iA_x^{\text{P}} (\bs{k}+\bs{g}_4 )-iA_x^{\text{P}}(\bs{k}+\bs{g}_2)\big)\notag \\
	&+ \sum_{\bs{g}, \bs{g}_2, \bs{g}_4, \bs{g}_5} \mathcal{P}_{\boldsymbol{g}_2\boldsymbol{g}}(\bs{k})\mathcal{P}_{\bs{g}_4 \bs{g}_2}  \mathcal{P}_{\bs{g} \bs{g}_5} \bra{ u_{\bs{k}},\bs{g}_2}\partial_{k_x}(\ket{ u_{\bs{k}},\bs{g}_2} \bra{ u_{\bs{k}},\bs{g}_4} )   \partial_{k_y}(\ket{ u_{\bs{k}},\bs{g}_5} \bra{ u_{\bs{k}},\bs{g}} ) \ket{u_{\bs{k}},\bs{g}}.
\end{align}

This structure closely resembles covariant derivatives of the density matrices, which we can make more complete by rewriting our expression as

\begin{align}
	\Tr (	P \partial_{k_x}P \partial_{k_y} P)	&= \sum_{\bs{g}, \bs{g}_2, \bs{g}_4} \mathcal{P}_{\boldsymbol{g}_2\boldsymbol{g}}(\bs{k})\big(\partial_{k_x}+iA_x^{\text{P}} (\bs{k}+\bs{g}_4 )-iA_x^{\text{P}}(\bs{k}+\bs{g}_2)\big)\mathcal{P}_{\bs{g}_4 \bs{g}_2} \big( \partial_{k_y}+ i A_y^{\text{P}}(\bs{k}+\bs{g})-i A_y^{\text{P}} (\bs{k}+\bs{g}_4)\big)\mathcal{P}_{\bs{g} \bs{g}_4} \notag \\
	&- \sum_{\bs{g}, \bs{g}_2, \bs{g}_4} \mathcal{P}_{\boldsymbol{g}_2\boldsymbol{g}}(\bs{k})\mathcal{P}_{\bs{g}_4 \bs{g}_2}  \mathcal{P}_{\bs{g} \bs{g}_4}\big(iA_x^{\text{P}} (\bs{k}+\bs{g}_4 )-iA_x^{\text{P}}(\bs{k}+\bs{g}_2)\big) \big(i A_y^{\text{P}}(\bs{k}+\bs{g})-i A_y^{\text{P}} (\bs{k}+\bs{g}_4))\notag \\
	&+ \sum_{\bs{g}, \bs{g}_2, \bs{g}_4, \bs{g}_5} \mathcal{P}_{\boldsymbol{g}_2\boldsymbol{g}}(\bs{k})\mathcal{P}_{\bs{g}_4 \bs{g}_2}  \mathcal{P}_{\bs{g} \bs{g}_5} \bra{ u_{\bs{k}},\bs{g}_2}\partial_{k_x}(\ket{ u_{\bs{k}},\bs{g}_2} \bra{ u_{\bs{k}},\bs{g}_4} )   \partial_{k_y}(\ket{ u_{\bs{k}},\bs{g}_5} \bra{ u_{\bs{k}},\bs{g}} ) \ket{u_{\bs{k}},\bs{g}}, \label{Eq_Tr_PXPYP_2}
\end{align}
by using an insertion of the identity. Then we consider the matrix element in the last term

\begin{align}
 \bra{ u_{\bs{k}},\bs{g}_2}\partial_{k_x}(\ket{ u_{\bs{k}},\bs{g}_2} \bra{ u_{\bs{k}},\bs{g}_4} )&   \partial_{k_y}(\ket{ u_{\bs{k}},\bs{g}_5} \bra{ u_{\bs{k}},\bs{g}} ) \ket{u_{\bs{k}},\bs{g}} \notag\\
  &= (-iA_x^{\text{P}}(\bs{k}+\bs{g}_2) \bra{ u_{\bs{k}},\bs{g}_4} + \partial_{k_x} \bra{ u_{\bs{k}},\bs{g}_4}  ) (\partial_{k_y}\ket{ u_{\bs{k}},\bs{g}_5} +iA_y^{\text{P}}(\bs{k}+\bs{g})\ket{ u_{\bs{k}},\bs{g}_5} )\notag \\
  &= (-iA_x^{\text{P}}(\bs{k}+\bs{g}_2)  )(-i \delta_{\bs{g}_4, \bs{g}_5} A_y^{\text{P}}(\bs{k}+\bs{g}_4) +i \delta_{\bs{g}_4, \bs{g}_5}A_y^{\text{P}}(\bs{k}+\bs{g})) \notag\\
  &\quad +(i \delta_{\bs{g}_4, \bs{g}_5} A_x^{\text{P}}(\bs{k}+\bs{g}_4))(iA_y^{\text{P}}(\bs{k}+\bs{g})) +(\partial_{k_x} \bra{ u_{\bs{k}},\bs{g}_4} ) \partial_{k_y}\ket{ u_{\bs{k}},\bs{g}_5}\notag \\
  &= \delta_{\bs{g}_4, \bs{g}_5} (iA_x^{\text{P}}(\bs{k}+\bs{g}_4)-iA_x^{\text{P}}(\bs{k}+\bs{g}_2)  )( +i A_y^{\text{P}}(\bs{k}+\bs{g})-i  A_y^{\text{P}}(\bs{k}+\bs{g}_4)) \notag \\
  &\quad +\partial_{k_x} (\bra{ u_{\bs{k}},\bs{g}_4}  \partial_{k_y}\ket{ u_{\bs{k}},\bs{g}_5}) - \bra{ u_{\bs{k}},\bs{g}_4}  \partial_{k_x} \partial_{k_y} \ket{ u_{\bs{k}},\bs{g}_5} \notag \\
  &= \delta_{\bs{g}_4, \bs{g}_5} (iA_x^{\text{P}}(\bs{k}+\bs{g}_4)-iA_x^{\text{P}}(\bs{k}+\bs{g}_2)  )( +i A_y^{\text{P}}(\bs{k}+\bs{g})-i  A_y^{\text{P}}(\bs{k}+\bs{g}_4)) \notag \\
  &\quad -i \delta_{\bs{g}_4, \bs{g}_5} \partial_{k_x} A_y^{\text{P}}(\bs{k}+\bs{g}_4)- \bra{ u_{\bs{k}},\bs{g}_4}  \partial_{k_x} \partial_{k_y} \ket{ u_{\bs{k}},\bs{g}_5}.
\end{align}

Inserting this into Eq.~\eqref{Eq_Tr_PXPYP_2}, we obtain
\begin{align}
	\Tr (	P \partial_{k_x}P \partial_{k_y} P)	=& \sum_{\bs{g}, \bs{g}_2, \bs{g}_4} \mathcal{P}_{\boldsymbol{g}_2\boldsymbol{g}}(\bs{k})\big(\partial_{k_x}+iA_x^{\text{P}} (\bs{k}+\bs{g}_4 )-iA_x^{\text{P}}(\bs{k}+\bs{g}_2)\big)\mathcal{P}_{\bs{g}_4 \bs{g}_2} \big( \partial_{k_y}+ i A_y^{\text{P}}(\bs{k}+\bs{g})-i A_y^{\text{P}} (\bs{k}+\bs{g}_4)\big)\mathcal{P}_{\bs{g} \bs{g}_4} \notag \\
	&-i \sum_{\bs{g}, \bs{g}_2, \bs{g}_4}  \partial_{k_x} A_y^{\text{P}}(\bs{k}+\bs{g}_4) \mathcal{P}_{\boldsymbol{g}_2\boldsymbol{g}}(\bs{k})\mathcal{P}_{\bs{g}_4 \bs{g}_2}  \mathcal{P}_{\bs{g} \bs{g}_4}- \sum_{\bs{g}, \bs{g}_2, \bs{g}_4, \bs{g}_5} \mathcal{P}_{\boldsymbol{g}_2\boldsymbol{g}}(\bs{k})\mathcal{P}_{\bs{g}_4 \bs{g}_2}  \mathcal{P}_{\bs{g} \bs{g}_5} \bra{ u_{\bs{k}},\bs{g}_4}  \partial_{k_x} \partial_{k_y} \ket{ u_{\bs{k}},\bs{g}_5}.
\end{align}

Next, note that the density matrices act as projectors: $\sum_{\bs{g}} \mathcal{P}_{\bs{g'} \bs{g}} \mathcal{P}_{\bs{g} \bs{g}''} = \mathcal{P}_{\bs{g}' \bs{g}''}$. Therefore we can simplify the last two terms to obtain

\begin{align}
	\Tr (	P \partial_{k_x}P \partial_{k_y} P)	=& \sum_{\bs{g}, \bs{g}_2, \bs{g}_4} \mathcal{P}_{\boldsymbol{g}_2\boldsymbol{g}}(\bs{k})\big(\partial_{k_x}+iA_x^{\text{P}} (\bs{k}+\bs{g}_4 )-iA_x^{\text{P}}(\bs{k}+\bs{g}_2)\big)\mathcal{P}_{\bs{g}_4 \bs{g}_2} \big( \partial_{k_y}+ i A_y^{\text{P}}(\bs{k}+\bs{g})-i A_y^{\text{P}} (\bs{k}+\bs{g}_4)\big)\mathcal{P}_{\bs{g} \bs{g}_4} \notag \\
	&-i \sum_{\bs{g}_4}  \partial_{k_x} A_y^{\text{P}}(\bs{k}+\bs{g}_4) \mathcal{P}_{\boldsymbol{g}_4\boldsymbol{g}_4}(\bs{k})- \sum_{ \bs{g}_4, \bs{g}_5}  \mathcal{P}_{\bs{g}_4 \bs{g}_5} \bra{ u_{\bs{k}},\bs{g}_4}  \partial_{k_x} \partial_{k_y} \ket{ u_{\bs{k}},\bs{g}_5}.
\end{align}

Using this result, we can obtain the Berry curvature 

\begin{align}
	\Omega(\bs{k})=&i\Tr (	P \partial_{k_x}P \partial_{k_y} P) - i \Tr (	P \partial_{k_y}P \partial_{k_x} P) \notag\\
	=&i\sum_{\bs{g}, \bs{g}_2, \bs{g}_4} \mathcal{P}_{\boldsymbol{g}_2\boldsymbol{g}}(\bs{k})\big(\partial_{k_x}+iA_x^{\text{P}} (\bs{k}+\bs{g}_4 )-iA_x^{\text{P}}(\bs{k}+\bs{g}_2)\big)\mathcal{P}_{\bs{g}_4 \bs{g}_2} \big( \partial_{k_y}+ i A_y^{\text{P}}(\bs{k}+\bs{g})-i A_y^{\text{P}} (\bs{k}+\bs{g}_4)\big)\mathcal{P}_{\bs{g} \bs{g}_4} - x \leftrightarrow y\notag \\
	&+ \sum_{\bs{g}} \mathcal{P}_{\bs{g} \bs{g}} (\partial_{k_x}A_y^{\text{P}}(\bs{k}+\bs{g}) - \partial_{k_y} A_x^{\text{P}}(\bs{k}+\bs{g})) -i\sum_{ \bs{g}_4, \bs{g}_5}  \mathcal{P}_{\bs{g}_4 \bs{g}_5} \bra{ u_{\bs{k}},\bs{g}_4} ( \partial_{k_x} \partial_{k_y} -\partial_{k_y} \partial_{k_x})\ket{ u_{\bs{k}},\bs{g}_5}.
\end{align}

The last term is zero from the commutation of derivatives, while the penultimate term can be recognized as the occupation-weighted Berry curvature of the parent band, $\sum_{\bs{g}} n(\bs{k}+\bs{g}) \Omega^{\text{P}}(\bs{k}+\bs{g})$, leading to the relatively simple expression

\begin{align}
	\Omega(\bs{k})=&i\Tr (	P \partial_{k_x}P \partial_{k_y} P) - i \Tr (	P \partial_{k_y}P \partial_{k_x} P) \notag\\
	=&i\sum_{\bs{g}_1, \bs{g}_2, \bs{g}_3} \mathcal{P}_{\boldsymbol{g}_2\boldsymbol{g}_1}(\bs{k})\big(\partial_{k_x}+iA_x^{\text{P}} (\bs{k}+\bs{g}_3 )-iA_x^{\text{P}}(\bs{k}+\bs{g}_2)\big)\mathcal{P}_{\bs{g}_3 \bs{g}_2} \big( \partial_{k_y}+ i A_y^{\text{P}}(\bs{k}+\bs{g}_1)-i A_y^{\text{P}} (\bs{k}+\bs{g}_3)\big)\mathcal{P}_{\bs{g}_1 \bs{g}_3} - (x \leftrightarrow y)\notag \\
	&+ \sum_{\bs{g}} n(\bs{k}+\bs{g}) \Omega^{\text{P}}(\bs{k}+\bs{g}).
\end{align}

Note that the combination $\big(\partial_{k_x}+iA_x^{\text{P}} (\bs{k}+\bs{g}_3 )-iA_x^{\text{P}}(\bs{k}+\bs{g}_2)\big)\mathcal{P}_{\bs{g}_3 \bs{g}_2} $ is covariant under changes to the gauge of the underlying parent band states (making the final expression invariant), as well as invariant under changes to the gauge of the crystal state. This is because, under the gauge transformation on the underlying states, where $\ket{\bs{k}+\bs{g}} \rightarrow e^{i \alpha_{\bs{k}+\bs{g}}} \ket{\bs{k}+\bs{g}}$ and $\bs{A}^p(\bs{k}+\bs{g}) \rightarrow \bs{A}^p(\bs{k}+\bs{g})- \nabla \alpha_{\bs{k}+\bs{g}}$, the density matrix transforms as $\mathcal{P}_{\boldsymbol{g}_3\boldsymbol{g}_2} \rightarrow e^{-i\alpha_{\bs{k}+\bs{g}_2} +i\alpha_{\bs{k}+\bs{g}_3} }\mathcal{P}_{\boldsymbol{g}_3\boldsymbol{g}_2}$ in order to describe the same crystal state. Then,
\begin{align}
\big(\partial_{k_x}&+iA_x^{\text{P}}(\bs{k}+\bs{g}_3 )-iA_x^{\text{P}}(\bs{k}+\bs{g}_2)\big)\mathcal{P}_{\bs{g}_3 \bs{g}_2} \notag \\
&\rightarrow  \big(\partial_{k_x}+iA_x^{\text{P}} (\bs{k}+\bs{g}_3 )-i\partial_{k_x} \alpha_{\bs{k}+\bs{g}_3}-iA_x^{\text{P}}(\bs{k}+\bs{g}_2)+i\partial_{k_x} \alpha_{\bs{k}+\bs{g}_2}\big)e^{-i\alpha_{\bs{k}+\bs{g}_2} +i\alpha_{\bs{k}+\bs{g}_3} }\mathcal{P}_{\bs{g}_3 \bs{g}_2} \notag\\
&=e^{-i\alpha_{\bs{k}+\bs{g}_2} +i\alpha_{\bs{k}+\bs{g}_3} }\big(\partial_{k_x}-i\partial_{k_x}\alpha_{\bs{k}+\bs{g}_2} +i\partial_{k_x}\alpha_{\bs{k}+\bs{g}_3} +iA_x^{\text{P}} (\bs{k}+\bs{g}_3 )-i\partial_{k_x} \alpha_{\bs{k}+\bs{g}_3}-iA_x^{\text{P}}(\bs{k}+\bs{g}_2)+i\partial_{k_x} \alpha_{\bs{k}+\bs{g}_2}\big)\mathcal{P}_{\bs{g}_3 \bs{g}_2}  \notag\\
&=e^{-i\alpha_{\bs{k}+\bs{g}_2} +i\alpha_{\bs{k}+\bs{g}_3} }\big(\partial_{k_x}+iA_x^{\text{P}} (\bs{k}+\bs{g}_3 )-iA_x^{\text{P}}(\bs{k}+\bs{g}_2)\big)\mathcal{P}_{\bs{g}_3 \bs{g}_2},
\end{align}
so the phase from the gauge transform passes through the derivative, as we expect for a covariant derivative. 

In some cases, it will be useful to express the Berry curvature in terms of the phase and amplitude of the density matrix. We write
\begin{align}
	\mathcal{P}_{\bs{g} \bs{g}'} = \sqrt{n(\bs{k}+\bs{g}) n(\bs{k}+\bs{g}')} e^{i \theta_{\bs{g}'\bs{g}}}
\end{align}
Then,
\begin{align}
\big(\partial_{k_x}+iA_x^{\text{P}} (\bs{k}+\bs{g}_3 )&-iA_x^{\text{P}}(\bs{k}+\bs{g}_2)\big)\mathcal{P}_{\bs{g}_3 \bs{g}_2}\notag \\
& = \big(i \partial_{k_x} \theta_{\bs{g}_2 \bs{g}_3} +iA_x^{\text{P}}(\bs{k}+\bs{g}_3)-iA_x^{\text{P}}(\bs{k}+\bs{g}_2) + \frac{1}{2}\partial_{k_x} \ln(n(\bs{k}+\bs{g}_2) n(\bs{k}+\bs{g}_3))\big) \mathcal{P}_{\bs{g}_3 \bs{g}_2},
\end{align}

and so the Berry curvature is given by

\begin{align}
	\Omega(\bs{k})&=-i \sum_{\bs{g}_1, \bs{g}_2, \bs{g}_3} \mathcal{P}_{\bs{g}_3 \bs{g}_2}\mathcal{P}_{\boldsymbol{g}_2\boldsymbol{g}_1}\mathcal{P}_{\bs{g}_1 \bs{g}_3} \big( \partial_{k_x} \theta_{\bs{g}_2 \bs{g}_3} +A_x^{\text{P}}(\bs{k}+\bs{g}_3)-A_x^{\text{P}}(\bs{k}+\bs{g}_2) - \frac{i}{2}\partial_{k_x} \ln(n(\bs{k}+\bs{g}_2) n(\bs{k}+\bs{g}_3))\big) \notag \\
	& \quad \big( \partial_{k_y} \theta_{\bs{g}_3 \bs{g}_1} +A_y^{\text{P}}(\bs{k}+\bs{g}_1)-A_y^{\text{P}}(\bs{k}+\bs{g}_3) - \frac{i}{2}\partial_{k_y} \ln(n(\bs{k}+\bs{g}_1) n(\bs{k}+\bs{g}_3))\big)  - x \leftrightarrow y \notag \\
	&+ \sum_{\bs{g}} n(\bs{k}+\bs{g}) \Omega^{\text{P}}(\bs{k}+\bs{g}).
\end{align}

Noting that $\mathcal{P}_{\bs{g}_3 \bs{g}_2}\mathcal{P}_{\boldsymbol{g}_2\boldsymbol{g}_1}\mathcal{P}_{\bs{g}_1 \bs{g}_3} = n(\bs{k}+\bs{g}_1) n(\bs{k}+\bs{g}_2)n(\bs{k}+\bs{g}_3)$, we can write this as

\begin{align}
	\Omega(\bs{k})&=-i \sum_{\bs{g}_1, \bs{g}_2, \bs{g}_3} n(\bs{k}+\bs{g}_1) n(\bs{k}+\bs{g}_2)n(\bs{k}+\bs{g}_3)\big[ \notag \\
	& \quad \big( \partial_{k_x} \theta_{\bs{g}_2 \bs{g}_3} +A_x^{\text{P}}(\bs{k}+\bs{g}_3)-A_x^{\text{P}}(\bs{k}+\bs{g}_2)\big) \big( \partial_{k_y} \theta_{\bs{g}_3 \bs{g}_1} +A_y^{\text{P}}(\bs{k}+\bs{g}_1)-A_y^{\text{P}}(\bs{k}+\bs{g}_3)\big) \notag \\
	&- \frac{1}{4} \partial_{k_x} \ln(n(\bs{k}+\bs{g}_2) n(\bs{k}+\bs{g}_3)) \partial_{k_y} \ln(n(\bs{k}+\bs{g}_1) n(\bs{k}+\bs{g}_3)) \notag  \\
	&-\frac{i}{2} \big( \partial_{k_x} \theta_{\bs{g}_2 \bs{g}_3} +A_x^{\text{P}}(\bs{k}+\bs{g}_3)-A_x^{\text{P}}(\bs{k}+\bs{g}_2)\big) \partial_{k_y} \ln(n(\bs{k}+\bs{g}_1) n(\bs{k}+\bs{g}_3)) \notag \\
	&-\frac{i}{2} \big( \partial_{k_y} \theta_{\bs{g}_3 \bs{g}_1} +A_y^{\text{P}}(\bs{k}+\bs{g}_1)-A_y^{\text{P}}(\bs{k}+\bs{g}_3)\big) \partial_{k_x} \ln(n(\bs{k}+\bs{g}_2) n(\bs{k}+\bs{g}_3))	\big] \notag \\
	&- x \leftrightarrow y +\sum_{\bs{g}} n(\bs{k}+\bs{g}) \Omega^{\text{P}}(\bs{k}+\bs{g}).
\end{align}

The first two terms are purely imaginary. However, they vanish because they are symmetric under the exchange $x \leftrightarrow y$ (as can be seen by swapping the dummy indices $\bs{g}_1$ and $\bs{g}_2$). On the other hand, the cross-terms are antisymmetric under this exchange. This leaves us with
\begin{align}
	\Omega(\bs{k})&=- \sum_{\bs{g}_1, \bs{g}_2, \bs{g}_3} n(\bs{k}+\bs{g}_1) n(\bs{k}+\bs{g}_2)n(\bs{k}+\bs{g}_3)\big[ 
	 \big( \partial_{k_x} \theta_{\bs{g}_2 \bs{g}_3} +A_x^{\text{P}}(\bs{k}+\bs{g}_3)-A_x^{\text{P}}(\bs{k}+\bs{g}_2)\big) \partial_{k_y} \ln(n(\bs{k}+\bs{g}_1) n(\bs{k}+\bs{g}_3)) \notag \\
	& \hspace{2cm} -\big( \partial_{k_y} \theta_{\bs{g}_2 \bs{g}_3} +A_y^{\text{P}}(\bs{k}+\bs{g}_3)-A_y^{\text{P}}(\bs{k}+\bs{g}_2)\big) \partial_{k_x} \ln(n(\bs{k}+\bs{g}_1) n(\bs{k}+\bs{g}_3))	\big] \notag \\
	& +\sum_{\bs{g}} n(\bs{k}+\bs{g}) \Omega^{\text{P}}(\bs{k}+\bs{g}).
\end{align}

Then we can resolve the sum over $\bs{g}_1$ by noting that
\begin{align*}
\sum_{\bs{g}_1} n(\bs{k}+\bs{g}_1) \partial_{k_x} \ln(n(\bs{k}+\bs{g}_1) n(\bs{k}+\bs{g}_3)) &= \partial_{k_x} \sum_{\bs{g}_1}  n(\bs{k}+\bs{g}_1) +  \partial_{k_x}\ln(n(\bs{k}+\bs{g}_3))  \sum_{\bs{g}_1} n(\bs{k}+\bs{g}_1)\\
&= \partial_{k_x} \ln(n(\bs{k}+\bs{g}_3)),
\end{align*}
where we used the fact that $\sum_{\bs{g}_1}n(\bs{k}+\bs{g}_1)=1$ for a crystalline state. As a result, the Berry curvature is simply
\begin{align}
	\Omega(\bs{k})&=- \sum_{\bs{g}_2, \bs{g}_3} n(\bs{k}+\bs{g}_2)\big[ 
	\big( \partial_{k_x} \theta_{\bs{g}_2 \bs{g}_3} +A_x^{\text{P}}(\bs{k}+\bs{g}_3)-A_x^{\text{P}}(\bs{k}+\bs{g}_2)\big) \partial_{k_y}  n(\bs{k}+\bs{g}_3) \notag \\
	& \hspace{2cm} -\big( \partial_{k_y} \theta_{\bs{g}_2 \bs{g}_3} +A_y^{\text{P}}(\bs{k}+\bs{g}_3)-A_y^{\text{P}}(\bs{k}+\bs{g}_2)\big) \partial_{k_x} n(\bs{k}+\bs{g}_3)	\big] \notag \\
	&\quad  +\sum_{\bs{g}} n(\bs{k}+\bs{g}) \Omega^{\text{P}}(\bs{k}+\bs{g}). \label{Eqn_Berry_curvature_generic}
\end{align}

\subsection{Weak potential limit}
\label{Section_weak_potential_projector}	
In the case where we apply a very weak potential (or interaction), the occupation in the bulk of the 1BZ is one for the parent band state with the lowest kinetic energy and zero for all others. As a result $\grad n(\bs{k}+\bs{g})=0$ and the Berry curvature from Eq.~\eqref{Eqn_Berry_curvature_generic} is just given by the occupation-weighted Berry curvature (which is also just the parent Berry curvature in the 1BZ):

\begin{align}
	\Omega^{\text{bulk}}(\bs{k})\approx \sum_{\bs{g}} n(\bs{k}+\bs{g}) \Omega^{\text{P}}(\bs{k}+\bs{g}) \approx  \Omega^{\text{P}}(\bs{k}).\label{Eq_Berry_curvature_bulk}
\end{align}	

\begin{figure}[h]
	\centering
	\includegraphics[width=0.5\linewidth]{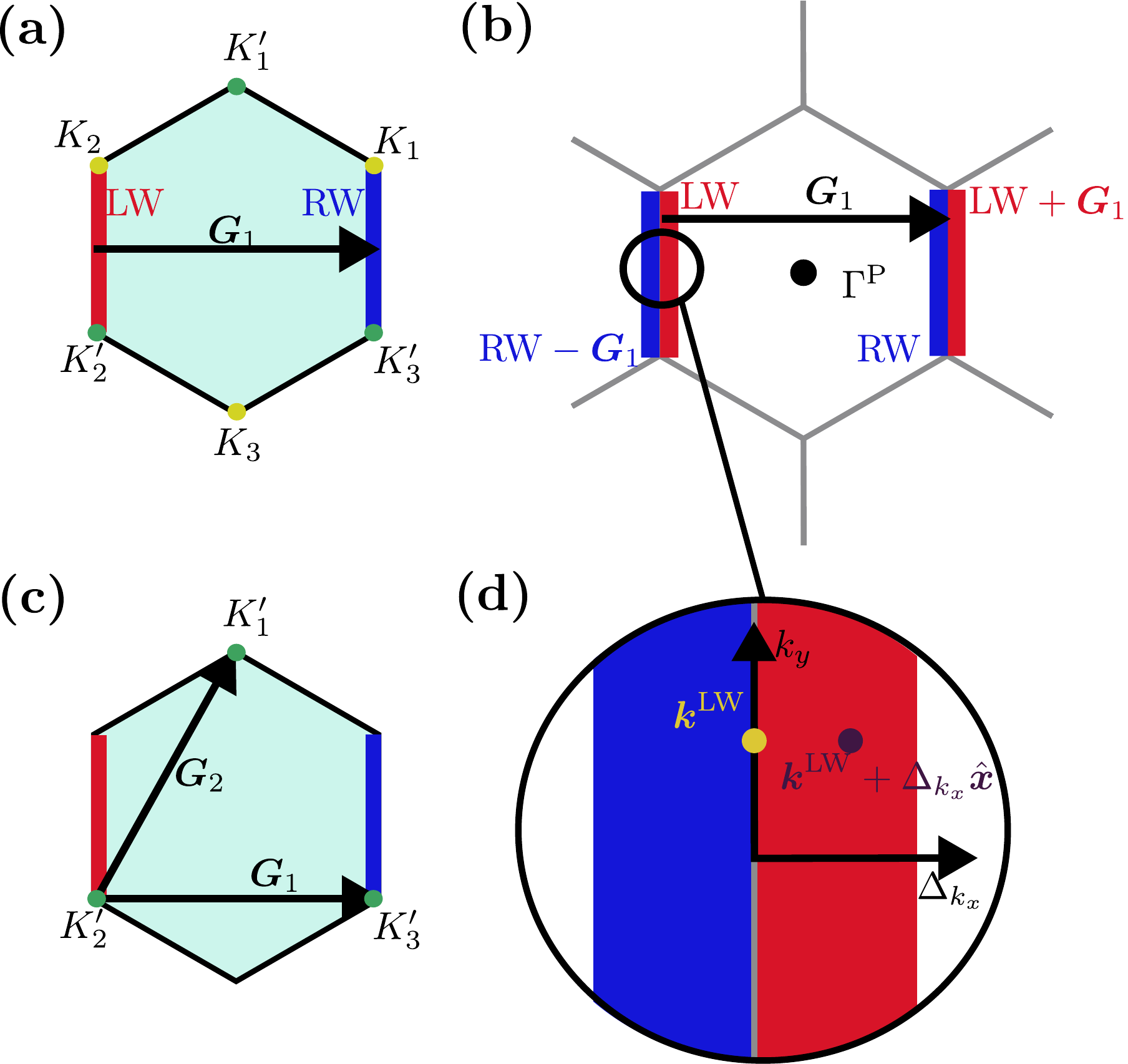}
	\caption{(a) We consider the Berry curvature at the boundary of the 1BZ, focusing on the left and right edges (LW and RW). These edges are adjacent due to the periodicity of the 1BZ. (b) This adjacency is reflected in the parent band when we consider the states mixed by the periodic potential or interaction, which are related by $\bs{G}_1$. (c) Near the corner, three states are mixed instead of just two, and the phase of this mixing is constrained by symmetry.}
	\label{Figure_wire}
\end{figure}

However, the occupation varies rapidly at the boundary, leading to a concentration of Berry curvature, which can dramatically change the Chern number. To examine this effect, we consider one part of the 1BZ boundary (with the other parts related by discrete rotation symmetry). We examine the regions near the left and right edges of the 1BZ, denoted by LW and RW, respectively, as shown in Fig. \ref{Figure_wire}. These regions are distant in the parent band, but are adjacent in crystal momentum due to the periodic nature of the crystal 1BZ, as shown in Fig.~\ref{Figure_wire}(b). To avoid issues associated with the folding of the crystal momentum at the boundary, we treat the two regions separately, and then show that we can combine the contributions from the two edges into a single smooth expression.

We first consider the left side, LW. In order to perform the area integral over the boundary region, we first integrate the Berry curvature along a line perpendicular to the boundary, starting at the edge and passing in to the interior across the boundary region. This integral gives the contribution to the Berry flux from the boundary, per unit length of the boundary. If we are far from the 1BZ corner, the only contributing states are $\bs{k}$ and $\bs{k}+\bs{G}_1$, where $\bs{k}+\bs{G}_1$ is just outside the 1BZ. Therefore, the Berry curvature is
\begin{align}
	\Omega^{\text{LW}}(\bs{k}) &= - n(\bs{k})\big[ 
	\big( \partial_{k_x} \theta_{\bs{0} \bs{G}_1} +A_x^{\text{P}}(\bs{k}+\bs{G}_1)-A_x^{\text{P}}(\bs{k})\big) \partial_{k_y}  n(\bs{k}+\bs{G}_1) -\big( \partial_{k_y} \theta_{\bs{0} \bs{G}_1} +A_y^{\text{P}}(\bs{k}+\bs{G}_1)-A_y^{\text{P}}(\bs{k})\big) \partial_{k_x} n(\bs{k}+\bs{G}_1)	\big] \notag \\
	&\quad - n(\bs{k}+\bs{G}_1)\big[ 
	\big( \partial_{k_x} \theta_{\bs{G}_1 \bs{0}} +A_x^{\text{P}}(\bs{k})-A_x^{\text{P}}(\bs{k}+\bs{G}_1)\big) \partial_{k_y}  n(\bs{k})  -\big( \partial_{k_y} \theta_{\bs{G}_1 \bs{0}} +A_y^{\text{P}}(\bs{k})-A_y^{\text{P}}(\bs{k}+\bs{G}_1)\big) \partial_{k_x} n(\bs{k})	\big] \notag \\
	&\quad  +\sum_{\bs{g}} n(\bs{k}+\bs{g}) \Omega^{\text{P}}(\bs{k}+\bs{g}) \notag\\
	&= - 	\big( \partial_{k_x} \theta_{\bs{0} \bs{G}_1} +A_x^{\text{P}}(\bs{k}+\bs{G}_1)-A_x^{\text{P}}(\bs{k})\big) \big(n(\bs{k})\partial_{k_y}  n(\bs{k}+\bs{G}_1) - n(\bs{k}+\bs{G}_1)\partial_{k_y}  n(\bs{k})\big) \notag\\
	&\quad +	\big( \partial_{k_y} \theta_{\bs{0} \bs{G}_1} +A_y^{\text{P}}(\bs{k}+\bs{G}_1)-A_y^{\text{P}}(\bs{k})\big) \big(n(\bs{k})\partial_{k_x}  n(\bs{k}+\bs{G}_1) - n(\bs{k}+\bs{G}_1)\partial_{k_x}  n(\bs{k})\big) \notag\\
	& \quad +\sum_{\bs{g}} n(\bs{k}+\bs{g}) \Omega^{\text{P}}(\bs{k}+\bs{g}). 
\end{align} 

Because we consider a small region near the boundary, we can ignore terms that remain finite as the width of the boundary region becomes zero. This includes the occupation-weighted Berry curvature (which we will include later anyway, as we will integrate this contribution over the entire 1BZ) and the term involving the derivative of the occupation parallel to the wire (along $y$). This leaves us with

\begin{align}
		\Omega^{\text{LW}}(\bs{k}) &\approx 	\big( \partial_{k_y} \theta_{\bs{0} \bs{G}_1} +A_y^{\text{P}}(\bs{k}+\bs{G}_1)-A_y^{\text{P}}(\bs{k})\big) \big(n(\bs{k})\partial_{k_x}  n(\bs{k}+\bs{G}_1) - n(\bs{k}+\bs{G}_1)\partial_{k_x}  n(\bs{k})\big).
\end{align}

Integrating this across the boundary region (the red region marked LW in Fig.~\ref{Figure_wire}(a)) gives us
\begin{align}
	\int_0^{\delta}d\Delta_{k_x}& \Omega^{\text{LW}}(\bs{k}^{\text{LW}}+\Delta_{k_x}) \notag \\
    &\approx  \int_0^{\delta} d\Delta_{k_x}	\big( \partial_{k_y} \theta_{\bs{0} \bs{G}_1} +A_y^{\text{P}}(\bs{k}^{\text{LW}}+\Delta_{k_x}+\bs{G}_1)-A_y^{\text{P}}(\bs{k}^{\text{LW}}+\Delta_{k_x})\big) \notag \\
	& \hspace{1cm} \times\big(n(\bs{k}^{\text{LW}}+\Delta_{k_x})\partial_{k_x}  n(\bs{k}^{\text{LW}}+\Delta_{k_x}+ \bs{G}_1) - n(\bs{k}^{\text{LW}}+\Delta_{k_x}+\bs{G}_1)\partial_{k_x}  n(\bs{k}^{\text{LW}}+\Delta_{k_x})\big), 
\end{align}
where $\Delta_{k_x}$ is the distance between the momentum at which the integrand is evaluated and the left wire (see Fig.~\ref{Figure_wire}(d)).

We can treat the parent band connection and derivative of $\theta$ as constant here, because they will not change appreciably over the small boundary region. Then we have

\begin{align}
	\int_0^{\delta}d\Delta_{k_x} &\Omega^{\text{LW}}(\bs{k}^{\text{LW}}+\Delta_{k_x}) \notag \\
    &\approx \big( \partial_{k_y} \theta_{\bs{0} \bs{G}_1} +A_y^{\text{P}}(\bs{k}^{\text{LW}}+\bs{G}_1)-A_y^{\text{P}}(\bs{k}^{\text{LW}})\big)\notag   \\
	&\quad \times \int_0^{\delta} d\Delta_{k_x}	 \big(n(\bs{k}^{\text{LW}}+\Delta_{k_x})\partial_{k_x}  n(\bs{k}^{\text{LW}}+\Delta_{k_x}+ \bs{G}_1) - n(\bs{k}^{\text{LW}}+\Delta_{k_x}+\bs{G}_1)\partial_{k_x}  n(\bs{k}^{\text{LW}}+\Delta_{k_x})\big) . \label{equation_Berry_LW}
\end{align}

Before we evaluate this, now consider the boundary region on the right side of the 1BZ. This area is close to the previous area in terms of crystal momentum, meaning that they form one connected region (as shown in Fig~\ref{Figure_wire}(b) and \ref{Figure_wire}(d), where the LW and RW regions, together with their region shifted by $\pm\vb{G}_1$, connect together to form the interior and exterior boundary region on either side of the 1BZ). We again want to integrate the Berry curvature over the boundary region RW, to get a complete contribution from the two edges. In this region, far from the corners, the only contributing states are $\bs{k}$ and $\bs{k}-\bs{G}_1$. We therefore have (using the same arguments from before)
\begin{align}
	\Omega^{\text{RW}}(\bs{k})\approx 	\big( \partial_{k_y} \theta_{ -\bs{G}_1 \bs{0}} +A_y^{\text{P}}(\bs{k})-A_y^{\text{P}}(\bs{k}-\bs{G}_1)\big) \big(n(\bs{k}-\bs{G}_1)\partial_{k_x}  n(\bs{k}) - n(\bs{k})\partial_{k_x}  n(\bs{k}-\bs{G}_1)\big),
\end{align}
and so the integral across the boundary region is
\begin{align}
	\int_{- \delta}^{0} d\Delta_{k_x}& \Omega^{\text{RW}}(\bs{k}^{\text{RW}}+\Delta_{k_x})\notag \\
    \approx& \big( \partial_{k_y} \theta_{ -\bs{G}_1 \bs{0}} +A_y^{\text{P}}(\bs{k}^{\text{RW}})-A_y^{\text{P}}(\bs{k}^{\text{RW}}-\bs{G}_1)\big)\notag   \\
	&\times \int_0^{\delta} d\Delta_{k_x}	 \big(n(\bs{k}^{\text{RW}}-\bs{G}_1+\Delta_{k_x})\partial_{k_x}  n(\bs{k}^{\text{RW}}+\Delta_{k_x}) - n(\bs{k}^{\text{RW}}+\Delta_{k_x})\partial_{k_x}  n(\bs{k}^{\text{RW}}-\bs{G}_1+\Delta_{k_x})\big) . 
\end{align}

Noting that $\bs{k}^{\text{RW}}-\bs{G}_1=\bs{k}^{\text{LW}}$, and $\theta_{-\bs{G}_1 \bs{0}}(\bs{k}^{\text{LW}}+\bs{G}_1) =\theta_{\bs{0} \bs{G}_1}(\bs{k}^{\text{LW}})$ from the continuity of Bloch states at the 1BZ boundary, this integral can be written as
\begin{align}
	\int_{- \delta}^{0} d\Delta_{k_x} &\Omega^{\text{RW}}(\bs{k}^{\text{RW}}+\Delta_{k_x}) \notag \\
    \approx &\big( \partial_{k_y} \theta_{ \bs{0}\bs{G}_1 } +A_y^{\text{P}}(\bs{k}^{\text{LW}}+\bs{G}_1)-A_y^{\text{P}}(\bs{k}^{\text{LW}})\big)\notag   \\
	&\times \int_0^{\delta} d\Delta_{k_x}	 \big(n(\bs{k}^{\text{LW}}+\Delta_{k_x})\partial_{k_x}  n(\bs{k}^{\text{LW}}+\bs{G}_1+\Delta_{k_x}) - n(\bs{k}^{\text{LW}}+\bs{G}_1+\Delta_{k_x})\partial_{k_x}  n(\bs{k}^{\text{LW}}+\Delta_{k_x})\big) . 
\end{align}

This allows us to combine this integral with the one from Eq.~\eqref{equation_Berry_LW}, into one smooth integral across the 1BZ boundary
\begin{align}
	\int_{-\delta}^{\delta}d\Delta_{k_x}&\Omega^{W_1}(\bs{k}+\Delta_{k_x}) \notag \\
    \approx&\big( \partial_{k_y} \theta_{ \bs{0}\bs{G}_1 } +A_y^{\text{P}}(\bs{k}^{\text{LW}}+\bs{G}_1)-A_y^{\text{P}}(\bs{k}^{\text{LW}})\big)\notag   \\
	&\times \int_{-\delta}^{\delta} d\Delta_{k_x}	 \big(n(\bs{k}^{\text{LW}}+\Delta_{k_x})\partial_{k_x}  n(\bs{k}^{\text{LW}}+\bs{G}_1+\Delta_{k_x}) - n(\bs{k}^{\text{LW}}+\bs{G}_1+\Delta_{k_x})\partial_{k_x}  n(\bs{k}^{\text{LW}}+\Delta_{k_x})\big),
\end{align}
where $W_1$ refers to the combination of the left and right boundaries of the BZ (there will be three such wires in total). This smooth combination is to be expected because the position of the crystal BZ boundary is somewhat arbitrary (the BZ should be thought of as a torus), and we just happened to choose our representation of the BZ to be the Wigner-Seitz 1BZ which coincides with the position where the occupation of the parent band changes. Now we can use the fact that $n(\bs{k}^{\text{LW}}+\Delta_{k_x})+ n(\bs{k}^{\text{LW}}+\bs{G}_1+\Delta_{k_x}) =1$ to write 
\begin{align}
\big(n(\bs{k}^{\text{LW}}+\Delta_{k_x})\partial_{k_x} & n(\bs{k}^{\text{LW}}+\bs{G}_1+\Delta_{k_x}) - n(\bs{k}^{\text{LW}}+\bs{G}_1+\Delta_{k_x})\partial_{k_x}  n(\bs{k}^{\text{LW}}+\Delta_{k_x})\big) \notag \\
	 &= n(\bs{k}^{\text{LW}}+\Delta_{k_x})(- \partial_{k_x}  n(\bs{k}^{\text{LW}}+\Delta_{k_x})) - (1-n(\bs{k}^{\text{LW}}+\Delta_{k_x}) )\partial_{k_x}  n(\bs{k}^{\text{LW}}+\Delta_{k_x})\notag \\
	&=-\partial_{k_x}  n(\bs{k}^{\text{LW}}+\Delta_{k_x}).
\end{align}

This can then be readily integrated across the boundary:
\begin{align}
	\int_{-\delta}^{\delta} d\Delta_{k_x}	\big(-\partial_{k_x}  n(\bs{k}^{\text{LW}}+\Delta_{k_x})\big) &= [-n(\bs{k}^{\text{LW}}+\Delta_{k_x})]^{\delta}_{-\delta} \notag\\
	&= -1.
\end{align}
Therefore, the integral of the Berry curvature across the wire is
\begin{align}
	\int_{-\delta}^{\delta}\Omega^{W_1}(\bs{k}^{\text{LW}}+\Delta_{k_x}) d\Delta_{k_x}=-\big(& \partial_{k_y} \theta_{ \bs{0}\bs{G}_1 } +A_y^{\text{P}}(\bs{k}^{\text{LW}}+\bs{G}_1)-A_y^{\text{P}}(\bs{k}^{\text{LW}})\big).
\end{align}

The total contribution to the Berry flux from the wire can then be obtained by integrating this along the wire (along $y$):
\begin{align}
	\Phi_{W_1}&=-\int_{W_1} dk_y \big( \partial_{k_y} \theta_{ \bs{0}\bs{G}_1 } +A_y^{\text{P}}(\bs{k}^{\text{LW}}+\bs{G}_1)-A_y^{\text{P}}(\bs{k}^{\text{LW}})\big)\\
	&= - \int_{W_1}  dk_y \partial_{k_y} \theta_{ \bs{0}\bs{G}_1 } + \int_{W_1} dk_y (A_y^{\text{P}}(\bs{k}^{\text{LW}}) - A_y^{\text{P}}(\bs{k}^{\text{LW}}+\bs{G}_1)).
\end{align}

Now, note that $\int dk_y (A_y^{\text{P}}(\bs{k}^{\text{LW}}) - A_y^{\text{P}}(\bs{k}^{\text{LW}}+\bs{G}_1))$ is the integral of the parent Berry connection along two of the six edges of the 1BZ, in a clockwise manner. By Stoke's theorem, together with the sixfold rotational symmetry, this is minus one third the total integral of the parent Berry flux through the 1BZ. That is, $\int dk_y (A_y^{\text{P}}(\bs{k}^{\text{LW}}) - A_y^{\text{P}}(\bs{k}^{\text{LW}}+\bs{G}_1))= -\frac{1}{3} \Phi^{\text{parent}}_{BZ}$.

From the sixfold symmetry, the total Berry flux contribution from $W_1$ is one-third the contribution from the entire boundary of the 1BZ. This means that the $-\frac{1}{3} \Phi^{\text{parent}}_{BZ}$ term becomes $- \Phi^{\text{parent}}_{BZ}$. This part of the boundary contribution then cancels the total Berry flux from the bulk of the 1BZ (because the Berry curvature in the bulk is the parent Berry curvature in the 1BZ, from Eq.~\eqref{Eq_Berry_curvature_bulk}). As such, the total Berry flux of the crystal is
\begin{align}
	\Phi&= 3 \Phi_{W_1} + \int d^2k \Omega_{\text{bulk}}\\
	&=-3 \int_{W_1}  dk_y \partial_{k_y} \theta_{ \bs{0}\bs{G}_1 } .
\end{align}
Here it is important to note that $\theta$ is a compact variable, and so we cannot immediately use the fact that this is the integral of an exact derivative.

We note that this is the same result obtained in Ref. \cite{bernevig2025berry} via the sliver-patch model, albeit with the need to pick a smooth gauge removed. On the other hand, Ref. \cite{bernevig2025berry} gives a more careful treatment of the 1BZ corners, illustrating that even as a third state becomes significant near the corner, the contribution to the flux only depends on the phase winding along the wire. This proof utilizes $C_6$ symmetry and $\mathcal{M}_x\mathcal{T}$ symmetry, which are also present here. We can therefore use the phase winding $\partial_{k_y} \theta_{ \bs{0}\bs{G}_1 }$ to determine the flux, even close to the corner. However, we do note that $\partial_{k_y} \theta_{\bs{0} \bs{G}_1}$ will diverge close to the corner, which will significantly affect the total Berry flux (and indeed, this feature is necessary to obtain a quantized Chern number). 

 So far, our expression holds for either a weak interaction or a weak periodic potential. The difference between these two cases is how the phase difference $\theta_{\bs{0} \bs{G}_1}$ is determined. In the interacting case, its derivative is related to the Berry connection of the parent band (assuming that the small momentum transfer component of the Fock term dominates) \cite{dong2024stability, bernevig2025berry}. On the other hand, for a weak periodic potential, the phase $\theta_{\bs{0} \bs{G}_1}$ is chosen to minimize the hopping energy in Eq.~\eqref{eq:pp_hopping}. When only two states are involved (i.e., away from the corners), this phase matches the form factor: $\theta_{\bs{0}\bs{G}} = \arg(\mathcal{F}(\bs{k}, \bs{k}+\bs{G}))$. However, at the corners there is an interaction between three states, which generically frustrates the hopping terms. From the $C_6$ symmetry, the phase difference $\theta_{\bs{0} \bs{G}_1}$ is locked to a multiple of $2 \pi/3$ (the phases $\theta_{\bs{0} \bs{G}_1}$, $\theta_{\bs{G}_1 \bs{G}_2}$, and $\theta_{\bs{G}_2 \bs{G}_0}$ are equal by symmetry, and must add to a multiple of $2 \pi$). As a result, the phase will round to the nearest multiple of $2 \pi/3$. This tells us that the phase winding along the wire is
\begin{align}
	W_{\text{capped}}&= \int_{W_1} \partial_{k_y}\arg(\mathcal{F}(\bs{k}, \bs{k}+\bs{G}_1)) +  \arg(\mathcal{F}(\bs{K}_2', \bs{K}_2'+\bs{G}_1)) - \lfloor \arg(\mathcal{F}(\bs{K}_2', \bs{K}_2'+\bs{G}_1))\rceil _{2 \pi /3} \notag \\
	&- \arg(\mathcal{F}(\bs{K}_2, \bs{K}_2+\bs{G}_1)) + \lfloor\arg(\mathcal{F}(\bs{K}_2, \bs{K}_2+\bs{G}_1))\rceil_{2 \pi /3}, \\
	&= \int_{W_1} \partial_{k_y}\arg(\mathcal{F}(\bs{k}, \bs{k}+\bs{G}_1)) +  \arg(\mathcal{F}(\bs{K}_2', \bs{K}_3')) - \arg(\mathcal{F}(\bs{K}_2, \bs{K}_1)) \notag \\
	&+ \lfloor\arg(\mathcal{F}(\bs{K}_2, \bs{K}_1))\rceil_{2 \pi /3} - \lfloor \arg(\mathcal{F}(\bs{K}_2', \bs{K}_3'))\rceil _{2 \pi /3} ,
\end{align}
where $\lfloor ...\rceil_{2\pi/3}$ denotes rounding to the nearest multiple of $2 \pi/3$, and $\bs{K}_2'$ and $\bs{K}_2$ are the corners of the wire (see Fig.\ref{Figure_wire}). The Chern number is

\begin{align} 
	\mathcal{C}&=-\frac{3}{2 \pi} W_{\text{capped}}\\
	&= -\frac{3}{2\pi} \big[ \int_{W_1} \partial_{k_y}\arg(\mathcal{F}(\bs{k}, \bs{k}+\bs{G}_1)) dk_y +  \arg(\mathcal{F}(\bs{K}_2', \bs{K}_3')) - \arg(\mathcal{F}(\bs{K}_2, \bs{K}_1)) \big] \notag \\
	&-\lfloor \frac{3}{2 \pi} \arg(\mathcal{F}(\bs{K}_2, \bs{K}_1))\rceil_{1} + \lfloor \frac{3}{2 \pi} \arg(\mathcal{F}(\bs{K}_2', \bs{K}_3'))\rceil_{1} . \label{Equation_Chern_low_U_capped}
\end{align}

We can see that this is quantized, because $\big[ \int_{W_1} \partial_{k_y}\arg(\mathcal{F}(\bs{k}, \bs{k}+\bs{G}_1)) +  \arg(\mathcal{F}(\bs{K}_2', \bs{K}_3') - \arg(\mathcal{F}(\bs{K}_2, \bs{K}_1)) \big]$ is the difference between the integral of a derivative and its values at the end points, which must be a multiple of $2\pi$ for an angle, and the latter two terms are explicitly rounded to integers.

\subsection{Form factor edge winding in the $\lambda_N$-jellium model}
	\label{Appendix_projector_lambda_N}
	We can now apply our expressions for the Berry curvature to the $\lambda_N$-jellium model. We consider the low $U_0$ case described in Section \ref{Section_weak_potential_projector}, with the form factor given by \eqref{eq_form_factor_explicit}
	\begin{align*}
		\mathcal{F}_N(\bs{k}, \bs{k}')= \mathcal{N}_{\bs{k}} \mathcal{N}_{\bs{k}'} \sum_{n=0}^{N-1} \frac{(\mathcal{B} k_z k'_{\bar{z}})^n}{n!} 
	\end{align*}
	where $k_z = \frac{1}{\sqrt{2}} (k_x+ik_y)$. We are only interested in the phase of the form factor, so we can drop the normalization constants. Then we note that $k_z = \frac{1}{\sqrt{2}}( -\frac{g}{2} +i k_y)$ and $(k+G)_z= \frac{1}{\sqrt{2}}( +\frac{g}{2} +i k_y)$. Therefore $(k+G)_{\bar{z}}= -k_z$. As a result, we can use the unnormalized form factor
	
	\begin{align*}
		\tilde{\mathcal{F}}_N(\bs{k}, \bs{k}+\bs{G}) = \sum_{n=0}^{N-1} \frac{\mathcal{B}^n}{n! \times 2^n}(-1)^n (-\frac{g}{2} + i k_y)^{2n}
	\end{align*}
	
	Now we can use $\frac{1}{\mathcal{F}_N}\partial_{k_y} |\mathcal{F}_N|e^{i \theta}= i \partial_{k_y} \theta +\frac{\partial_{k_y}|\mathcal{F}_N|}{|\mathcal{F}_N|}$ (assuming the form factor never becomes zero on the wire) to obtain
	
	$$ \partial_{k_y} \arg(\mathcal{F}_N(\bs{k}, \bs{k}+\bs{G})) = \frac{1}{	i\tilde{\mathcal{F}}_N(\bs{k}, \bs{k}+\bs{G})} \partial_{k_y} \tilde{\mathcal{F}}_N(\bs{k}, \bs{k}+\bs{G}) - \frac{1}{2i |\tilde{\mathcal{F}}_N(\bs{k}, \bs{k}+\bs{G})|^2} \partial_{k_y}|\tilde{\mathcal{F}}_N(\bs{k}, \bs{k}+\bs{G})|^2 $$
Here,
	\begin{align*}
		\partial_{k_y} \tilde{\mathcal{F}}_N(\bs{k}, \bs{k}+\bs{G}) &=  i\sum_{n=1}^{N-1} \frac{\mathcal{B}^n}{(n-1)! \times 2^{n-1}}(-1)^n(- \frac{g}{2} + i k_y)^{2n-1}\\
		&=  -i\mathcal{B} (- \frac{g}{2} + i k_y)\sum_{n=1}^{N-1} \frac{\mathcal{B}^{n-1}}{(n-1)! \times 2^{n-1}}(-1)^{n-1}(- \frac{g}{2} + i k_y)^{2n-2}\\
		&=-i\mathcal{B}(-\frac{g}{2}+i k_y) \sum_{n=0}^{N-2}\frac{\mathcal{B}^n}{n! \times 2^n}(- \frac{g}{2} + i k_y)^{2n}\\
		&=-i\mathcal{B}(-\frac{g}{2}+i k_y)  \tilde{\mathcal{F}}_{N-1}(\bs{k}, \bs{k}+\bs{G}).
	\end{align*}

Note that taking $k_y \rightarrow -k_y$ conjugates the form factor. This makes $|\mathcal{F}_N(\bs{k}, \bs{k}+\bs{G})|$ symmetric under mirroring in the $x$-axis, so the contribution to the integral from the derivative of $|\tilde{\mathcal{F}}_N(\bs{k}, \bs{k}+\bs{G})|^2$ will vanish. As a result,	
$$\int_{W_1} \partial_{k_y} \arg(\mathcal{F}_N(\bs{k}, \bs{k}+\bs{G})) = \int_{K_2'}^{K_2} dk_y \mathcal{B}(\frac{g}{2}-i k_y) \frac{\tilde{\mathcal{F}}_{N-1}(\bs{k}, \bs{k}+\bs{G})}{\tilde{\mathcal{F}}_{N}(\bs{k}, \bs{k}+\bs{G})}.$$

We can perform this integral numerically to predict the Chern number for a range of $N$. The results of this agree well with the exact diagonalization results shown in Figure \ref{fig:phase_diagram_periodic_potential}. We can also gain some insights into particular features of the phase diagram. One notable point is that the Chern number sometimes changes sign, without going through an intermediate phase with $\mathcal{C}=0$. As an example of this, consider the simplest case, with $N=2$. In this case, $\tilde{\mathcal{F}}_{N-1}=1$ and 

$$\tilde{\mathcal{F}}_N(\bs{k}, \bs{k}+\bs{G})=1-\frac{\mathcal{B}}{2}(\frac{g}{2}-ik_y)^2= 1-\frac{\mathcal{B}}{2}(g^2/4 -k_y^2)+i\frac{\mathcal{B}}{2} k_y g$$

Therefore,
\begin{align*}
    \frac{\tilde{\mathcal{F}}_{N-1}(\bs{k}, \bs{k}+\bs{G})}{\tilde{\mathcal{F}}_{N}(\bs{k}, \bs{k}+\bs{G})} = \frac{1-\frac{\mathcal{B}}{2}(g^2/4 -k_y^2)-i\frac{\mathcal{B}}{2} k_y g}{ (1-\frac{\mathcal{B}}{2}(g^2/4 -k_y^2))^2 +\frac{\mathcal{B}^2 g^2k_y^2}{4}}
\end{align*}

When we integrate, we only need to keep the parts that are even in $k_y$, so we have
\begin{align}
    \int_{\text{wire 1}} \partial_{k_y} \arg(\mathcal{F}_N(\bs{k}, \bs{k}+\bs{G})) &=  \int dk_y \frac{\mathcal{B}g}{2} \frac{1-\frac{\mathcal{B}}{2}(g^2/4 -k_y^2)}{ (1-\frac{\mathcal{B}}{2}(g^2/4 -k_y^2))^2 +\frac{\mathcal{B}^2 g^2k_y^2}{4}} -  \frac{\mathcal{B}k_y}{2} \frac{\frac{\mathcal{B}}{2} k_y g}{ (1-\frac{\mathcal{B}}{2}(g^2/4 -k_y^2))^2 +\frac{\mathcal{B}^2 g^2k_y^2}{4}}\notag\\
    &=\int dk_y \frac{\mathcal{B}g}{2} \frac{1-\frac{\mathcal{B}}{2}(g^2/4)}{ (1-\frac{\mathcal{B}}{2}(g^2/4 -k_y^2))^2 +\frac{\mathcal{B}^2 g^2k_y^2}{4}}. 
\end{align}

The numerator switches sign when $\mathcal{B}g^2/8 =1$, and there is a pole in the same position of the denominator. Using $A_{1BZ} = \frac{\sqrt{3}}{2} g^2$, this corresponds to $\mathcal{B}A_{1BZ}=4 \sqrt{3} \approx 6.9$, which agrees well with the point at which the Chern number switches in the phase diagram, Figure \ref{fig:phase_diagram_periodic_potential}.

Another example to consider is the infinite Chern band, which is obtained in the $N \rightarrow \infty$ limit. In this case $\tilde{\mathcal{F}}_{N-1}/ \tilde{\mathcal{F}}_N =1$, giving us

$$-3 \int_{W_1} dk_y  \partial_{k_y} \arg(\mathcal{F}_{\infty}(\bs{k}, \bs{k}+\bs{G})) = -3 \int dk_y \mathcal{B}(g/2 - ik_y)= -3 \int dk_y \mathcal{B}g/2=-3 \mathcal{B}g/2 \cdot L= - \frac{\sqrt{3}}{2}\mathcal{B}g^2 = -\mathcal{B}A_{1BZ}.$$

Then we must include the additional terms in Eq.~\eqref{Equation_Chern_low_U_capped} from the phase rounding at the corners. The phase at the start of the wire is given by 

\begin{align*}
    \arg(\mathcal{F}_{\infty}(\bs{K}_2', \bs{K}_2'+\bs{G})) &= \arg(\mathcal{F}_{\infty}(-\frac{g}{2}\hat{x} -\frac{g}{2 \sqrt{3}}\hat{y}, \frac{g}{2}\hat{x} -\frac{g}{2 \sqrt{3}}\hat{y} ))\\
    &=\arg( e^{-i\frac{\mathcal{B}}{2} (-\frac{g}{2}\hat{x} -\frac{g}{2 \sqrt{3}}\hat{y}) \times(\frac{g}{2}\hat{x} -\frac{g}{2 \sqrt{3}}\hat{y}   )})\\
    &=-\frac{\mathcal{B}}{2} \frac{g^2}{2 \sqrt{3}}\\
    &=-\frac{\mathcal{B}A}{6}.
\end{align*}

Similarly, the phase at the end of the wire is $+\frac{\mathcal{B}A}{6}$. Therefore, the total winding is
\begin{align*}
    W_{\text{capped}}&= \mathcal{B}A_{1BZ}/3 + (-\mathcal{B}A_{1BZ}/6 + \lfloor \mathcal{B}A_{1BZ}/6\rceil_{2 \pi/3}) - (\mathcal{B}A_{1BZ}/6 - \lfloor \mathcal{B}A_{1BZ}/6\rceil_{2 \pi/3})\\
    &=2\lfloor \mathcal{B}A_{1BZ}/6\rceil_{2 \pi/3}.
\end{align*}

As a result, the Chern number is
\begin{align}
    C&= -\frac{3}{2 \pi}W_{\text{capped}}\notag\\
    &= - \frac{6}{2 \pi}\lfloor \mathcal{B}A_{1BZ}/6\rceil_{2 \pi/3}\notag\\
    &= -\lfloor\frac{\mathcal{B}A_{1BZ}}{2 \pi}\rceil_{\frac{2 \pi}{3} \cdot \frac{6}{2 \pi}}\notag\\
    &=-\lfloor\frac{\mathcal{B}A_{1BZ}}{2 \pi}\rceil_{2}.
\end{align}

We therefore see that the Chern number is given by rounding the Berry flux enclosed. However, unlike the interacting case, the negative of the Berry flux is rounded, and it is rounded to the nearest even integer (in units of $2 \pi$). This matches the results from Ref. \cite{tan2024parent}. We note that this type of flux rounding is not generically present for $\lambda_N$-jellium, because the result depends on the winding of the form factor, not the total Berry flux.

%%%%%%%%%%%%%%%%%%%%%%%%%%%%%%%%%%%%%%
\section{Generalized occupation-weighted flux rounding}\label{app:generalized_flux_rounding}

%%%%%%%%%%%%%%%%%%%%%%%%%%%%%%%%%%%%%%%%%%%%%%%%%%%%%%%%%%%
\begin{figure}
    \centering
    \includegraphics[width=1.0\textwidth]{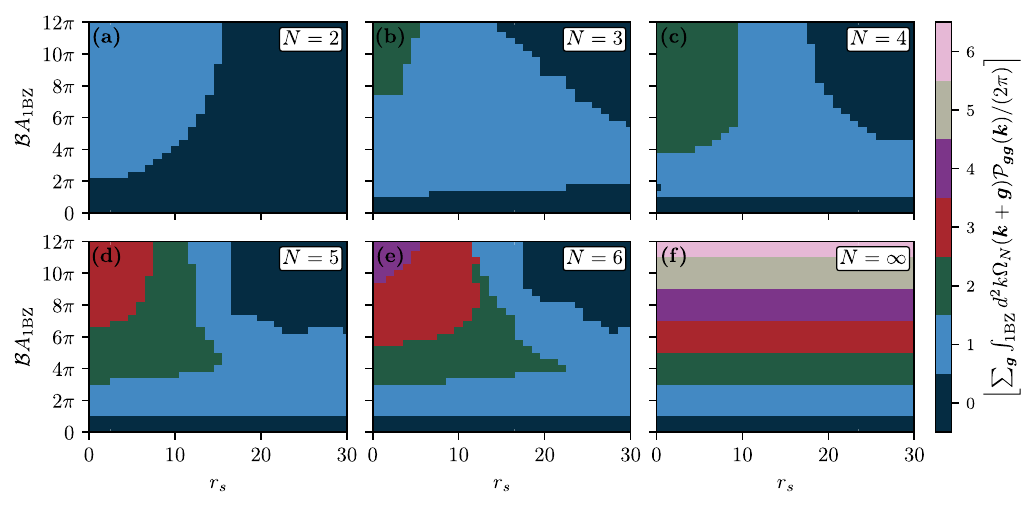}\vspace{-3mm}
    \caption{Numerical evaluation of the rounded occupation-weighted Berry curvature \eqref{eq:occupation_weighted_flux_rounding} using our Hartree-Fock results (Fig.~\ref{fig:hf_phase_diagram}) for the $\lambda_N$ jellium model~\eqref{eq:lambda_n_int} with (a) $N=2$, (b) $N=3$, (c) $N=4$, (d) $N=5$, (e) $N=6$, (f) $N\to\infty$ as a function of $\mathcal{B}A_{\text{1BZ}}$ and $r_s$.}
\label{fig:generalized_flux_rounding}
\end{figure}
%%%%%%%%%%%%%%%%%%%%%%%%%%%%%%%%%%%%%%%%%%%%%%%%%%%%%%%%%%%

Fig.~\ref{fig:fig_symmetry_indicators} shows the numerical evaluation of Eq.~\eqref{eq:occupation_weighted_flux_rounding} using our Hartree-Fock results presented in Fig.~\ref{fig:hf_phase_diagram}. Comparing with the Chern number of the electronic crystals in Fig.~\ref{fig:hf_phase_diagram}, we see that the Chern number predicted by the generalized occupation-weighted flux rounding~\eqref{eq:occupation_weighted_flux_rounding} is not in particularly good agreement with observations for $N=2$, but works well for $N\ge 3$. 

\end{document}